\documentclass{aa}
\usepackage{txfonts}
\usepackage{natbib}
\bibpunct{(}{)}{;}{a}{}{,}
\usepackage{graphicx}
\usepackage{placeins}

\newcommand{\teff}{$T_{\rm{eff}}$}
\newcommand{\reff}{$R_{\rm{eff}}$}

\newcommand{\lL}{\ifmmode \log \frac{L}{L_{\sun}} \else $\log \frac{L}{L_{\sun}}$\fi}
\newcommand{\mdot}{$\dot{M}$}
\newcommand{\lmdot}{$\log \dot{M}$}
\newcommand{\myr}{M$_{\sun}$ yr$^{-1}$}

\newcommand{\vinf}{$\varv_{\infty}$}

\newcommand{\kms}{km~s$^{-1}$}
\newcommand{\rsun}{R$_{\sun}$}
\newcommand{\msun}{M$_{\sun}$}
\newcommand{\lsun}{L$_{\sun}$}

\newcommand{\ha}{H$\alpha$}
\newcommand{\hb}{H$\beta$}
\newcommand{\hg}{H$\gamma$}
\newcommand{\heiiuv}{\ion{He}{ii}~1640}
\newcommand{\heiiopt}{\ion{He}{ii}~4686}

\newcommand{\nivopta}{\ion{N}{iv}~4058}

\newcommand{\nvopt}{\ion{N}{v}~4604-20}

\newcommand{\niiibb}{\ion{N}{iii}~4634-40-42}

\begin{document}

\title{Surface chemical composition of single WNh stars}
\author{Fabrice Martins\inst{1}  
}
\institute{LUPM, Universit\'e de Montpellier, CNRS, Place Eug\`ene Bataillon, F-34095 Montpellier, France  \\
           \email{fabrice.martins@umontpellier.fr}
}

\offprints{Fabrice Martins\\ \email{fabrice.martins@umontpellier.fr}}

\date{Received / Accepted }

\abstract
{Wolf-Rayet (WR) stars of the WNh category contain a significant fraction of hydrogen at their surface. They can be hydrogen-burning, very massive stars or stars in a post-main sequence phase of evolution. Also, WNh stars are sometimes not included in population synthesis models. }
{We aim to better characterise the properties of single WNh stars in the Galaxy and the Magellanic Clouds. In particular, we want to constrain their surface chemistry beyond the hydrogen content by determining the helium, carbon, and nitrogen surface abundances.}
{We perform a spectroscopic analysis of 22 single WNh stars. We fit their ultraviolet and/or optical spectra using synthetic spectra computed with the code CMFGEN. We determine the main stellar parameters (temperature, luminosity, mass-loss rates) and the surface H, He, C, and N mass fractions. We investigate the ability of current evolutionary models to reproduce all parameters at the same time.}
{We find that all WNh stars show the signatures of CNO-cycle material at their surface: they are carbon-depleted and nitrogen-rich. A clear trend of higher nitrogen content at higher metallicity is observed, as expected. The amount of hydrogen (X) varies significantly from one star to another, independently of luminosity. Values of X larger than 0.4 are not exceptional. The majority of Galactic WNh stars can be explained by evolutionary models, provided sufficient fine-tuning of the input parameters of evolutionary calculations. At lower metallicity, most stars escape predictions from evolutionary models. This has been noted in the literature but constraints on the surface nitrogen content exacerbate this severe issue. }
{Our study highlights the need to refine the treatment of WR stars in both stellar evolution and population synthesis models.}

\keywords{Stars: massive -- Stars: Wolf-Rayet  -- Stars: atmospheres -- Stars: fundamental parameters -- Stars: abundances -- Stars: evolution}

\authorrunning{F. Martins}
\titlerunning{Surface abundances of WNh stars}

\maketitle

\section{Introduction}
\label{s_intro}

Wolf-Rayet stars (WR) are the descendants of stars born as O and early B stars on the zero-age main sequence (ZAMS; see \citealt{crowther07} for a review). They are characterised primarily by their spectroscopic appearance. WR stars show numerous emission lines from the ultraviolet (UV) to the optical and the infrared (IR) wavelength range. Their strong stellar winds are responsible for these emissions. Radiative acceleration is the primary mechanism driving these winds but a complex physics is at work, involving the proximity to the Eddington limit, multiple scattering, opacity variations across the atmosphere, and the presence of inhomogeneities (clumping); see \citet{gh05}, \citet{gh08}, \citet{graefener11}, \citet{sv20}, and \citet{sander22}.

WR stars are classified in three different spectroscopic categories, referred to as WN, WC, and WO. These are distinguished by the dominant emission lines in their spectra: mostly helium and nitrogen for WN stars, helium and carbon for WC stars, and finally strong oxygen lines in WO stars \citep{beals38,hs66,smith68,ssm90,ssm96}. The general understanding of these sequences is that they correspond to different degrees of chemical processing appearing at the surface of the star. WN stars show mostly the products of hydrogen burning through the CNO cycle, and are therefore nitrogen rich, while WC stars display helium-burning products at their surface. WO stars may correspond to an even more advanced phase \citep{aadland22} but their very hot temperatures also play a role in their appearance \citep{tramper13}.

The exposure of nucleosynthesis products at the surface of WR stars is caused by two main processes. First, WR stars are in an advanced evolutionary state compared to OB stars and therefore products of nucleosynthesis in the core have been transported outward by internal mixing, such as that induced by stellar rotation \citep[e.g.][]{mm05}. Second, their external layers have been removed, meaning that the stellar surface and photosphere are pushed inward to more chemically processed layers. Strong stellar winds can be the cause, and this evolution is referred to as the Conti scenario \citep{conti76,mc94}. Alternatively, stripping of the external layers by a close companion in binary systems can lead to the same outcome \citep{vanb98}. Both effects can be at work in advanced phases of binary evolution \citep{shenar19}. 

In addition to these general properties, some WR stars of the WN category display hydrogen lines in their spectra. A WNh spectral type is used to identify them \citep{ssm96,smithWNH}. No WC or WO stars show hydrogen. Given our understanding of WR star physics, this is consistent with WNh stars being objects showing the products of CNO cycle nucleosynthesis. Depending on the formation process (stellar winds, binary stripping, or both) and the extent to which the external layers have been removed, different degrees of chemical processing can be exposed to the surface. WNh stars can therefore be classical WR stars in a relatively early phase of evolution \citep{hamann06,hainich14}. However, another category of WNh objects is observed in very luminous and massive stars. These latter are found in young massive clusters such as R136 in the Large Magellanic Cloud (LMC) or the Arches in the Galactic center \citep{mh98,figer02}. Spectroscopic analysis of some of these objects indicates that they are very likely main sequence objects, that is, they are hydrogen-burning, with strong stellar winds \citep{martins08,besten20}. The proximity to the Eddington limit increases their mass-loss rates so that they develop spectra dominated by emission lines typical of WR stars \citep{gh08,vink11,besten20a,graef21}. However, in terms of evolution, they remain similar to lower-mass main sequence OB stars. The most massive stars known, whether single or in binary systems, are all classified as WNh \citep{schnurr08,schnurr09,crowther10,besten11}. The rare class of WN3/O3 objects discovered by \citet{neugent12} and \citet{massey17} and analysed by \citet{neugent17} also show relatively high hydrogen mass fractions. 

Stellar evolution models deal with WR stars mostly by applying different mass-loss rates when calculations enter this phase of evolution. The switch between normal OB-type winds \citep[usually from][]{vink01} and WR-type winds relies on the criteria defining the WR phases. A major criterion is the amount of hydrogen at the surface. The vast majority of evolutionary models consider a maximum hydrogen mass fraction of 0.3 or 0.4 for WR stars \citep{chen15,choi16,lc18,martinet23}. \citet{hainich14,hainich15} showed that, in the Magellanic Clouds, a significant fraction of WN stars have more hydrogen than this limit. This raises the question of the treatment of WR phases by stellar evolution codes. Hainich et al. also showed that evolutionary models faced serious problems in accounting for the position of WN stars in the Hertzsprung-Russell diagram (HRD). At low metallicity, models usually do not account for the position in the HRD and the surface hydrogen content. 
A related issue is encountered in population synthesis models. WR spectra are only assigned to stars that have relatively low hydrogen mass fractions \citep{leitherer14,bpass}. H-rich WN stars are therefore left out, which can lead to severe limitations, especially when very massive stars (VMSs) are present. \citet{mp22} showed that such stars must be properly taken into account in order to reproduce key features of young massive clusters, especially \heiiuv. 

Understanding WNh stars is therefore important for stellar evolution and population studies. Their surface chemistry tells us about their nature and past history. If the hydrogen content has been determined for relatively large samples \citep{hamann06,hainich14,hainich15,shenar19}, constraints on other elements are rather limited. \citet{martins08} provided helium, carbon, and nitrogen abundances for stars in the Arches cluster and showed that they are objects in the H-burning phase. \citet{brands22} found similar results for the WNh stars in R136. \citet{shenar16,shenar19} studied the surface chemistry of binary stars in the Magellanic Clouds. In the present study, we extend the determination of surface abundances to a larger sample of single WNh stars in the Galaxy and the Magellanic Clouds. We aim to better understand their nature and the ability of current stellar evolution models to account for them. We focus on single stars. Our sample and the observational data are presented in Sect.~\ref{s_obs}. We explain our analysis method in Sect.~\ref{s_analysis} and present our results in Sect.~\ref{s_res}.  We discuss our findings in Sect.~\ref{s_disc} and outline our conclusions in Sect.~\ref{s_conc}.

\section{Sample and observational data}
\label{s_obs}

In this study, we focused on (presumably) single WNh stars. We selected objects in the Galaxy, the LMC, and the SMC. For the former we relied on the work of \citet{hamann06,hamann19} to identify WNh stars that lack evidence for the presence of a companion. In the LMC, the single WN stars have been studied by \citet{hainich14} and binarity has been investigated by \citet{foellmi03} and \citet{shenar19}. We cross-checked both studies to select our single LMC sources. Finally, for the SMC, we relied on the analysis of \citet{hainich15}, \citet{foellmi03b}, and \citet{shenar16} to select single WNh stars. We did not include the three stars analysed by \citet{martins09}, because these authors used the same method as the one used in the present study. We simply include the findings of these latter authors when discussing our results. In total, our sample is made up of 8 Galactic stars, 11 LMC stars, and 3 SMC stars.

Spectroscopic data all came from science archives or private communication for those not publicly available. The UV spectra were extracted from the IUE database hosted at MAST\footnote{\url{https://archive.stsci.edu/iue/search.php}}. Low- (6~\AA) or high-resolution (2~\AA) spectra exist, depending on the target. We favoured the latter when possible. SWP (1150-2000~\AA) and LWP (1850-3000~\AA) spectroscopy was used for spectral energy distribution (SED) fitting. SWP spectra were also used for spectroscopic analysis and parameter determination. Optical spectra came from various sources. The ESO database provided several FEROS and GIRAFFE spectra with a spectral resolution of about 0.1~\AA. FEROS data cover the whole optical range, while GIRAFFE data cover portions of it. We also relied on the data published by \citet{foellmi03b,foellmi03}, which are fully described in those papers. The spectral resolution is lower, typically 6~\AA. We stress that such a low resolution is not an issue for our analysis because the lines of the WNh stars are mostly formed in the stellar wind and are broad. All spectroscopic data were manually normalised to the continuum. The details of the spectroscopic data are given in Table~\ref{tab_obs}.

The spectroscopic data are complemented by optical and IR photometry for SED fitting. The data are taken from the \emph{SIMBAD} astronomical database\footnote{http://simbad.u-strasbg.fr/simbad/} \citep{simbad00}.

\section{Spectroscopic analysis}
\label{s_analysis}

We performed our spectroscopic analysis with the code CMFGEN \citep{hm98}, which solves the radiative transfer under non-Local Thermodynamical Equilibrium (non-LTE) conditions in a spherical geometry, and takes stellar winds into account. The calculations are done in two main steps. First the atmospheric structure is modelled. Line profiles are assumed to be Gaussian and have a width that is independent of the position in the atmosphere. In addition to the natural width, a microturbulent velocity of 20 \kms\ is assumed for all lines. Resolution of the statistical equilibrium and radiative transfer equations lead to the radiation field and the level populations. Radiative acceleration is calculated and used to solve the equation of motion in the inner atmosphere, below the critical point. The velocity structure in that part of the atmosphere results from the equation of mass conservation. It is connected to a $\beta$-velocity law\footnote{$v =$ \vinf\ $\times\ (1-R/r)^{\beta}$ with $R$ the stellar radius, $r$ the distance to the centre of the star, and \vinf\ the maximum velocity at the top of the atmosphere} in the outer atmosphere. We adopt $\beta$=1 for the analysis of all stars (see \citealt{lefever23} for a discussion on the effect of $\beta$). In a second step, a formal solution of the radiative transfer is performed with the level populations fixed. This provides the emergent spectrum. More realistic line profiles are used and a depth-dependent microturbulent velocity is used, which varies from 10~\kms\ at the stellar radius to 10\% of \vinf\ at the top of the atmosphere. Stark broadening is taken into account. In all computations, we used solar abundances from \citet{asplund09} for Galactic stars, or solar-scaled abundances for LMC and SMC stars. This is mainly relevant for elements with atomic numbers larger than that of oxygen, because for lighter elements we vary the abundances to find the best-fit values (see below). 

The final synthetic spectrum is convoluted with a Gaussian profile to take into account the finite spectral resolution of the observations. We also convolute the spectrum by a rotational profile of 40~\kms\ but stress that this does not affect the shape of the spectrum because (1) spectral resolution is low for most data and (2) spectral lines are mostly formed in the wind and are consequently broadened up to higher velocities than the rotational velocity.

To determine the main stellar parameters, we proceeded as follows:

\begin{itemize}

\item \textit{Effective temperature}: \teff\ was constrained by the ionisation balance method; that is, the relative strength of lines of successive ions of the same element. When both \ion{He}{i} and \ion{He}{ii} lines were present, we used them as the primary diagnostic. For the hottest stars with \teff\ above about 45000~K, \ion{He}{i} lines disappear and we switched to nitrogen lines. Nitrogen lines were also used as secondary indicators when \ion{He}{i} lines are detected. In practice, we relied on \ion{He}{i}~4026, \ion{He}{i}~4471, \ion{He}{i}~4713, \ion{He}{i}~4920, \ion{He}{ii}~4200, \ion{He}{ii}~4542, \ion{He}{ii}~5412, \niiibb, \nivopta,\ and \nvopt. \heiiopt\ was not considered for \teff\ determination, because in addition to temperature, its strength is widely driven by the wind density and the helium content (see below). Uncertainties on \teff\ depend on \teff\ itself. In the low temperature range, where several indicators are present, 2000~K is a typical error determination. In the high \teff\ range, especially for the SMC targets, only \ion{He}{ii} and \ion{N}{v} are available and the range of acceptable \teff\ is much wider before these lines become too weak or lower ionisation lines appear. In that range, an error of 20~000~K is possible.

\item \textit{Luminosity}: The SED of models was compared to flux-calibrated UV spectra and optical-infrared photometry to provide the stellar luminosity. In the process, an extinction had to be applied to the synthetic SEDs. The law of \citet{ccm89} is used for all stars, with R$_V$ set to 3.1 and E(B-V) as a free parameter. The extinction of LMC and SMC stars is low and mostly due to foreground Galactic extinction. The choice of a different law would barely affect the final luminosity, which depends mostly on effective temperature. For stars in the Large and Small Magellanic Cloud, we used distance moduli of 18.48 and 18.98, respectively \citep{pietr19,gra20}. For Galactic stars, we relied on the parallaxes from Gaia DR3 \citep{gaiadr3}. The uncertainty on \lL\ is mainly driven by that on \teff, meaning that an error of 0.05 to 0.15 dex was estimated for low and high temperatures, respectively.

\item \textit{Mass-loss rate and terminal velocity}: \vinf\ was determined from the blueward extent of UV P-Cygni profiles, when available. If not, the width of the main optical emission lines (\ha\ and \heiiopt) was used instead. In that case, \vinf\ was less accurately constrained than with the blueward extension of P-Cygni profiles. Mass-loss rate affects the intensity of emission lines; we therefore
used their strength ---measured by peak intensity--- as the primary diagnostic for \mdot. Optical lines from hydrogen and helium are also sensitive to the He/H abundance ratio (see below), meaning that the determination of \mdot\ was made together with the abundance analysis. \mdot\ was adjusted so that the synthetic spectrum reproduced the intensity of the largest number of emission lines. In practice, the strongest lines, that is, \heiiopt\ and \ha, dominated our  determination of the mass-loss rate.

\item \textit{Clumping factor}: We considered optically thin clumping using a volume-filling-factor formalism: $f = f_{\infty} + (1-f_{\infty})e^{-\frac{v}{v_{cl}}}$, with $f_{\infty}$ the filling factor at the top of the atmosphere (the terminal clumping factor), $v$ the velocity in the atmosphere, and $v_{cl}$ a velocity characterising the region in which clumping becomes significant (there is no clumping at the photosphere at low velocities). We adopted $v_{cl} = 100$ \kms. The terminal clumping factor $f_{\infty}$ is initially set to 0.1. The shape of the strongest optical emission lines, especially \heiiopt\ and \ha, was used to refine this parameter. As demonstrated by \citet{hil91}, the electron scattering red wing of these lines is sensitive to the degree of inhomogeneity (see also \citealt{martins09} for practical applications). For clarity, we stress that $f_{\infty}$ is a lower limit on the clumping factor at the formation depth of most lines, because they form over an extended region of the atmosphere where $f$ varies. 
  
\end{itemize}

\begin{figure*}[t]
\centering
\includegraphics[width=0.49\textwidth]{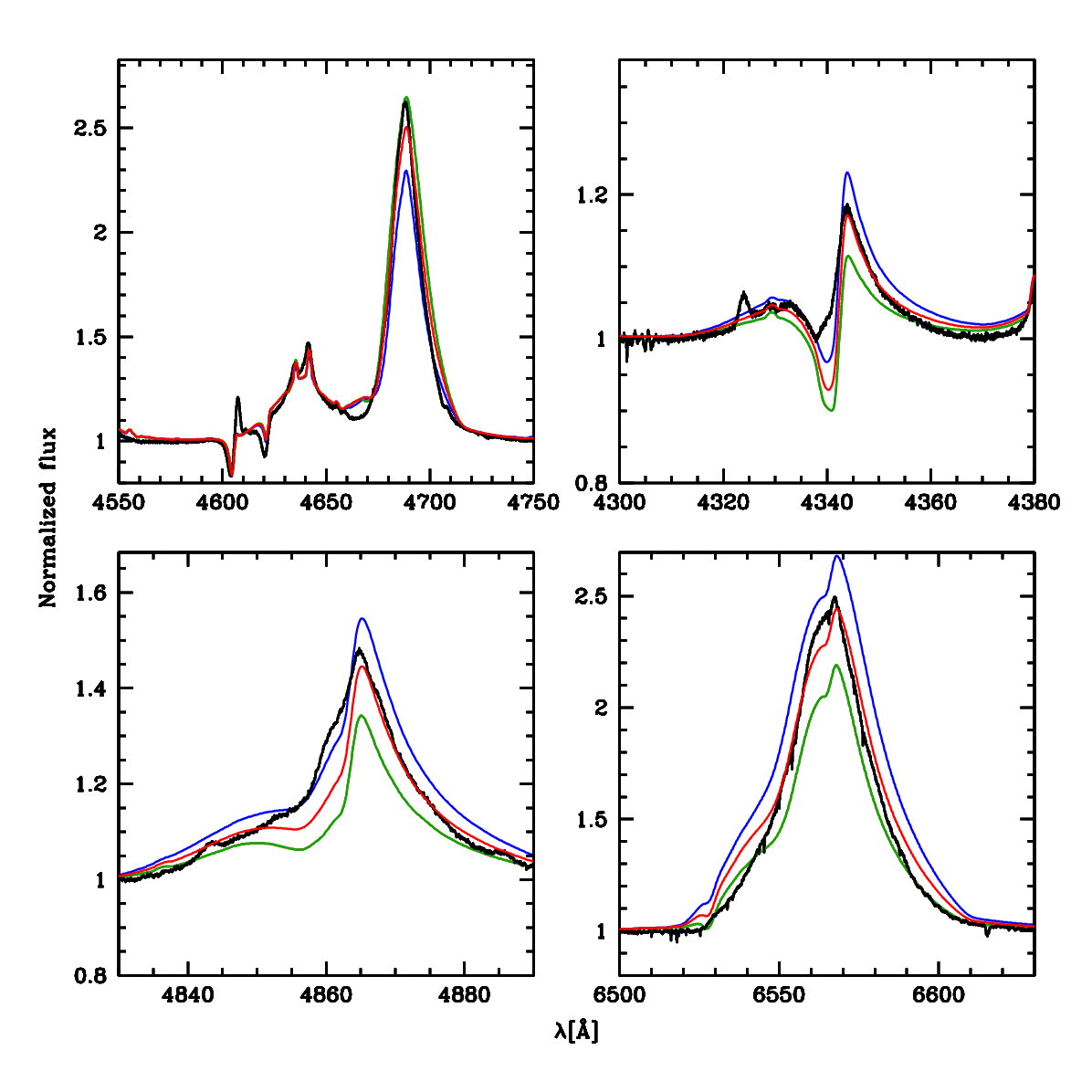}
\includegraphics[width=0.49\textwidth]{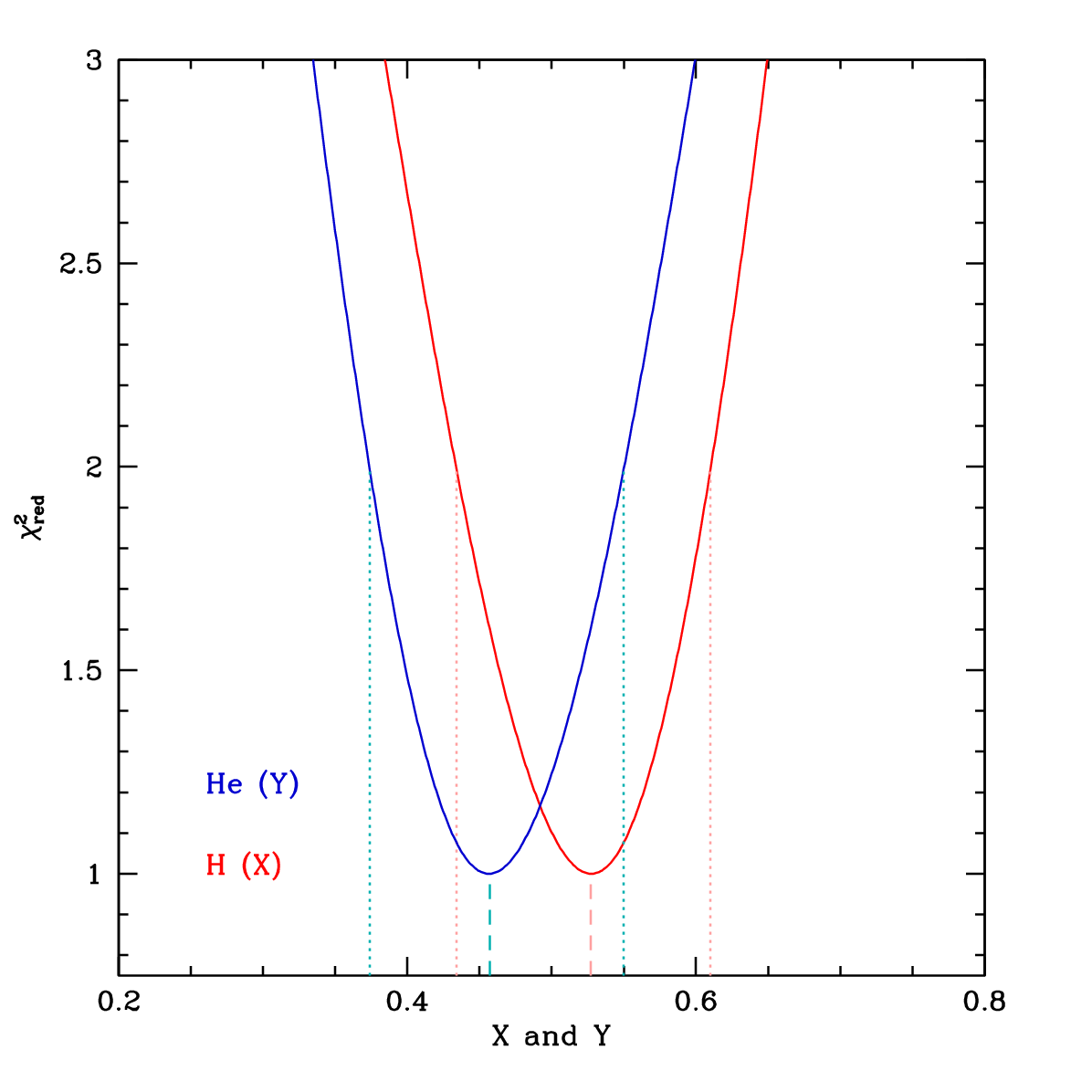}
\caption{Determination of the hydrogen and helium mass fractions in star WR24. \textit{Left panel:} The black line is the observed spectrum. The coloured lines are models in which the amounts of H and He are varied: the ratio He/H (by number) is 0.1 (0.2, 0.4) in the blue (red, green) model. \textit{Right panel:} $\chi^2$ function renormalised to the minimum value for hydrogen (red) and helium (blue), as a function of mass fraction. The vertical dashed lines indicate the best-fit values, while the dotted lines mark the error bars. }
\label{fit_HHe}
\end{figure*}

To determine the hydrogen and helium mass fractions, we proceeded as in \citet{martins09}. The method is illustrated in Fig.~\ref{fit_HHe}. We adopted the stellar and wind parameters determined as described above, and ran new models with various He/H ratios. We focused on the lines shown in the left panel of Fig.~\ref{fit_HHe}: \heiiopt, \hg, \hb,\ and \ha. These are the strongest H and He lines of the optical spectrum. When possible, we used the four lines. In some cases, H$\gamma$ and H$\beta$ are weak and their normalisation to the continuum is uncertain, especially in case of low-signal-to-noise-ratio(S/N) data. In those circumstances, the two lines were left out and only \heiiopt\ and \ha\ were used for the abundance determination. We see how the lines behave under variations of the He/H ratio. We performed a $\chi^2$ analysis of these features. The $\chi^2$ function was renormalised to its minimum value $\chi^2_{min}$ and the error bars were taken at the mass fractions for which $\chi^2$=$\chi^2_{min}$+1. The results are shown in the right panel of Fig.~\ref{fit_HHe}. 

We proceeded in a similar way for the determination of the carbon and nitrogen contents. We ran additional models with the best-fit stellar parameters and the hydrogen and helium mass fractions determined as above. In these models, we varied the C/H and N/H ratios. We then inspected the behaviour of the following nitrogen lines: \ion{N}{iv}~4058, \ion{N}{iii}~4511-15-18, \ion{N}{iv}~4512, \ion{N}{v}~4604-20, \ion{N}{iii}~4634-40-42, \ion{N}{iv}~5200-05, \ion{N}{iv}~5736, and \ion{N}{iv}~7103-10-22. Depending on the quality of the spectra and the normalisation, we selected all or a subset of these lines for a $\chi^2$ analysis.
As for H and He, the best-fit N/H value and the associated error bar were determined from the renormalised  $\chi^2$ function and the values corresponding to $\chi^2_{min}$+1. For carbon, the only lines strong enough to be analysed in a quantitative way are \ion{C}{iv}~5801-12. When present, we used them as a diagnostic for the C/H ratio; otherwise the carbon content was not determined.

The final stellar parameters and surface abundances are gathered in Table~\ref{tab_param}. We give two sets of values for the temperature and stellar radius: one set estimated at an optical depth of two-thirds and designated by the \textit{eff} subscript, and one set at an optical depth of 20 (subscript *). In Appendix~\ref{ap_bf}, we show the best-fit models compared to the observed spectra for all stars. 

\FloatBarrier

\small
\begin{sidewaystable*}[!h]
\begin{center}
  \caption{Stellar parameters determined in the present study.} \label{tab_param}
  \small
\begin{tabular}{llllllllllllllllllllll}
\hline
Star        &  ST     & \teff  &        T*   &  \reff   & R*    & E(B-V)  & \lL   & \lmdot & \vinf & $f_{\infty}$ & X  & Y        & X(C) & X(N)   \\
            &         & [kK]  & [kK] & [\rsun] & [\rsun] &        &      & [\myr] & [\kms] &             & & & & \\
\hline
WR16        &  WN8h   & 42.0  & 42.3 & 13.5  &  13.2    & 0.67    & 5.7   & -5.3 &  800  & 0.01     & 0.28$\pm$0.03 & 0.71$\pm$0.03  & 5.0$\pm$3.0 10$^{-5}$ & 11.2$\pm$3.0 10$^{-3}$ \\
WR24        &  WN6ha  & 42.0  & 43.1 & 21.3  &  20.2    & 0.28    & 6.1   & -5.3        &  2100 & 0.01     & 0.53$\pm$0.09 & 0.46$\pm$0.08  & 4.6$\pm$2 10$^{-5}$ & 11.5$\pm$4 10$^{-3}$  \\
WR40        &  WN8h   & 36.2  & 41.7 & 18.1  &  13.6    & 0.45    & 5.7   & -4.6        &  800  & 0.1      & 0.19$\pm$0.03 & 0.80$\pm$0.03  & 7.0$\pm$5.0 10$^{-5}$ & 5.1$\pm$1.1 10$^{-3}$ \\
WR78        &  WN7h   & 41.0  & 41.5 & 20.0  &  19.5    & 0.60    & 6.0   & -4.7        &  1600 & 0.05     & 0.17$\pm$0.05 & 0.82$\pm$0.05 & 4.7$\pm$3.0 10$^{-5}$ & 7.4$\pm$2.0 10$^{-3}$ \\
WR89        &  WN8h+abs &35.0 & 35.1 & 34.4  &  34.3    & 1.85    & 6.2   & -4.5         &  1500 & 0.1      & 0.26$\pm$0.04 & 0.73$\pm$0.04 & 1.0$\pm$0.7 10$^{-4}$ & 12.3$\pm$10.0 10$^{-3}$ \\
WR108       &  WN9ha  & 35.0  & 35.2 & 21.7  &  21.5    & 1.15    & 5.8   & -5.0        &  900  & 0.1      & 0.17$\pm$0.04 & 0.82$\pm$0.03 & 5.9$\pm$5.4 10$^{-4}$ & 9.3$\pm$4.3 10$^{-3}$ \\
WR124       &  WN8h   & 34.0  & 41.5 & 20.5  &  13.8    & 1.20    & 5.7   & -4.5        &  800  & 0.1      & 0.19$\pm$0.08 & 0.80$\pm$0.03 &  --                   & 2.5$\pm$1.0 10$^{-3}$ \\
WR128       &  WN4(h) & 65.0  & 65.4 & 4.0   &  3.9     & 0.40    & 5.4   & -5.5        &  1800 & 0.1      & 0.23$\pm$0.18 & 0.75$\pm$0.19 & 6.2$\pm$2.5 10$^{-5}$ & 21.6$\pm$6 10$^{-3}$  \\
\hline
BAT99-35    &  WN3(h) & 74.0  & 77.3 &  3.9   & 3.5    &0.13      & 5.6   & -5.15       &  1600 & 0.1      & 0.11$\pm$0.10 & 0.88$\pm$0.10 & 3.7$\pm$0.4 10$^{-5}$ & 10.0$\pm$2.6 10$^{-3}$ \\
BAT99-44    &  WN8ha  & 38.9  & 39.7 &  11.1  & 10.6     &0.10    & 5.4   & -5.2 &  600  & 0.1      & 0.31$\pm$0.05 & 0.69$\pm$0.04 & --                   & 4.2$\pm$1.7 10$^{-3}$ \\
BAT99-50    &  WN5h   & 53.0  & 56.1 &  6.7   & 6.0      &0.0     & 5.5   & -5.6 &  1600 & 0.1      & 0.39$\pm$0.16 & 0.60$\pm$0.16 & 5.4$\pm$3.0 10$^{-5}$ & 9.0$\pm$5.0 10$^{-3}$ \\
BAT99-63    &  WN4ha  & 64.0  & 64.5 &  5.5   & 5.4      &0.15    & 5.65  & -5.35        &  2000 & 0.1      & 0.31$\pm$0.02 & 0.68$\pm$0.03 & 1.3$\pm$0.2 10$^{-4}$ & 10.6$\pm$9.0 10$^{-3}$  \\
BAT99-66    &  WN3(h) & 85.0  & 85.6 &  3.3   & 3.2      &0.10    & 5.7   & -5.6 &  1600 & 0.1      & 0.27$\pm$0.10 & 0.72$\pm$0.11 & $<$3.0 10$^{-4}$     & 7.8$\pm$1.0 10$^{-3}$ \\ 
BAT99-67    &  WN5ha  & 42.0  & 51.7 &  16.9   & 11.1    &0.40    & 5.90  & -5.5 &  1800 & 0.01     & 0.45$\pm$0.10 & 0.53$\pm$0.10 & 2.7$\pm$1.1 10$^{-5}$ & 8.8$\pm$7.6 10$^{-3}$  \\
BAT99-73    &  WN5ha  & 51.0  & 52.7 &  7.2   & 6.8      &0.30    & 5.5   & -5.2 &  1300 & 0.01     & 0.42$\pm$0.08 & 0.57$\pm$0.07 & 3.3$\pm$1.1 10$^{-5}$ & 6.6$\pm$2.9 10$^{-3}$ \\
BAT99-74    &  WN3(h)a& 80.0  & 82.2 &  2.9   & 2.8      &0.10    & 5.5   & -5.8 &  2200 & 0.1      & 0.30$\pm$0.05 & 0.68$\pm$0.04 & $<$1.0 10$^{-4}$     & 8.1$\pm$0.6 10$^{-3}$ \\
BAT99-81    &  WN5h   & 47.0  & 47.8 &  6.8   & 6.5      &0.45    & 5.3   & -5.6        &  1250 & 0.1      & 0.61$\pm$0.13 & 0.38$\pm$0.23 & 2.3$\pm$0.5 10$^{-4}$ & 7.3$\pm$3.8 10$^{-3}$ \\
BAT99-89    &  WN7h   & 39.2  & 50.5 &  15.5  & 9.3      &0.40    & 5.7   & -4.5 &  1000 & 0.1      & 0.22$\pm$0.04 & 0.77$\pm$0.06 & --                   & 3.2$\pm$0.7 10$^{-3}$ \\
BAT99-122   &  WN5h   & 42.0  & 44.3 &  21.3  & 19.1     &0.37    & 6.1   & -4.5 &  1800 & 0.1      & 0.17$\pm$0.10 & 0.82$\pm$0.10 & 4.2$\pm$0.9 10$^{-5}$ & 3.7$\pm$1.3 10$^{-3}$ \\
\hline
AB9         &  WN3ha  & 80.0  & 81.6 & 3.7   &  3.6     & 0.1     & 5.7   & -5.9 &  2000 & 0.1      & 0.28$\pm$0.08  & 0.71$\pm$0.10     &$<$1.0 10$^{-4}$  & 3.8$\pm$0.8 10$^{-3}$ \\
AB10        &  WN3ha  & 80.0  & 82.4 & 2.9   &  2.8     & 0.1     & 5.5   & -5.8 &  1600 & 0.1      & 0.35$\pm$0.08  & 0.65$\pm$0.11     &$<$2.0 10$^{-5}$  & 3.5$\pm$1.3 10$^{-3}$ \\
AB11        &  WN4ha  & 80.0  & 81.9 & 2.9   &  2.8     & 0.0     & 5.5   & -6.15        &  2000 & 0.1      & 0.26$\pm$0.05  & 0.73$\pm$0.06     &$<$7.0 10$^{-5}$ & 7.6$\pm$1.6 10$^{-3}$ \\
\hline
\end{tabular}
\normalsize
\tablefoot{Columns are source ID, spectral type, temperature at an optical depth of two-thirds and 20, radius at an optical depth of two-thirds and 20, colour excess, luminosity, mass-loss rate, terminal velocity, maximum clumping factor, hydrogen mass fraction, helium mass fraction, carbon mass fraction and nitrogen mass fraction.}
\end{center}
\end{sidewaystable*}

\normalsize

\section{Results}
\label{s_res}

\subsection{Comparison with literature data}

All of our sample stars have been previously studied. As stated in Sect. 1, earlier analyses did not provide the surface carbon and nitrogen abundances. We gather the parameters found in the literature in Table~\ref{tab_comp}. Most results come from the analysis carried out by the Potsdam group \citep{hamann06,hainich14,hainich15}. Inspection of the different values of the parameters for a given star indicates very good agreement in most, if not all, cases. This conclusion is valid for temperatures, luminosities, and surface hydrogen mass fraction. These parameters can therefore be considered as rather robust.

For two stars, WR128 and BAT99-63, carbon and nitrogen abundances were presented in \citet{martins13}. The former are in excellent agreement with the values presented here. For nitrogen, we note the systematically larger values in the current study, although the results are consistent within the error bars. The largest difference is for star WR128. We find a high nitrogen content, about a factor of two larger than in \citet{martins13}. Inspection of the best fit of Martins et al. (their Fig.4) and comparison to the present Fig.~\ref{fig_wr128} indicate that \nivopta\ and \nvopt\ are better reproduced by the new model. This may partly explain the difference.

\subsection{Surface chemical composition}
\label{s_surfchem}

\begin{figure}[t]
\centering
\includegraphics[width=0.49\textwidth]{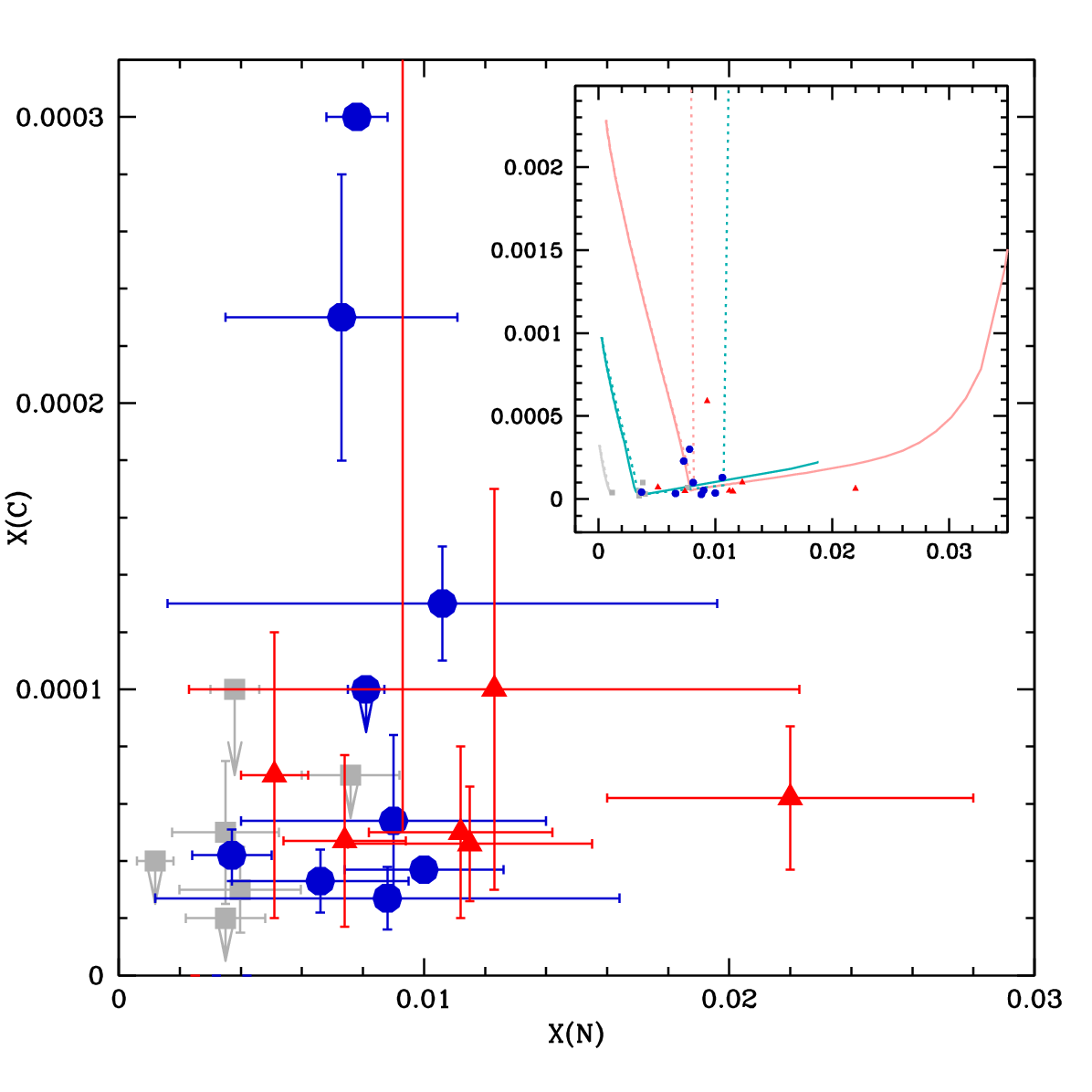}
\caption{Carbon mass fraction as a function of nitrogen mass fraction for the sample stars. Red triangles, blue circles, and grey squares refer to Galactic, LMC, and SMC stars, respectively. One Galactic star is out of the frame and is seen in the insert where we show the same data points  on a wider scale together with 40~\msun\ (solid line) and 60~\msun\ (dotted line) Geneva models at solar (pink), LMC (cyan), and SMC (grey) metallicity. The three SMC stars analysed by \citep{martins09} are included.}
\label{xnxc}
\end{figure}

\begin{figure}[t]
\centering
\includegraphics[width=0.49\textwidth]{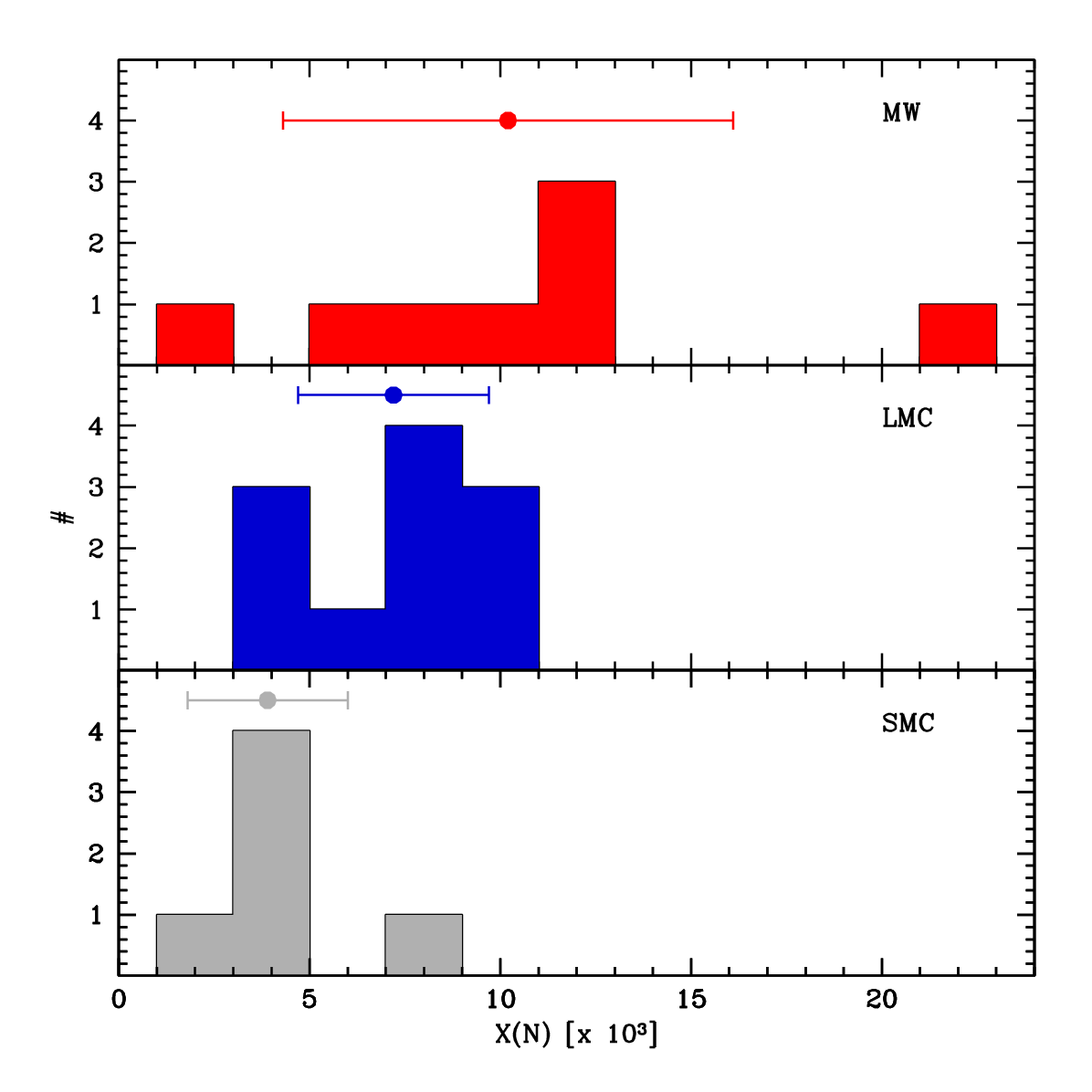}
\caption{Number distribution of the surface nitrogen mass fraction for the Galaxy, LMC, and SMC  (top, middle, and bottom panels, respectively). In each panel, the filled circle indicates the average X(N) and the error bar shows the dispersion.}
\label{histoxn}
\end{figure}

Figure~\ref{xnxc} shows the carbon mass fraction as a function of nitrogen mass fraction for the sample stars. In the insert, the stellar tracks illustrate the evolutionary status of the sample sources. On the main sequence, tracks start with low nitrogen mass fraction and high carbon mass fraction in the left part of the figure. With nucleosynthesis and mixing processes at work, nitrogen is brought to the surface while carbon is depleted, in agreement with CNO burning. In later phases, carbon is produced from helium burning. We see that all stars are clearly chemically evolved, most of them lying near the minimum carbon mass fraction. These objects therefore show C and N consistent with CNO equilibrium, as seen in other WNh stars \citep{martins08}.
At CNO equilibrium, the sum of almost all the initial C, N, and O is found in the form of nitrogen. The nitrogen mass fraction therefore directly scales with initial C+N+O, and thus with metallicity. This is predicted by models (see insert) and also confirmed by the present results. The average nitrogen mass fraction of WNh stars is 0.0039$\pm$0.0021, 0.0072$\pm$0.0025, and 0.0102$\pm$0.0059 in the SMC, LMC, and the Galaxy, respectively; that is, a clear increase for higher metallicity is detected. This is seen in Fig.~\ref{histoxn}, which shows the number distribution of the nitrogen mass fraction in the three galaxies. 

\begin{figure}[t]
\centering
\includegraphics[width=0.49\textwidth]{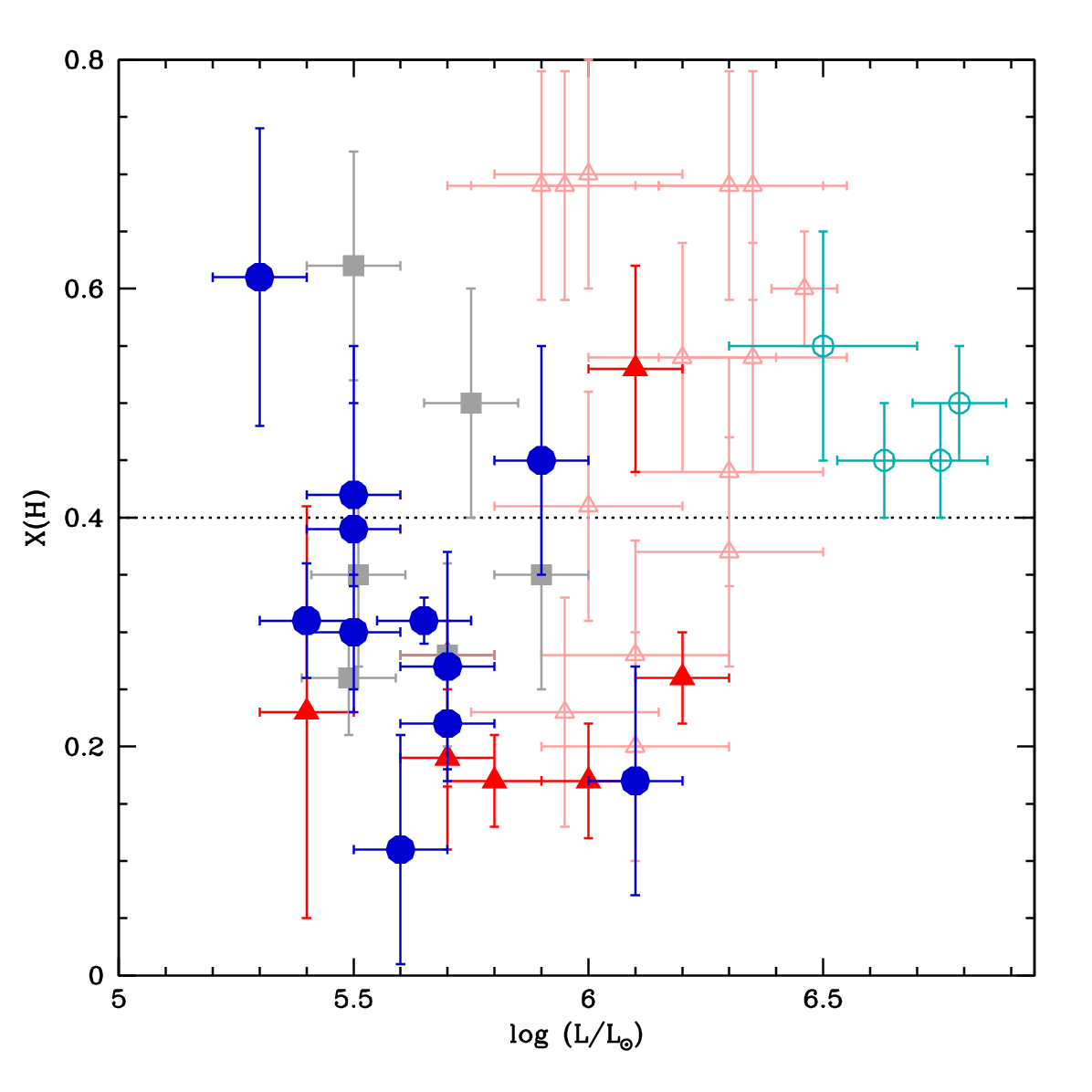}
\caption{Hydrogen mass fraction as a function of luminosity for the sample stars. Red triangles, blue circles, and grey squares refer to Galactic, LMC, and SMC stars, {respectively}. The horizontal dotted line marks the value X=0.4. The figure also includes WNh stars from the literature: stars in R136 \citep{besten20}, star VFTS~682 \citep{besten11}, star B in NGC~3603 \citep{crowther10}, and stars in the Arches cluster \citep{martins08}. Literature objects in the LMC (Galaxy) are shown in cyan (pink) and by open symbols.}
\label{X_L}
\end{figure}

Figure~\ref{X_L} shows the hydrogen mass fraction X as a function of luminosity for the sample sources and additional literature objects. WNh stars cover a wide range of H mass fractions, from barely any hydrogen depletion (X $\sim$ 0.6) to almost exhaustion (X $\sim$ 0). There is no obvious trend with luminosity and/or metallicity. Of the 42 points in Fig.~\ref{X_L}, 16 are strictly above the 0.4 limit, including their uncertainty. Alternatively, some stars are located across the limit when the error bar on their X is considered. If we count all objects that can potentially populate the X$>$0.4 region given the uncertainties, we find 24 stars. In that sample, the fraction of stars with X larger than 0.4 is therefore between 40\%\ and 60\%, indicating that WNh stars with a large hydrogen content are relatively common objects.

\subsection{HR diagram and stellar evolution}

\subsubsection{Single star tracks}
\label{s_sing}

\begin{figure}[t]
\centering
\includegraphics[width=0.49\textwidth]{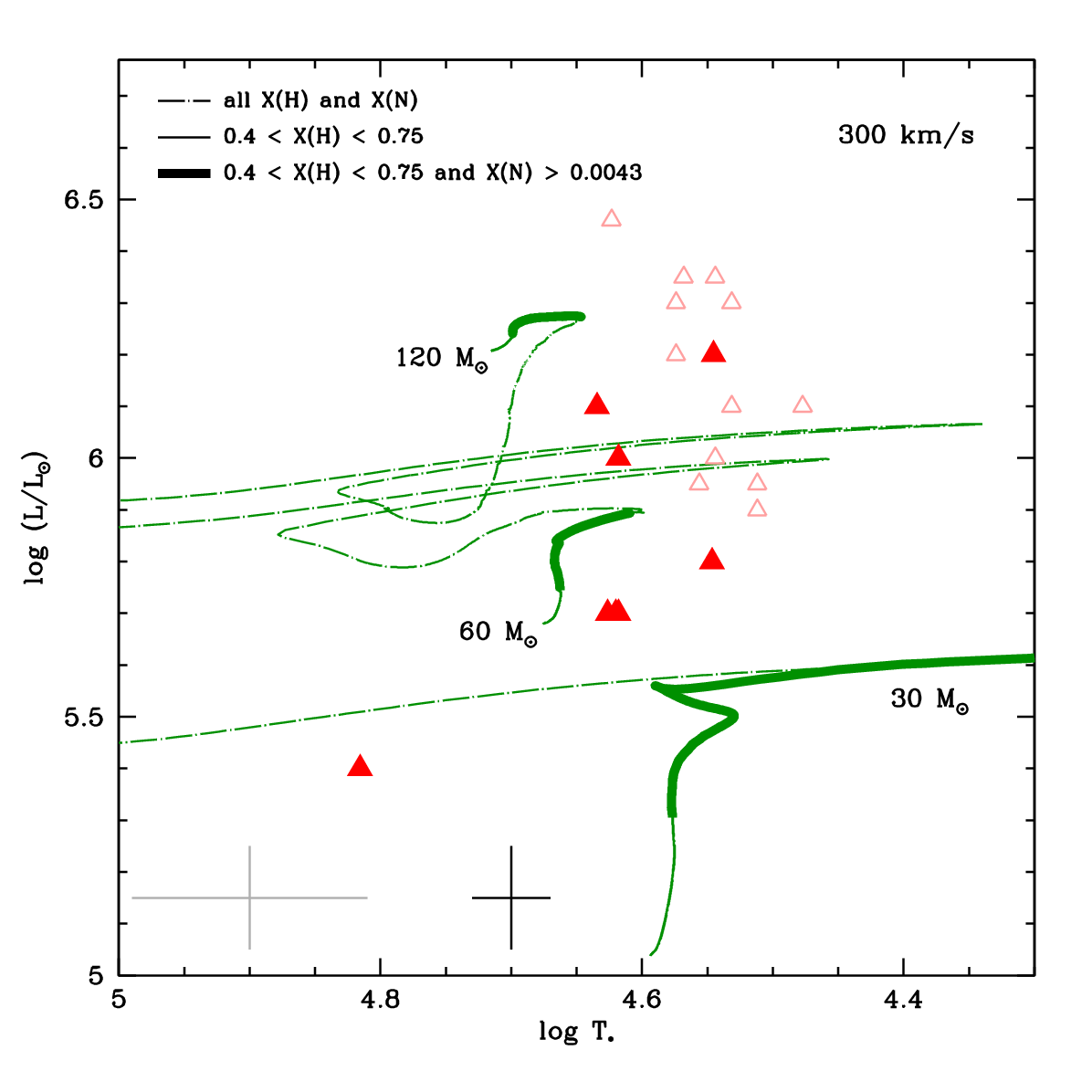}
\caption{HR diagram for Galactic stars. Tracks are from \citet{lc18} and have an initial rotational velocity of 300~\kms. The dot-dashed lines show the full tracks. The thin solid lines show the parts for which 0.4$<$X$<$0.75. The thick solid lines show the parts for which in addition to 0.4$<$X$<$0.75, X(N) is larger than 0.0043. Symbols are the sample stars (filled triangles) and literature objects (open symbols); see text.}
\label{hrd}
\end{figure}

\begin{figure*}[t]
\centering
\includegraphics[width=0.49\textwidth]{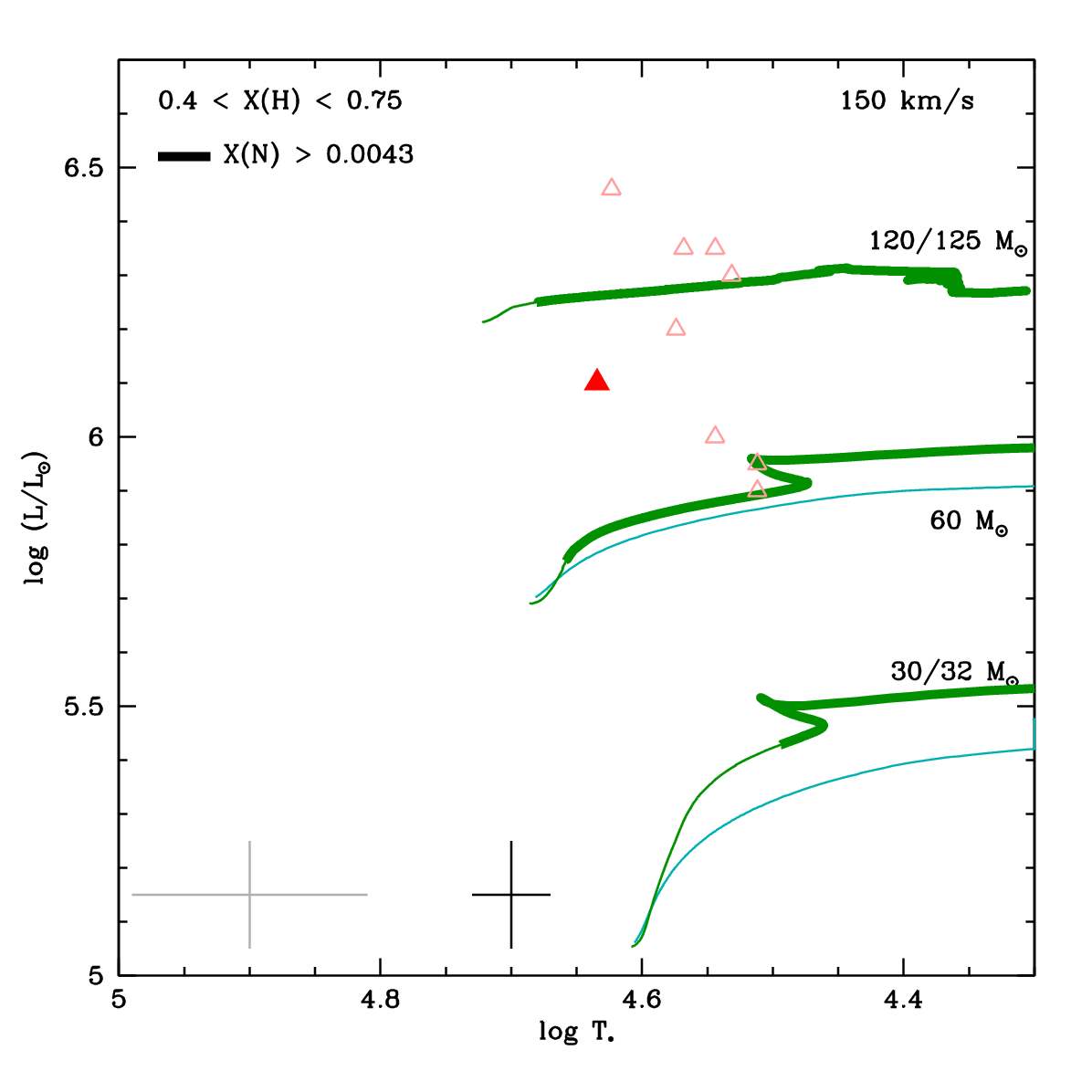}
\includegraphics[width=0.49\textwidth]{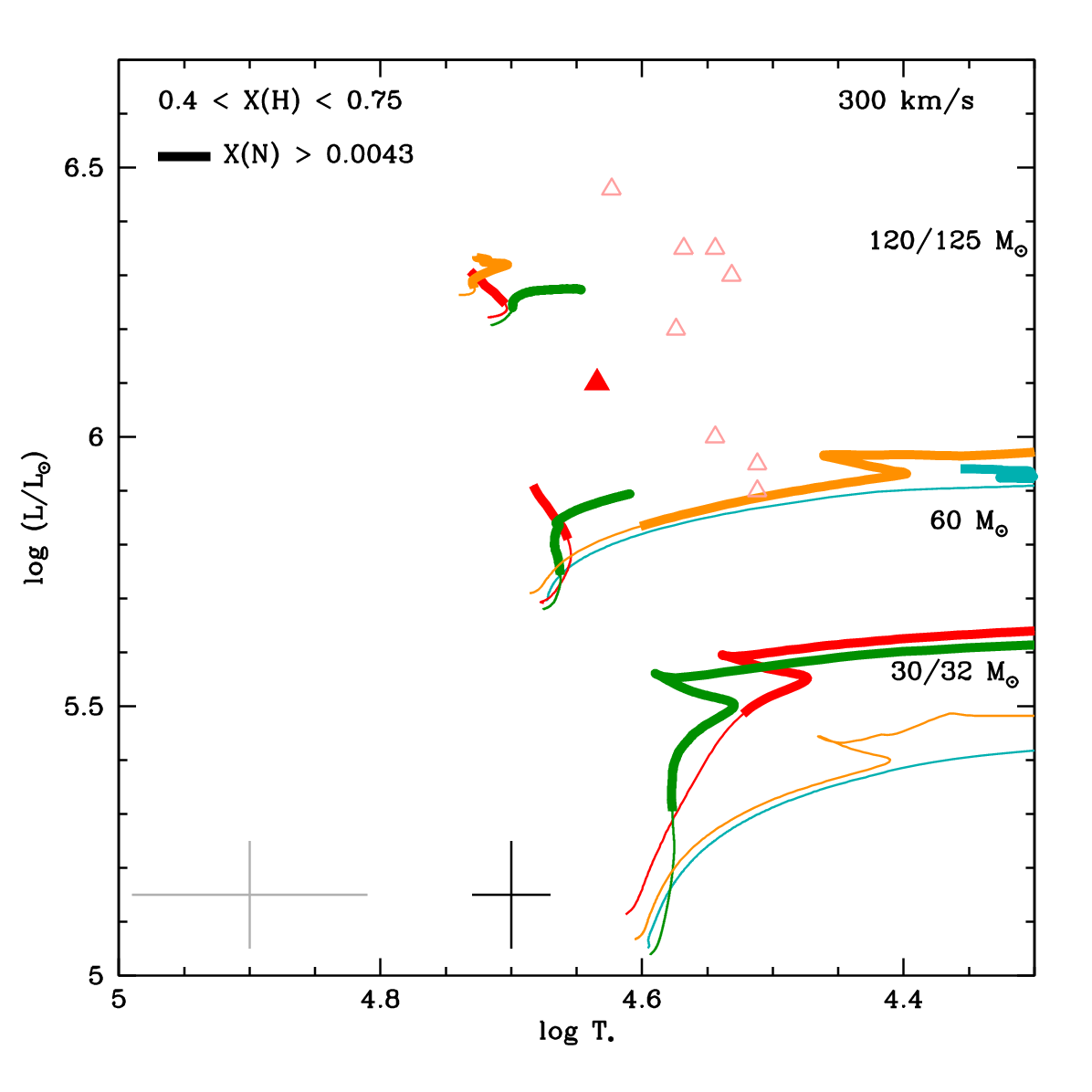}

\includegraphics[width=0.49\textwidth]{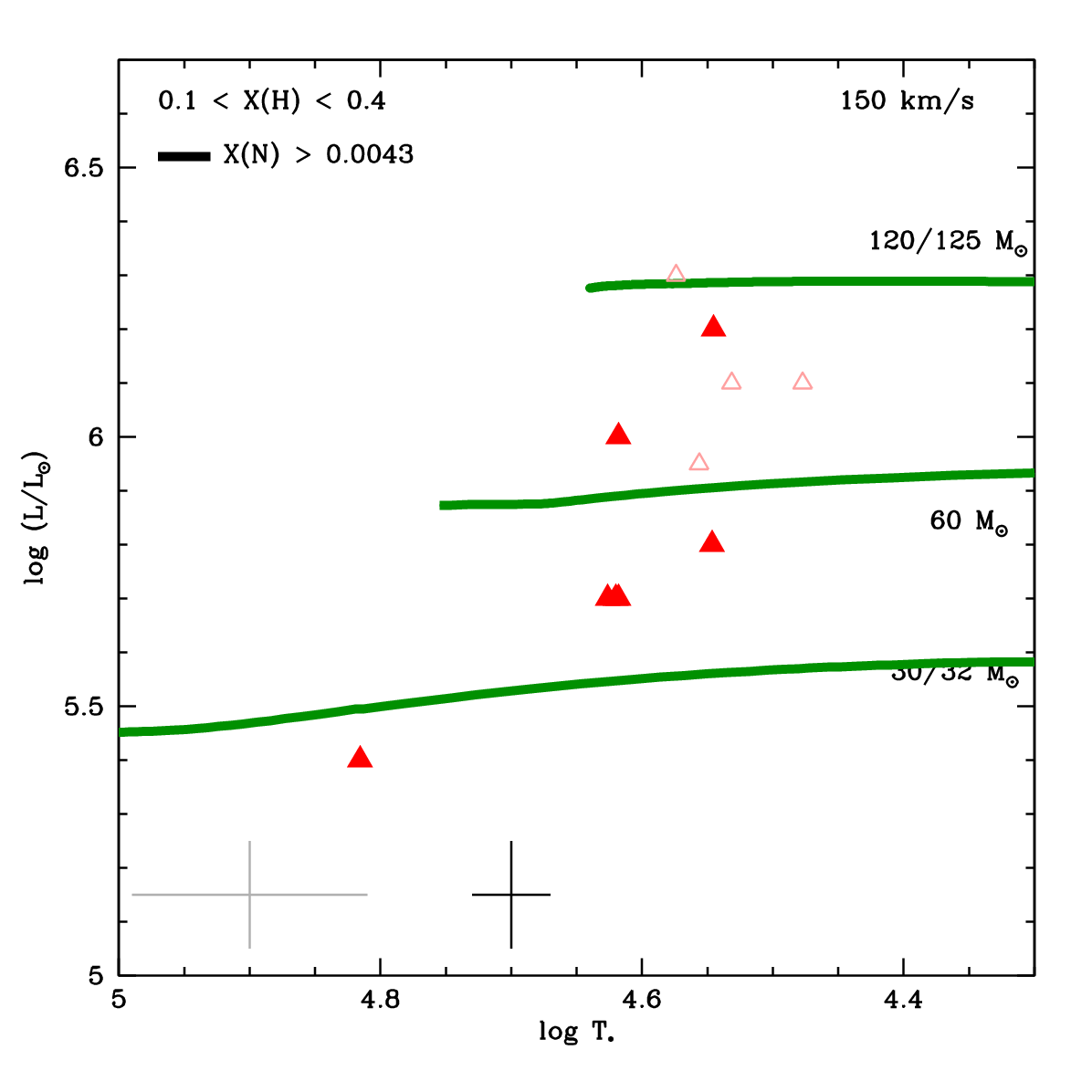}
\includegraphics[width=0.49\textwidth]{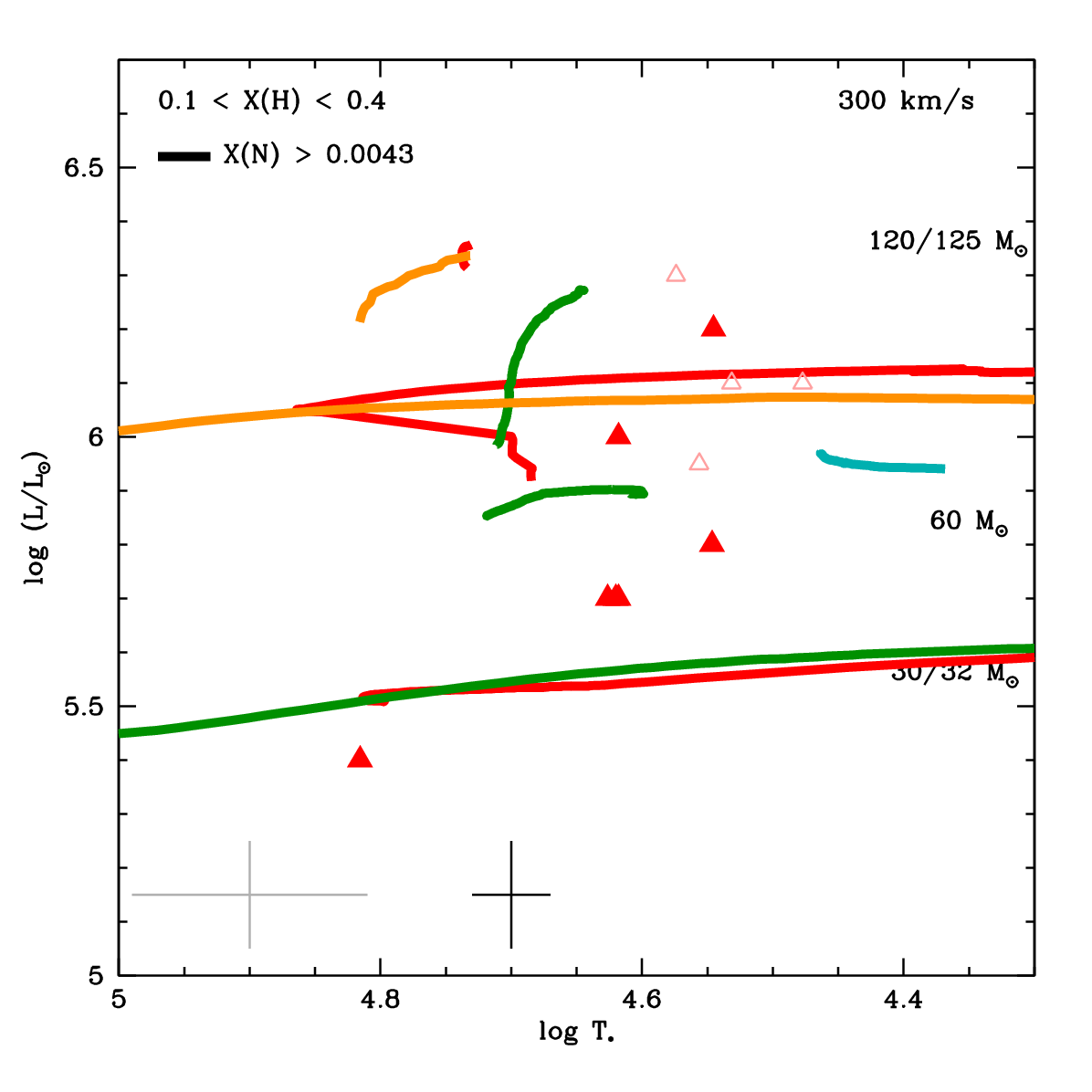}
\caption{HR diagram for Galactic stars. The top and bottom panels show stars with a hydrogen mass fraction of greater or smaller than 0.4{, respectively}. Filled triangles are sources analysed in the present study, while open triangles are literature data: WNh stars in the Arches cluster \citep{martins08} and NGC3603-B \citep{crowther10}. Evolutionary tracks shown in red, green, cyan, and orange are from \citet{ek12}, \citet{lc18}, \citet{brott11}, and \citet{grasha21}, respectively. The left and right panels show tracks with initial velocities of 150 and 300\kms{, respectively}. In each panel, the range of hydrogen mass fraction is the same for the tracks and the observed stars. The bold part of the tracks is for a nitrogen mass fraction of greater than 0.0043, as explained in Fig.~\ref{hrd}.}
\label{hrd_gal}
\end{figure*}

\begin{figure*}[t]
\centering
\includegraphics[width=0.33\textwidth]{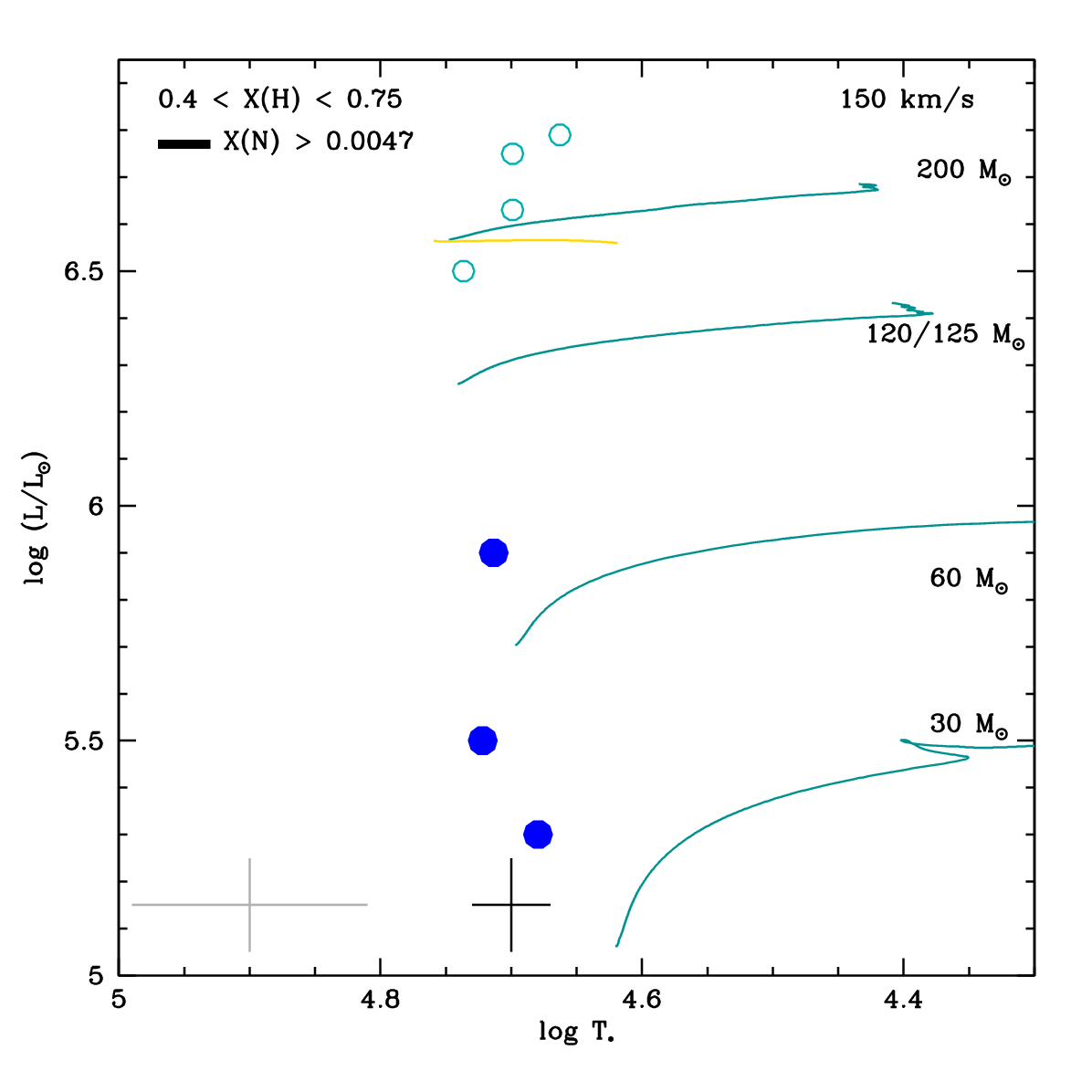}
\includegraphics[width=0.33\textwidth]{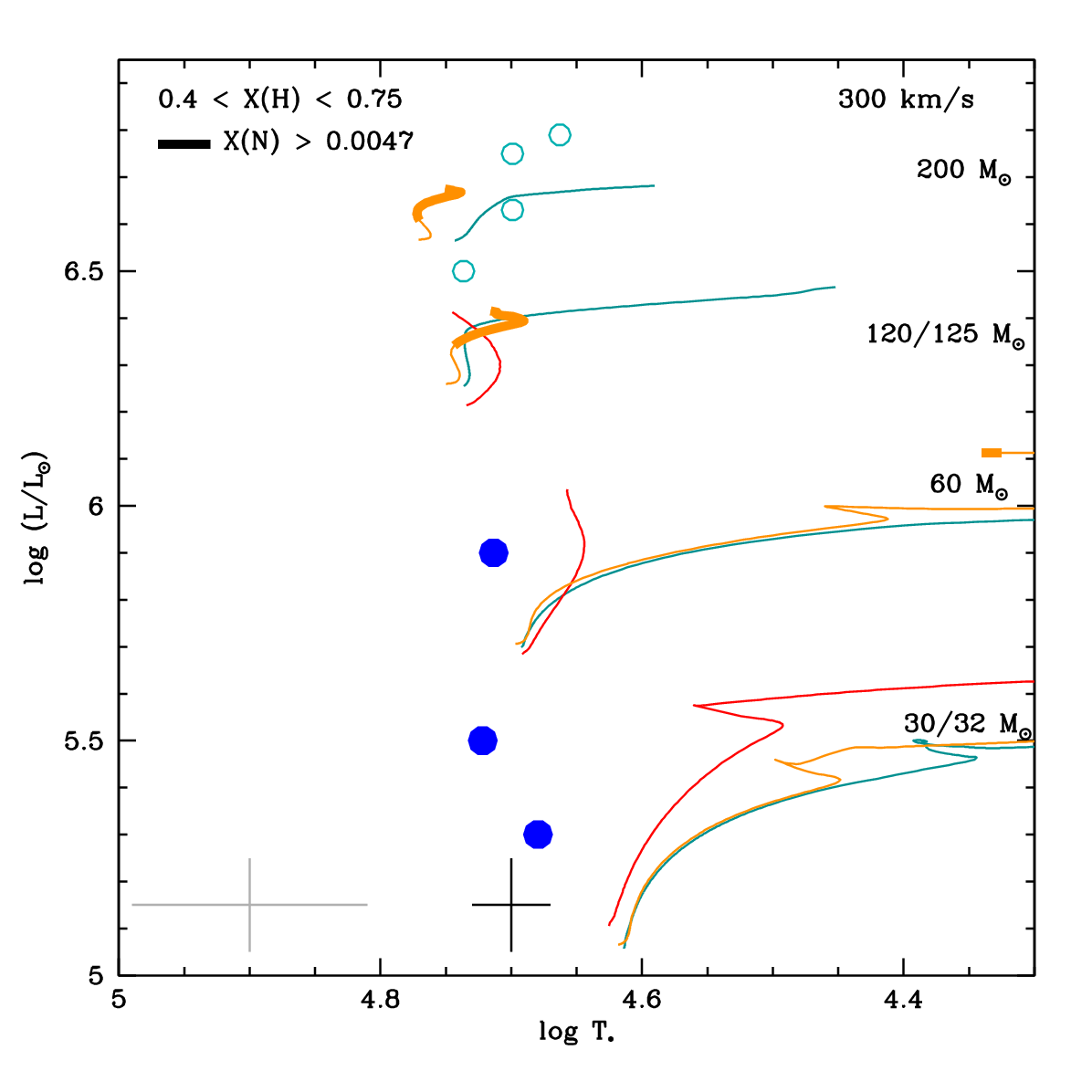}
\includegraphics[width=0.33\textwidth]{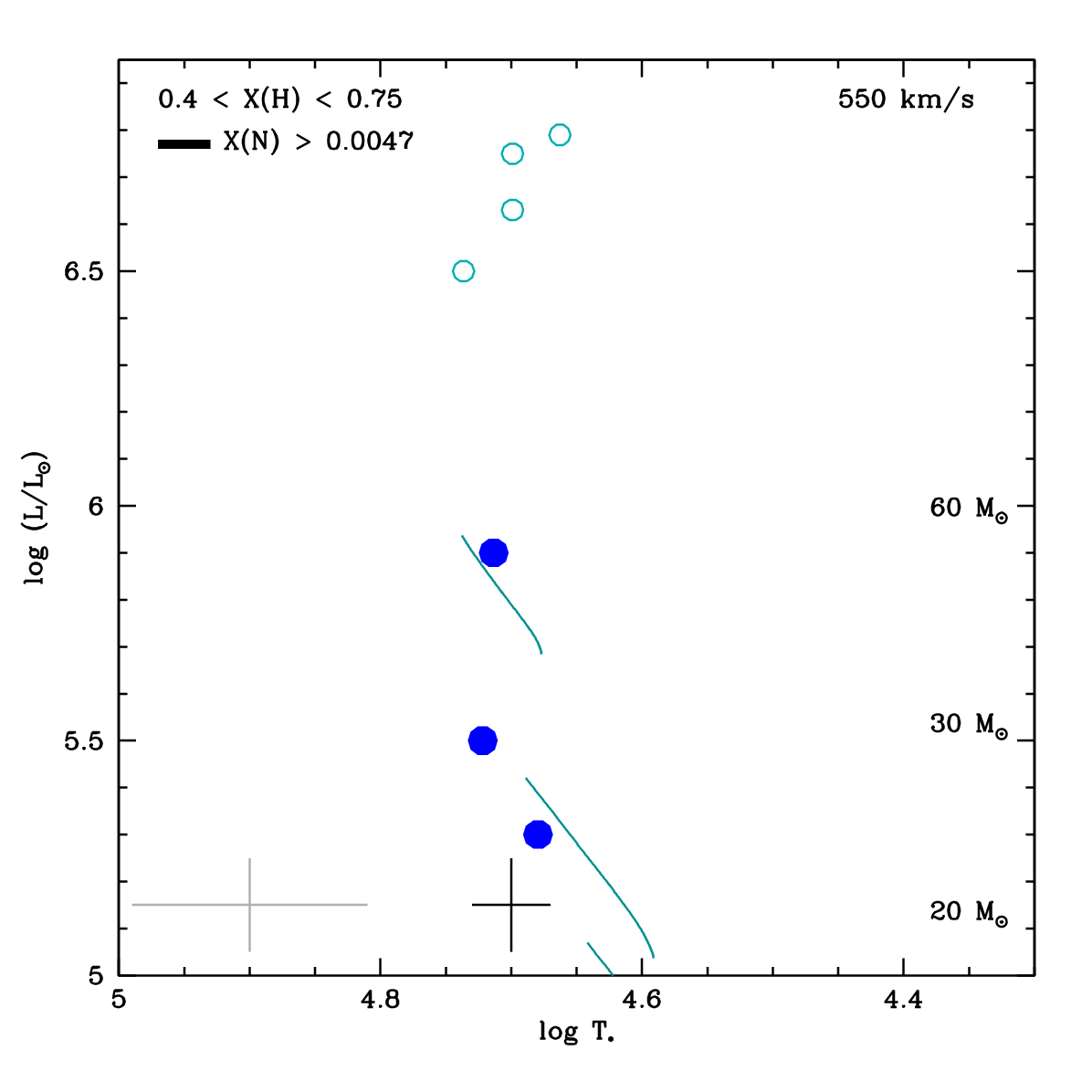}

\includegraphics[width=0.33\textwidth]{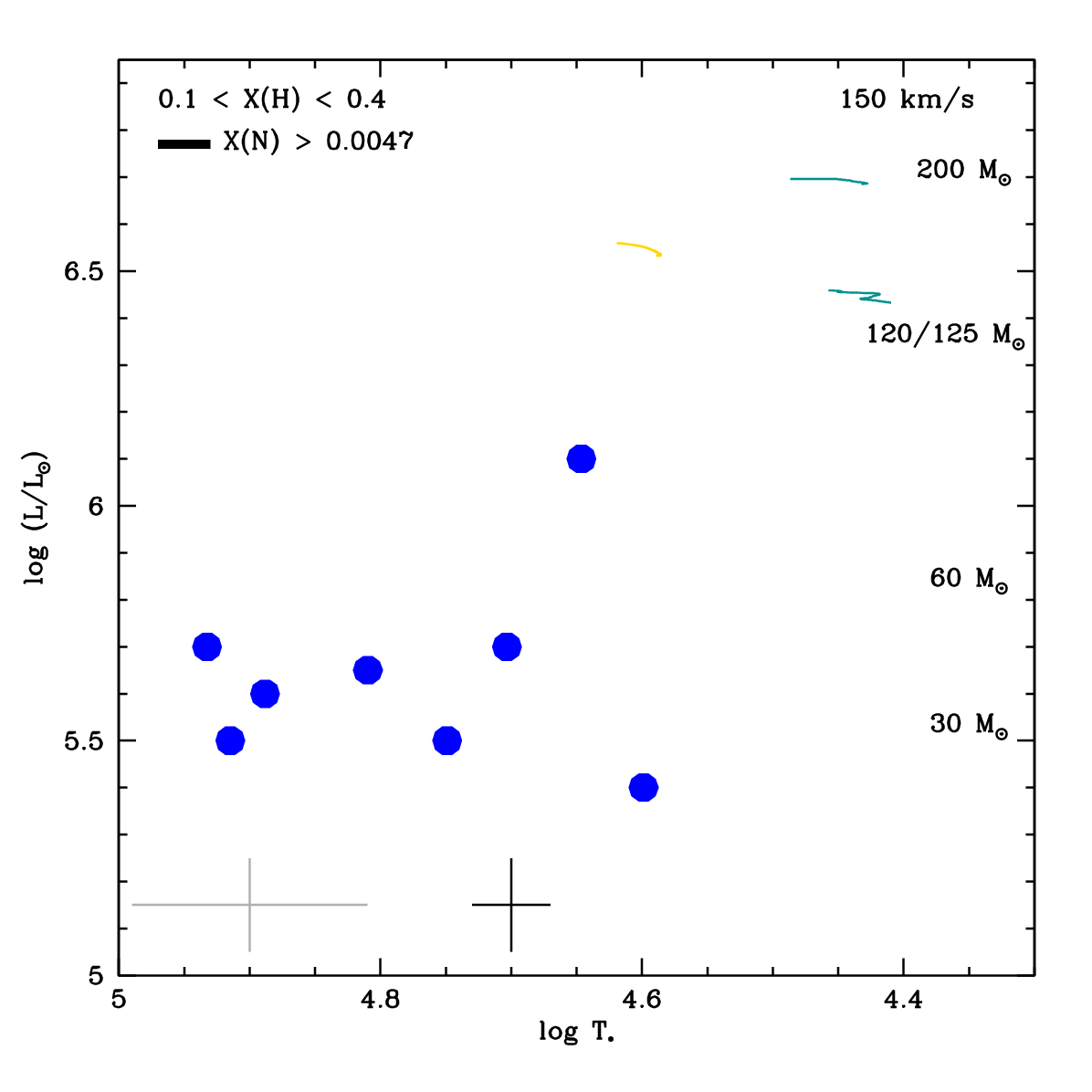}
\includegraphics[width=0.33\textwidth]{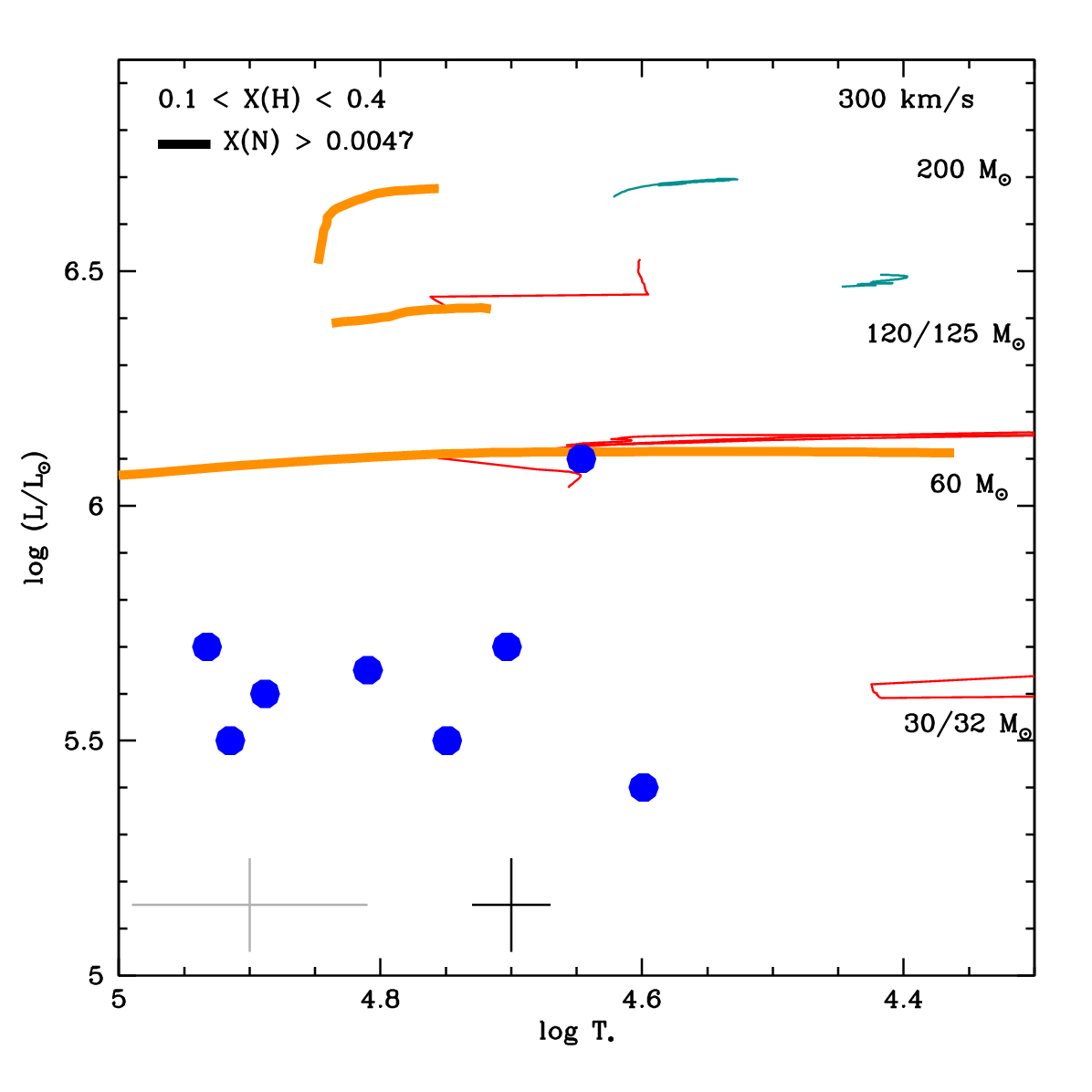}
\includegraphics[width=0.33\textwidth]{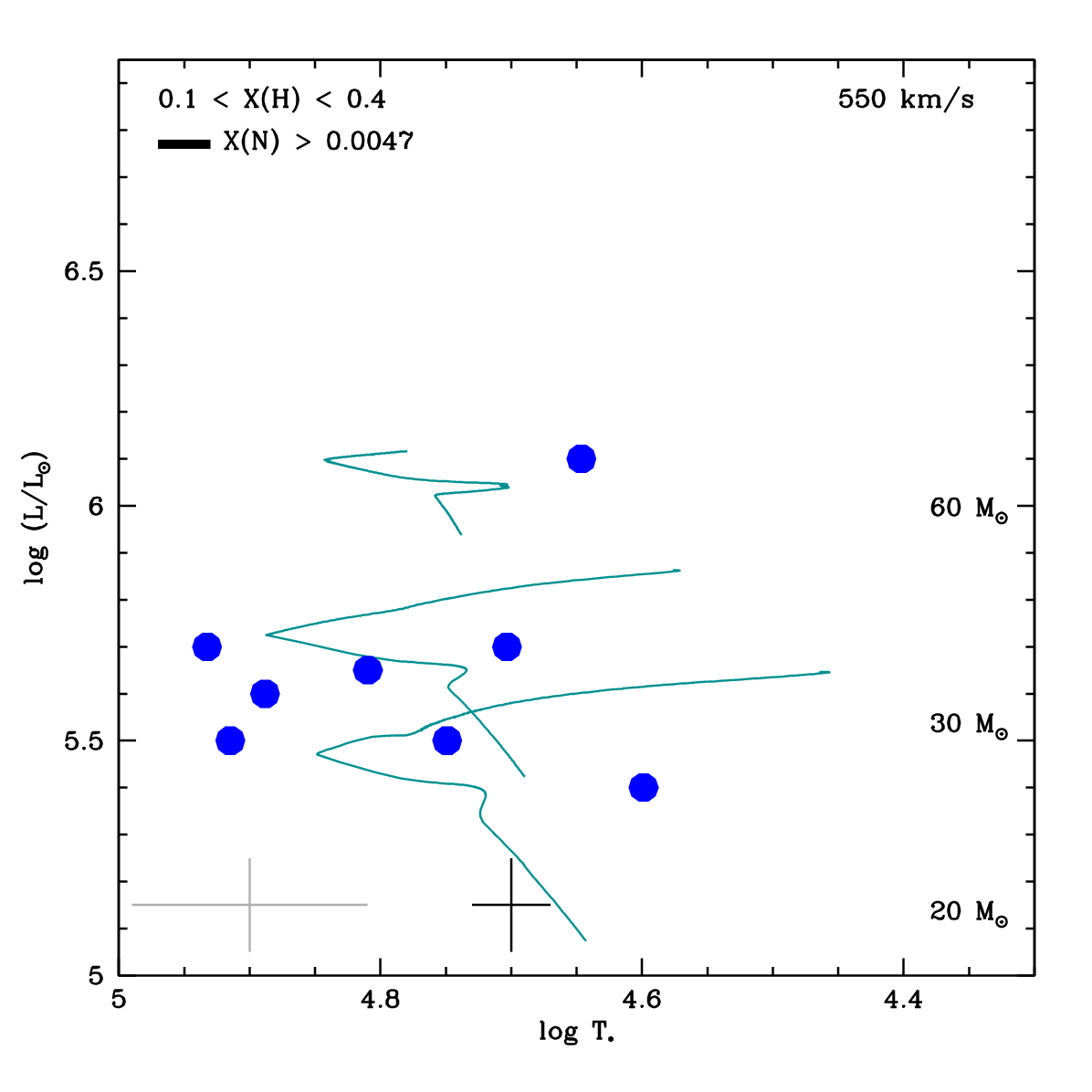}
\caption{Same as Fig.~\ref{hrd_gal} but for the LMC. The tracks of \citet{brott11} are complemented with those of \citet{kohler15} for masses above 100~\msun. The left panel shows the 200~\msun\ non-rotating tracks of \citet{mp22} in yellow. The right panels show models for an initial rotational velocity of 550~\kms. The red tracks are from \citet{eggen21}. The bold part of the tracks is for a nitrogen mass fraction of greater than 0.0047. Open circles are stars in 30~Dor from \citet{besten11,besten20}.}
\label{hrd_lmc}
\end{figure*}

\begin{figure*}[t]
\centering
\includegraphics[width=0.33\textwidth]{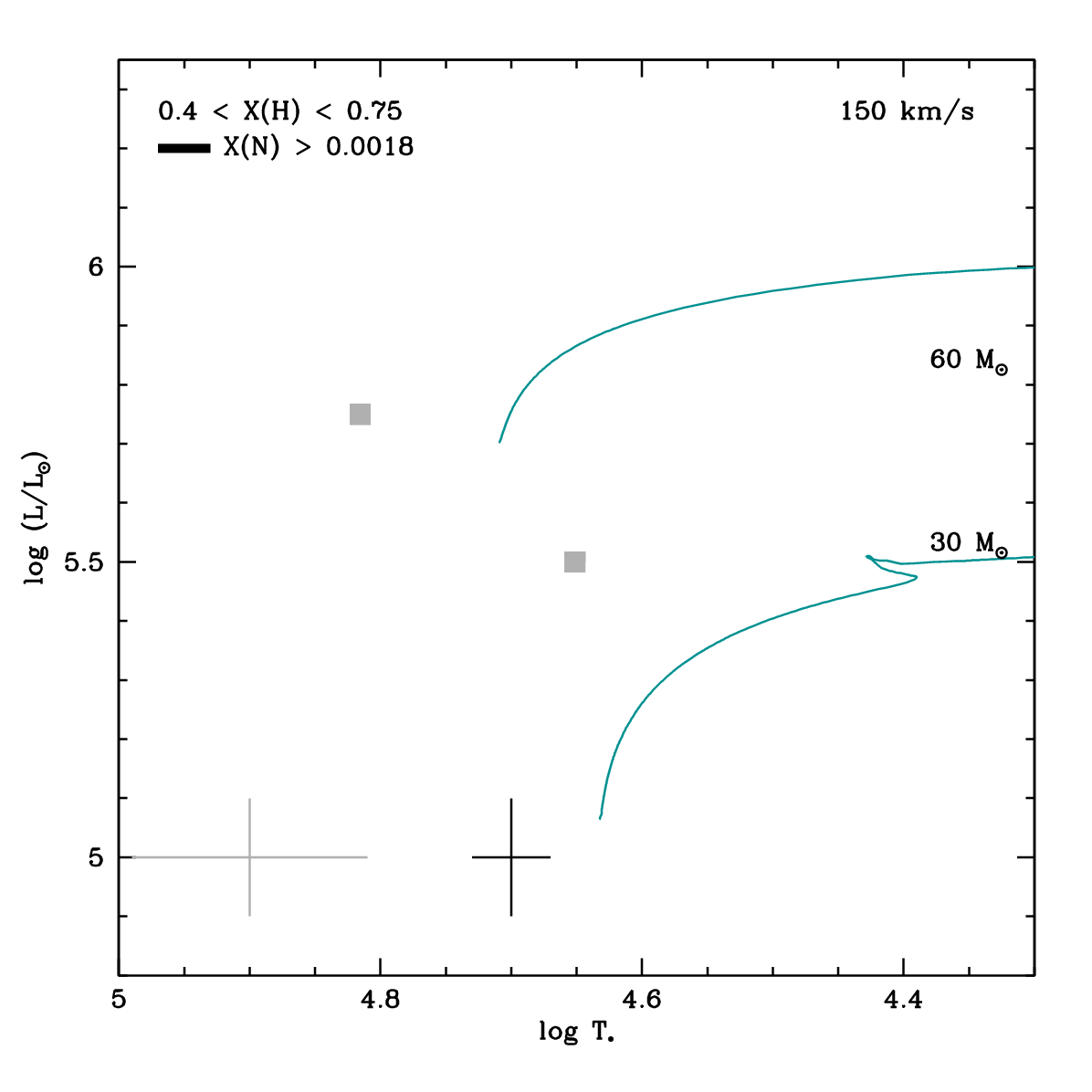}
\includegraphics[width=0.33\textwidth]{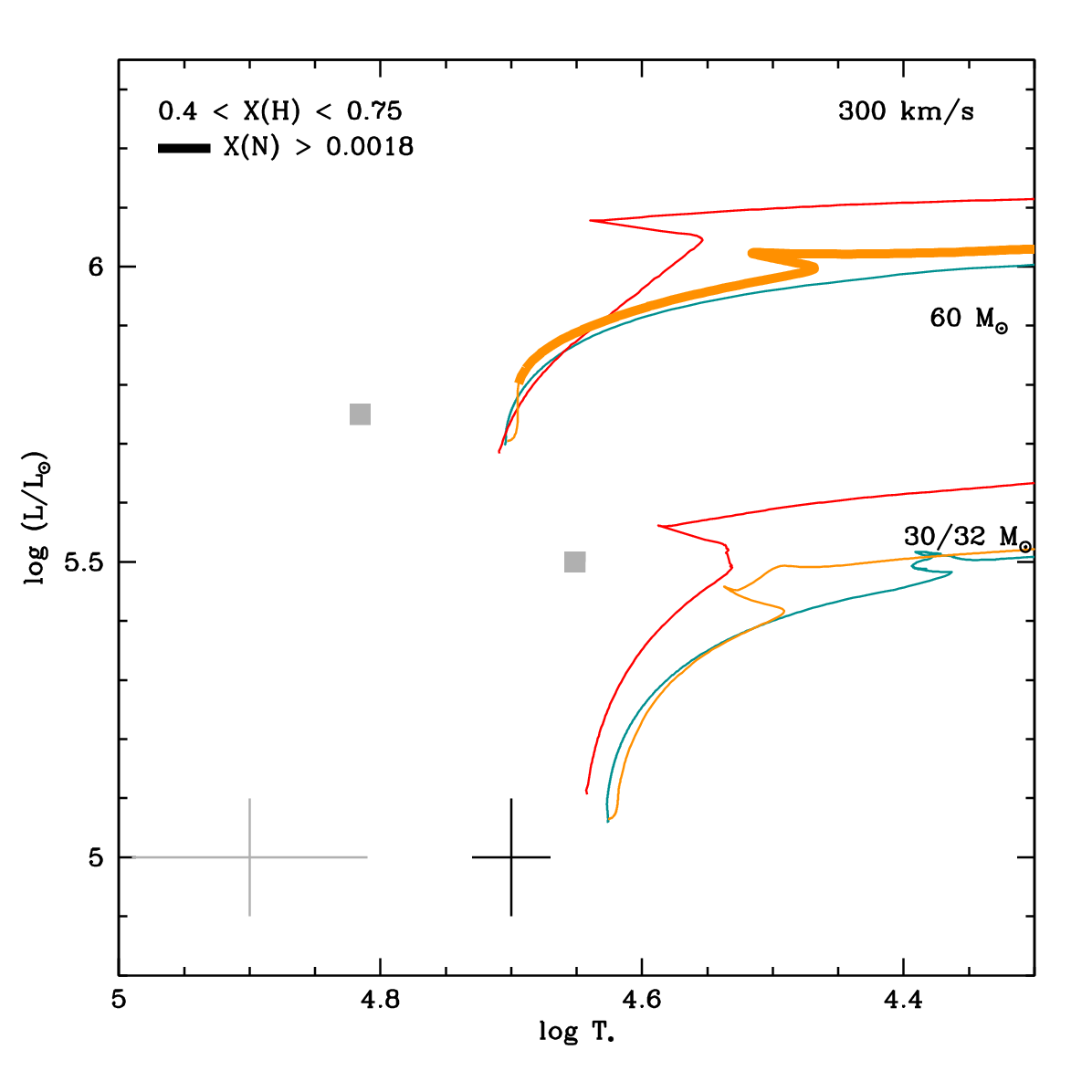}
\includegraphics[width=0.33\textwidth]{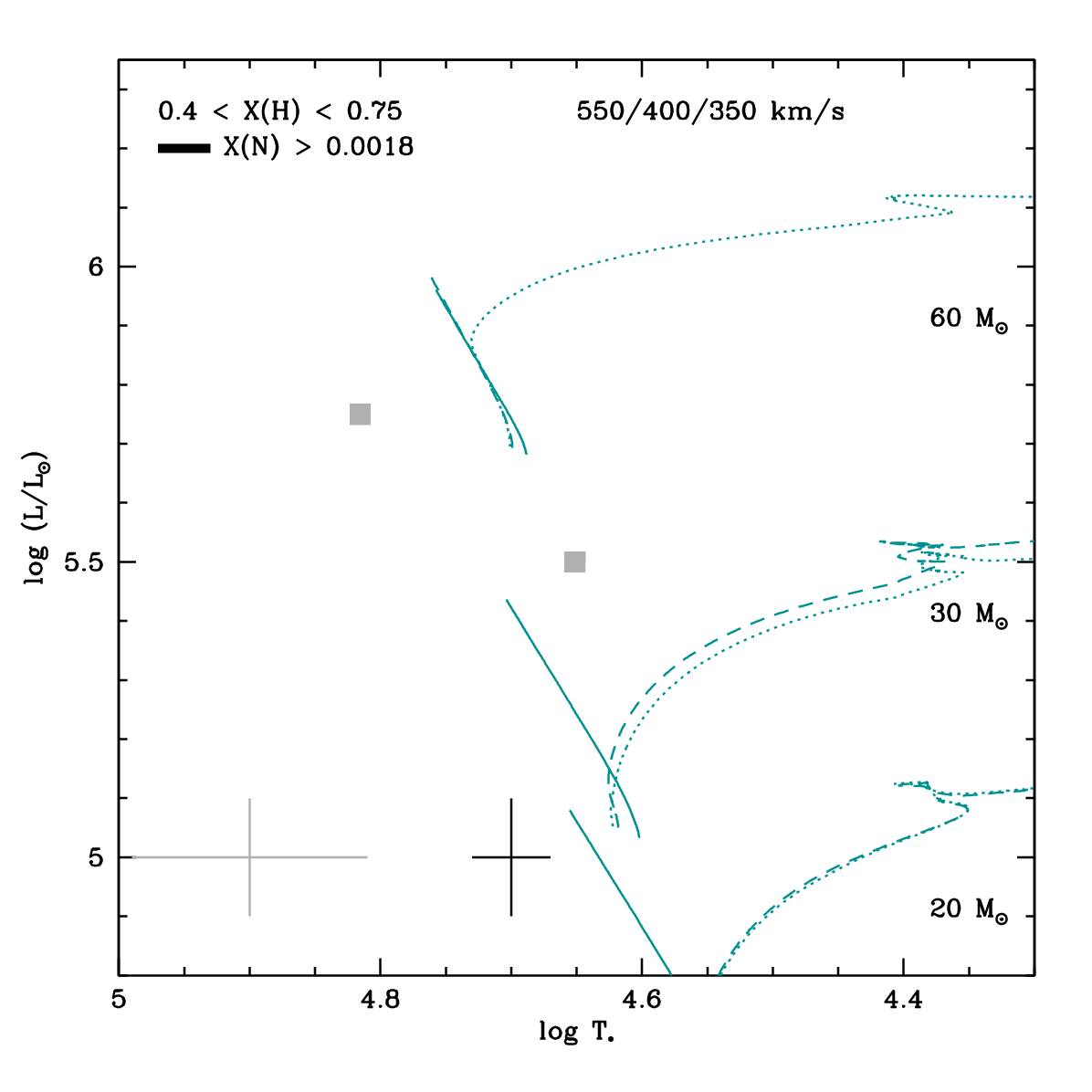}

\includegraphics[width=0.33\textwidth]{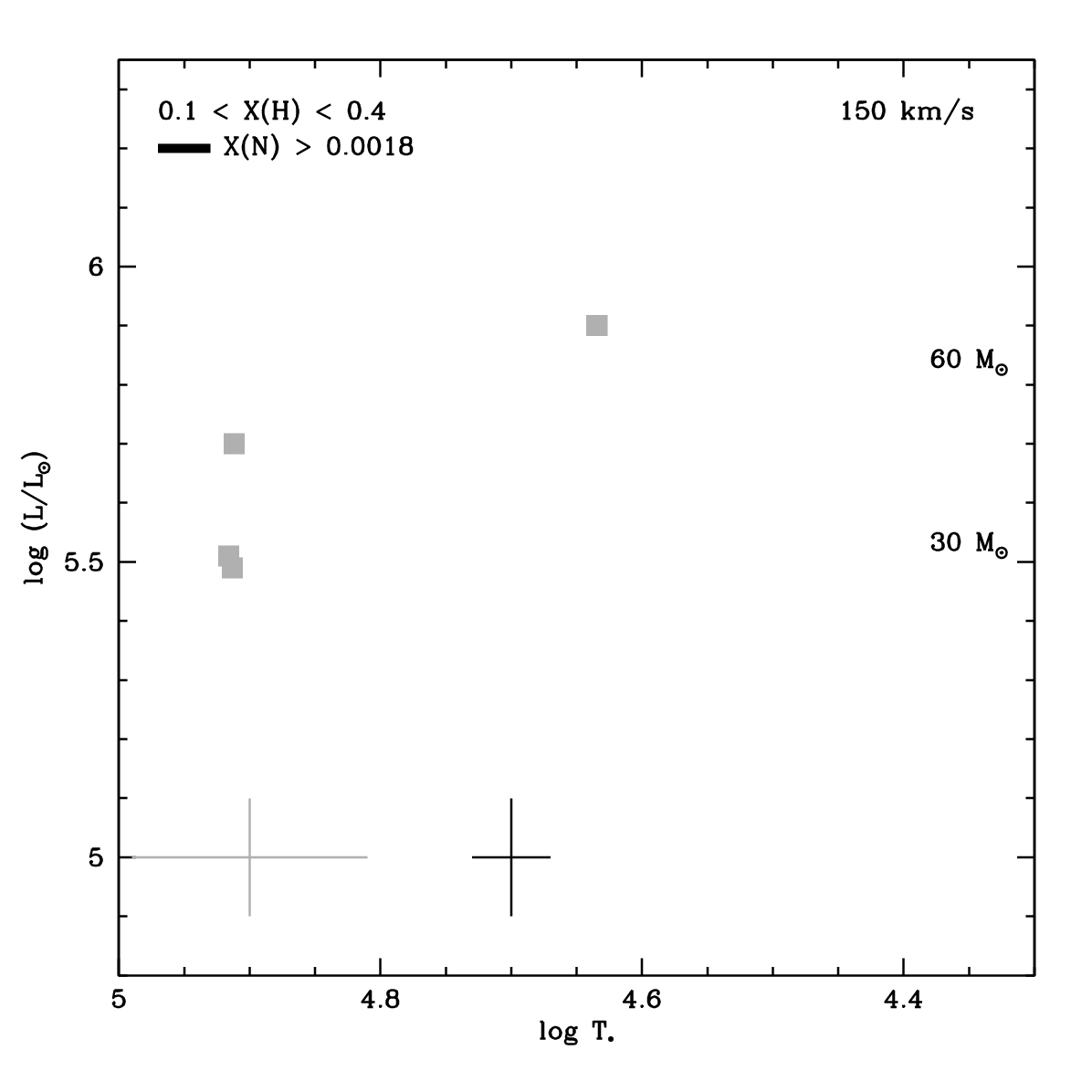}
\includegraphics[width=0.33\textwidth]{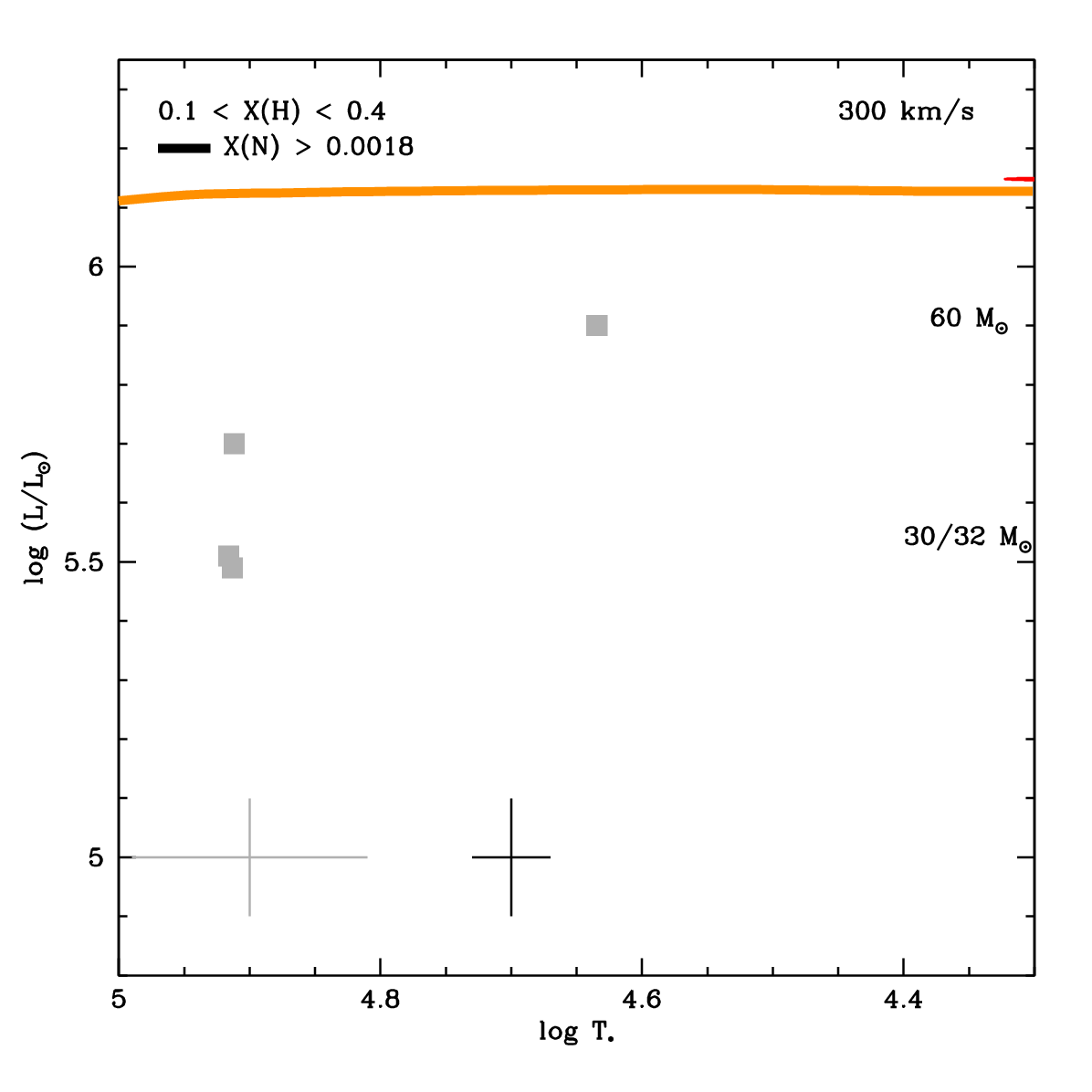}
\includegraphics[width=0.33\textwidth]{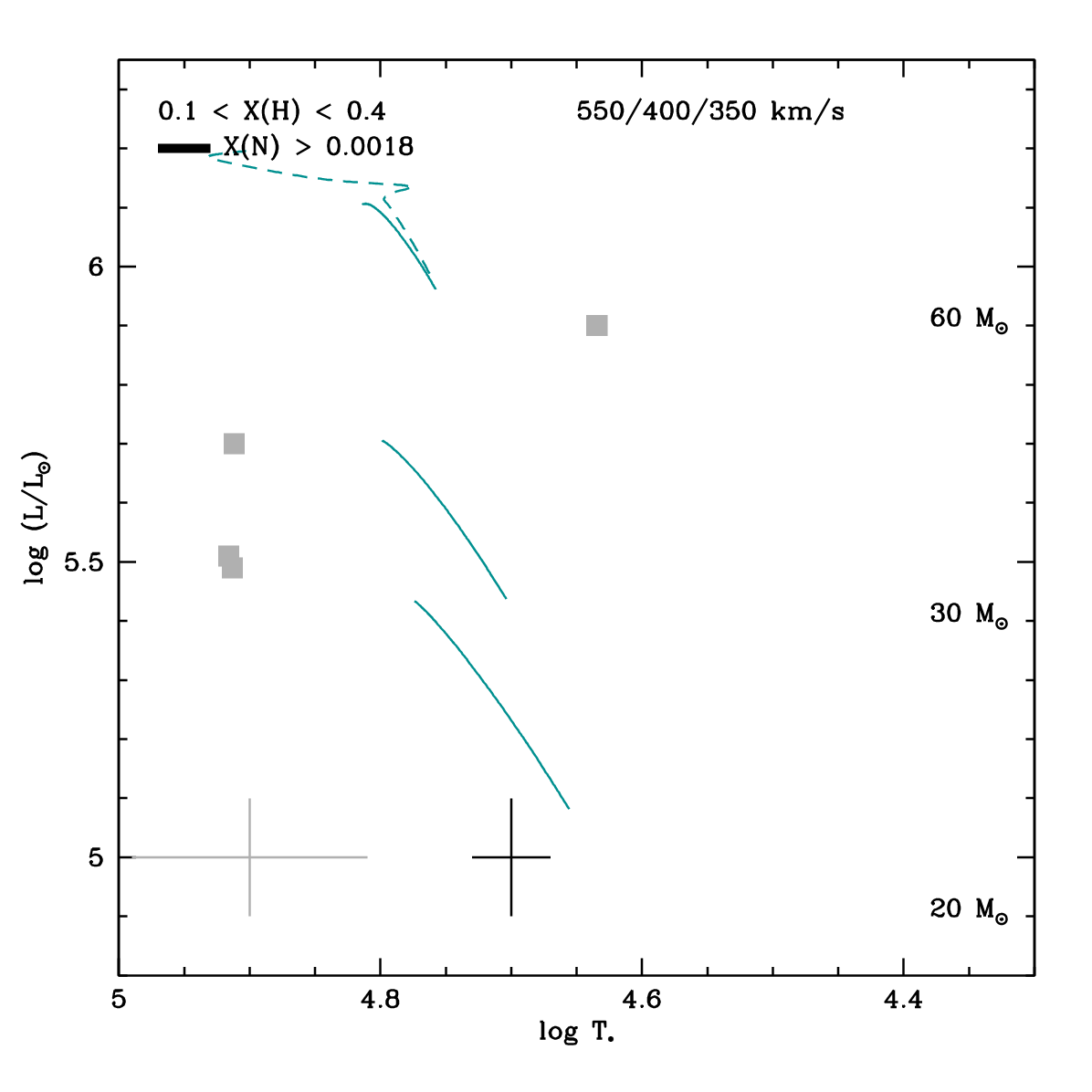}
\caption{Same as Fig.~\ref{hrd_lmc} but for the SMC. In the right panels the dotted and dashed lines are tracks for initial rotational velocities of 350 and 400 \kms{, respectively}. No tracks populate the part of the HRD shown in the top and middle right panels. The red tracks are from \citet{georgy13}. The bold part of the tracks is for a nitrogen mass fraction of greater than 0.0018.}
\label{hrd_smc}
\end{figure*}

Our main goal is to investigate the ability of evolutionary models to simultaneously reproduce the position in the HRD (i.e. the temperature and luminosity) and the surface chemical composition of the sample stars. We focus on hydrogen, helium, and nitrogen here because carbon is not determined for all targets. 
To this end, we compared the position of the sample WNh stars to predictions of various evolutionary models from \citet{brott11}, \citet{ek12}, \citet{kohler15}, \citet{lc18}, \citet{grasha21}, and \citet{mp22}; see Figs~\ref{hrd_gal} to \ref{hrd_smc}.
In all figures of this section, we plot evolutionary tracks as illustrated in Fig.~\ref{hrd}. The broken line shows the full track. The thin solid line is the part for which a certain range of hydrogen mass fraction is predicted. In Fig.~\ref{hrd} this is set to 0.4$<$X$<$0.75. The bold part of the tracks adds another constraint in addition to that on X: it considers only surface nitrogen mass fraction higher than a given threshold, 0.0047 in the figure. In the following, we do not show the full tracks, but only the parts with constraints on hydrogen and nitrogen mass fractions. We consider two ranges for hydrogen: X$>$0.4 and 0.1$<$X$<$0.4. The threshold for nitrogen mass fraction depends on the galaxy: it is set to X(N)$_{ave}$-1$\sigma$ with X(N)$_{ave}$ and $\sigma$ being the average and dispersion{, respectively}, reported in Sect.~\ref{s_surfchem}. Consequently, the bold part of the tracks corresponds to X(N)$>$0.0043, 0.0047, and 0.0018 for the Galaxy, LMC, and SMC, respectively. These assumptions imply that we compare the global properties of the observed stars, and not the individual ones, to predictions of evolutionary models. We discuss how our results depend on these assumptions at the end of this section.

Figure~\ref{hrd_gal} shows the HR diagram for the Galactic stars. Additional single WNh sources for which the surface composition of at least hydrogen is known have been added: sources in the Arches cluster and NGC3603-B. Focusing first on the stars with X$>$0.4 (top panels), moderately rotating models from \citet{lc18} are able to account for the most luminous objects, indicating their very massive nature. The bold part of the tracks, which has H and N surface mass fraction similar to the values determined for the sources, also reproduce the position of these sources in the HR diagram. For relatively fast rotation, the relevant (in terms of HRD position and surface composition) part of the tracks remain confined to near the ZAMS (Fig.~\ref{hrd_gal}, top right panel). The models of \citet{brott11} do not produce enough surface N-enrichment when they pass through the area of the HR diagram occupied by the sample stars. We stress that VMSs require a special treatment of their mass-loss rate that is not implemented in the tracks shown in Fig.~\ref{hrd_gal}. The calculations of \citet{mp22} (see also \citealt{graef21} and \citealt{saba22}) show that tracks evolve to cooler temperatures for VMSs while displaying hydrogen and helium chemical processing at their surface. However, no calculation at solar metallicity is yet available.  

Moving to the subsample with  low H mass fraction (bottom panels), a large variety of tracks can account for the observed stars. In particular, the predictions of \citet{ek12}, \citet{lc18}, and \citet{grasha21} consistently predict the surface chemistry and the HRD position. However, different conditions are needed for different sets of models. The tracks of \citet{lc18} favour moderate rotation, while those of \citet{ek12} and \citet{grasha21} require relatively high rotational velocities to explain the Galactic sources.

Fine-tuning of the implementation of rotationally induced mixing and of the amount of rotation is therefore required to reproduce all stars, but to first order the models appear to be capable of accounting for the Galactic WNh stars of all masses. Further observational constraints on the amount of rotation would be needed to identify which physics implementation in evolutionary models is the most relevant.

The LMC stars are shown in Fig.~\ref{hrd_lmc}, complemented with stars in 30~Dor. Compared to Fig.~\ref{hrd_gal}, an extra value of rotation (550 \kms) is shown in the right panels. \citet{lc18} do not provide tracks at LMC metallicity.
For X$>$0.4 (top panels), the position of the observed stars and their surface H mass fraction are reproduced by the various sets of tracks. This favours an interpretation of WNh stars being young objects near the ZAMS. However, the surface nitrogen mass fraction remains too small in almost all tracks (almost no bold tracks are seen in the top panels). The most massive models of \citet{grasha21} can reproduce the VMSs in 30~Dor with the correct range of surface N composition. The dedicated VMS models of \citet{mp22} are shown in the left panels. These models are for no rotation. We see that they reproduce the position of the most massive objects but underestimate the nitrogen content (see also discussion below).
The less massive objects analysed in the present study escape prediction when the constraint on surface N composition is added to surface H composition and HRD position. In the top right panel, even the very fast rotating models of \citet{brott11} ---which correspond to quasi-homogeneous evolution (QHE; see \citealt{maeder87,langer92,yl06})--- fail to produce as much surface nitrogen as observed.

For stars with 0.1$<$X$<$0.4 (bottom panels), the situation is worse. With the exception of the object with \lL$\sim$6.1, which is explained by standard rotation models (middle panel), none of the other stars can be accounted for by any model with low to moderate rotation. Very fast rotation is required to produce tracks that populate the region occupied by the objects at low luminosity (bottom right panel). However, the amount of surface nitrogen is underpredicted (no bold part exists for the tracks).
In conclusion, for LMC WNh stars, whether they have high or low hydrogen mass fraction, none of the currently available models in the literature can reproduce the observed properties below \lL$\lesssim$6.0.

In the SMC, the most H-rich stars are shown in the top panels of Fig.~\ref{hrd_smc}. The two objects are relatively close to the ZAMS considering uncertainties on their temperatures. The tracks of \citet{brott11} and \citet{georgy13} for low and moderate rotation populate the regions where these two stars are located with the correct range of surface hydrogen, but not of surface nitrogen (no bold track close to the stars). The 60~\msun\ track of \citet{grasha21} with moderate rotation accounts marginally for all the observational constraints. The tracks that best reproduce the HRD position and surface hydrogen mass fractions are those of \citet{brott11} for fast rotation (top right panel of Fig.~\ref{hrd_smc}). Unfortunately, these tracks do not produce enough surface nitrogen: they reach a maximum of 0.0012, while the observational minimum value defined in Sect.~\ref{s_surfchem} is 0.0018.

The objects with 0.1$<$X$<$0.4 shown in the bottom panels of Fig.~\ref{hrd_smc} escape all predictions. Models with low or moderate rotation do not cover the relevant portions of the HRD. The QHE models of \citet{brott11} do not reach the high effective temperatures of the four of the observed stars. All stars have X(N) $>$ 0.0035, much larger than the maximum value predicted by the models (0.0012). These shortcomings are further discussed in Sect.~\ref{s_disc_evol}.

\vspace{0.5cm}

The results we present immediately above partly rely on the choice of the minimum surface nitrogen mass fraction used to highlight some parts of the tracks in Figs.~\ref{hrd_gal} to \ref{hrd_smc}. As discussed above, we chose to set this minimum value to X(N)$_{ave}$-1$\sigma$ for the three galaxies. Inspection of Fig.~\ref{histoxn} and Table.~\ref{tab_param} reveals that a few objects actually have X(N) lower than these limits. They represent the extreme of the distributions of Fig.~\ref{histoxn}. In order to test whether our results are affected by the definition of the minimum surface nitrogen fraction in evolutionary tracks, we show similar figures in Appendix \ref{ap_hrdxn} to Figs. \ref{hrd_gal} to \ref{hrd_smc} but now lowering the minimum nitrogen mass fraction used to code the bold part of the tracks. We now adopt the minimum value of all stars in each galaxy: 0.0025 in the Galaxy, 0.0032 in the LMC, and 0.0012 in the SMC \citep[see][]{martins09}. The results are presented in Figs.~\ref{hrd_gal2} to \ref{hrd_smc2}. In the Galaxy and the LMC, the conclusions are almost unchanged. Very little difference in the various HRDs is observed, except that the VMS models of \citet{mp22} now account not only for the position but also the surface nitrogen content of the most massive stars (the literature objects shown by open circles have the same range of X(N) as that of the objects studied here; see \citealt{brands22}). In the SMC, a fraction of the the 32~\msun\ track of \citet{grasha21} now accounts for the new constraint on X(N), allowing the model to explain the WNh stars with the largest surface hydrogen mass fraction. The most striking changes are for the very fast rotating models of \citet{brott11} that now spend a fraction of their time with X(N)$>$0.0012. These models can therefore marginally explain the properties of star AB1 \citep{martins09}. Besides this, the conclusions remain unchanged. In particular, the objects with X$<$0.4 are located too far away from the tracks and remain more N-rich than predicted by models.

\vspace{0.5cm}

We can therefore summarise our main conclusions as follows. Overall, WNh stars can be reasonably reproduced by standard evolutionary models that include rotation at solar metallicity. Different models succeed differently in accounting for the observed properties, but it is possible to find at least one model that produces the observed HRD position together with the surface H and N mass fractions.
At lower metallicity, this is not the case. VMSs can be accounted for but WNh stars with luminosities below 10$^6$ \lsun\ usually escape predictions. QHE may be a solution, but constraints on the surface nitrogen content limit this interpretation to one object in the SMC. These results are discussed in Sect.~\ref{s_disc_evol}, but before this, we examine binary models. 

\subsubsection{Binary tracks}
\label{s_bin}

Our sample selection was based on the absence of evidence for binarity, and so we do not expect the sample stars to be part of binary systems. However, for the sake of completeness, we perform a comparison of the observed properties with models from the LMC grid of \citet{pauli22}, which covers a range of mass ratios and orbital periods. 

\begin{figure*}[t]
\centering
\includegraphics[width=0.33\textwidth]{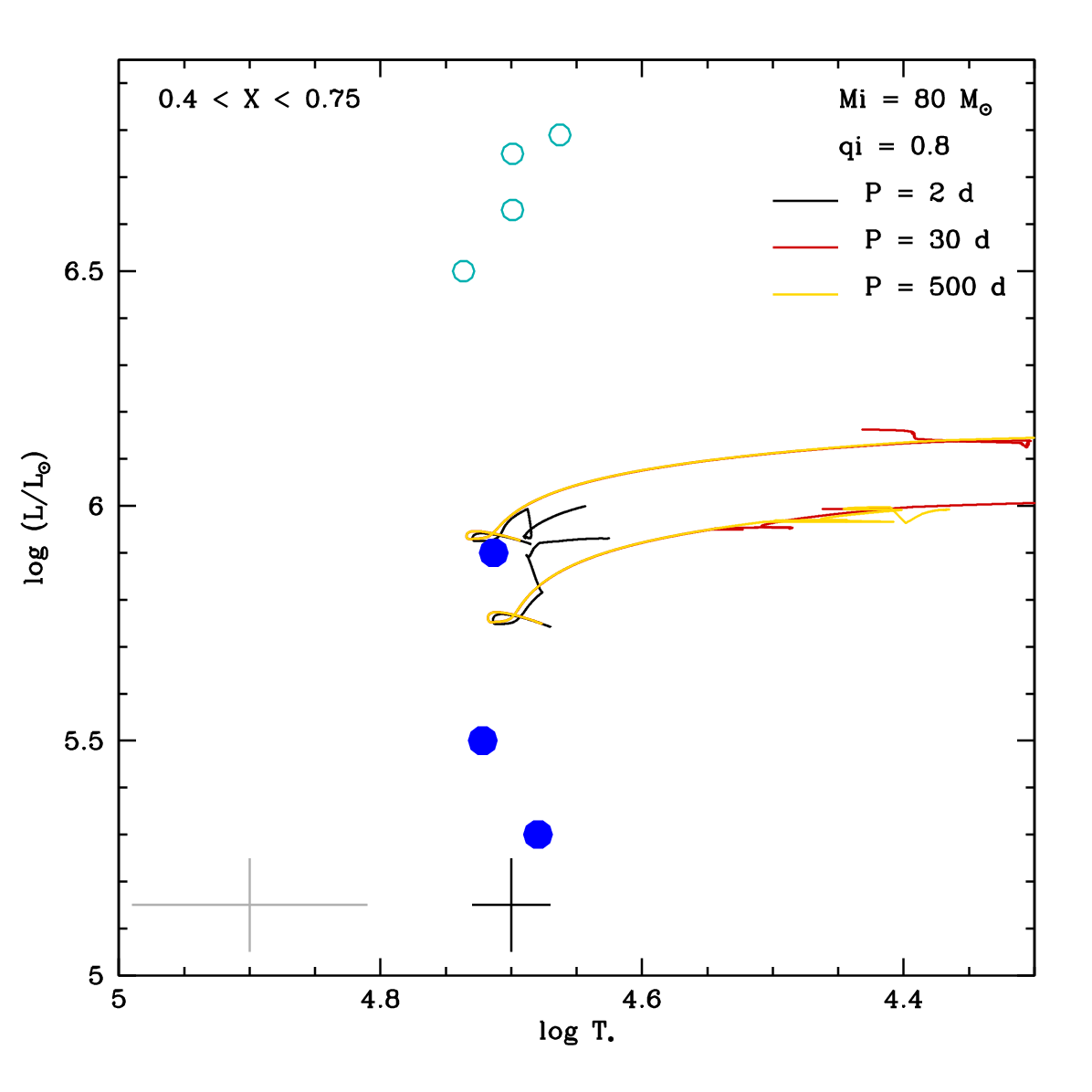}
\includegraphics[width=0.33\textwidth]{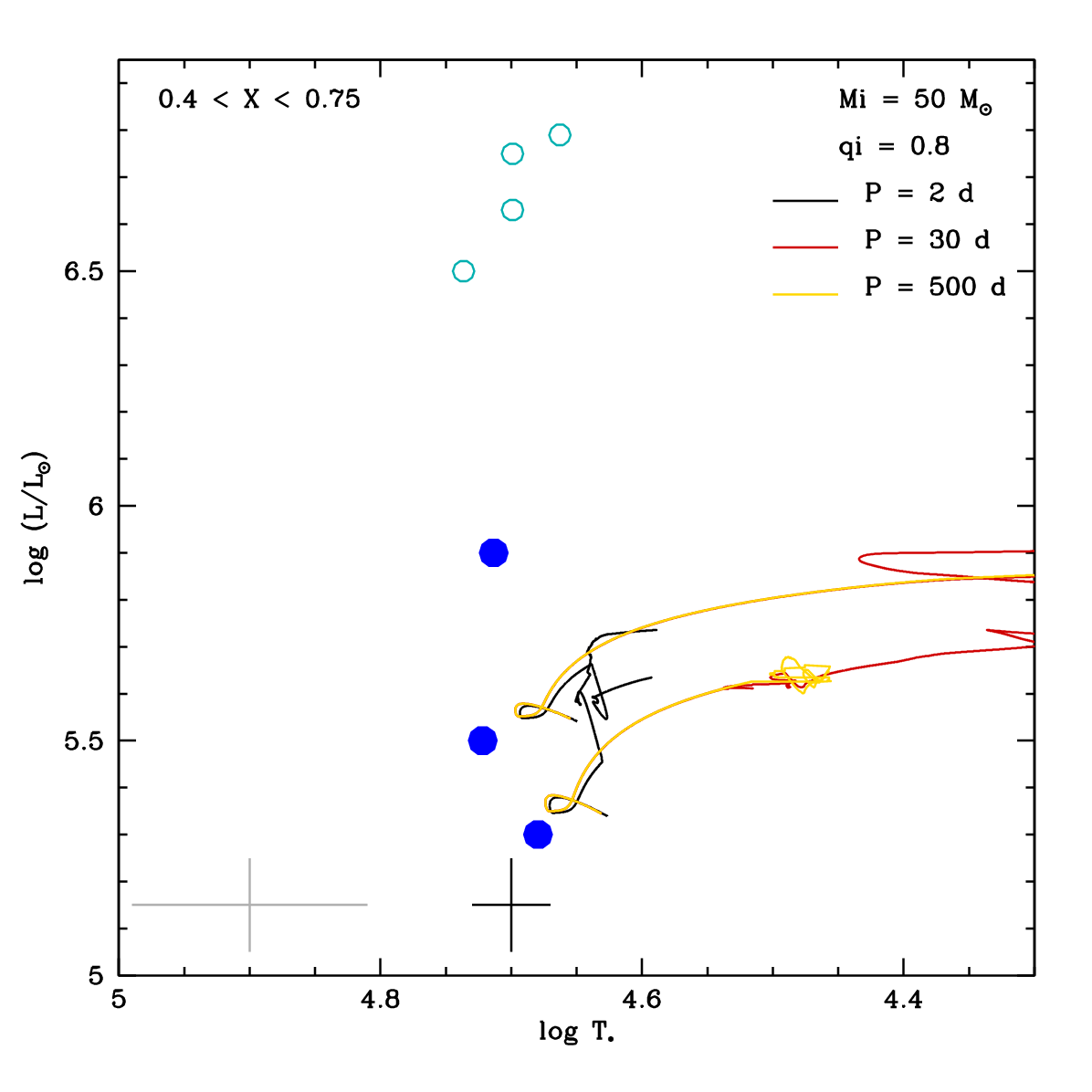}
\includegraphics[width=0.33\textwidth]{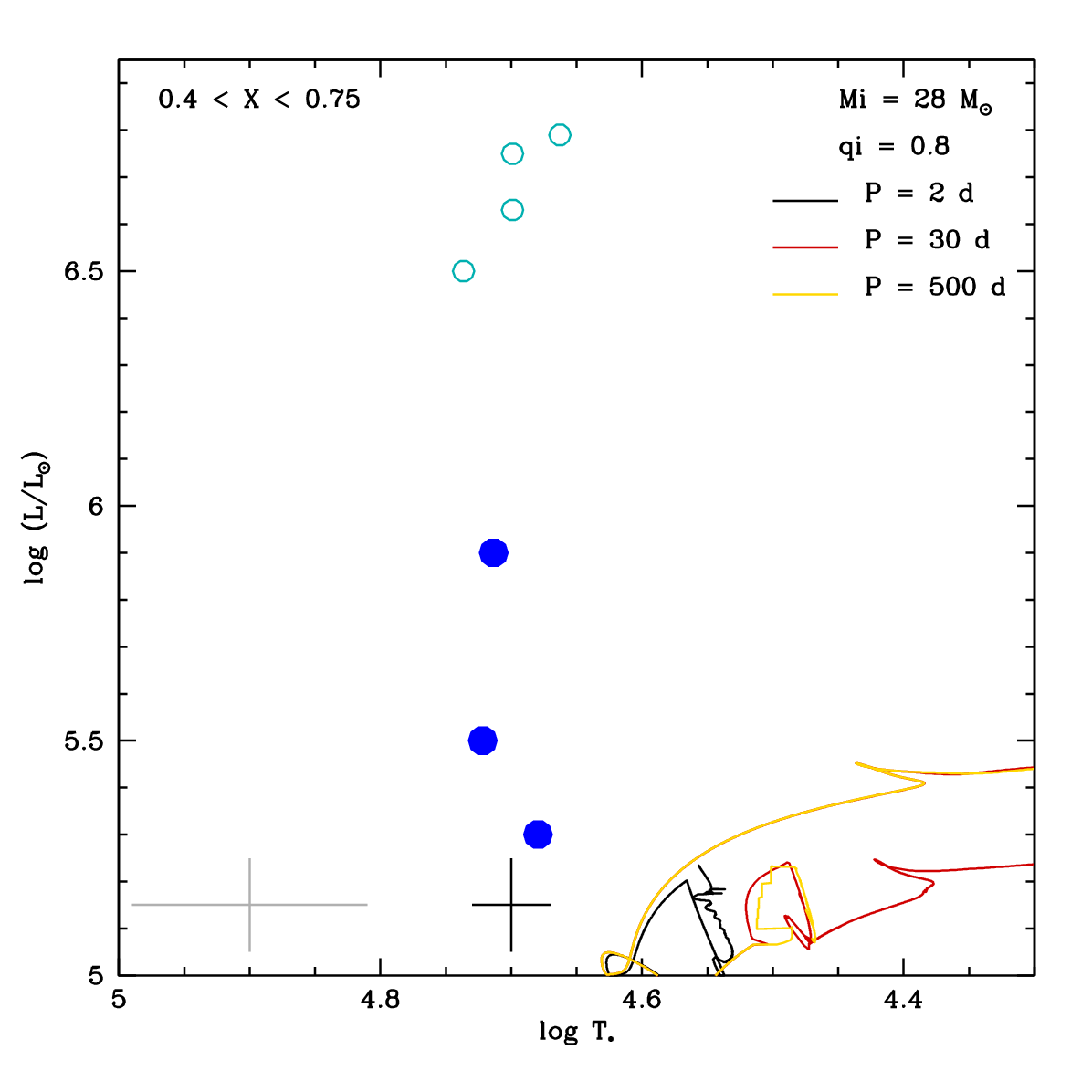}

\includegraphics[width=0.33\textwidth]{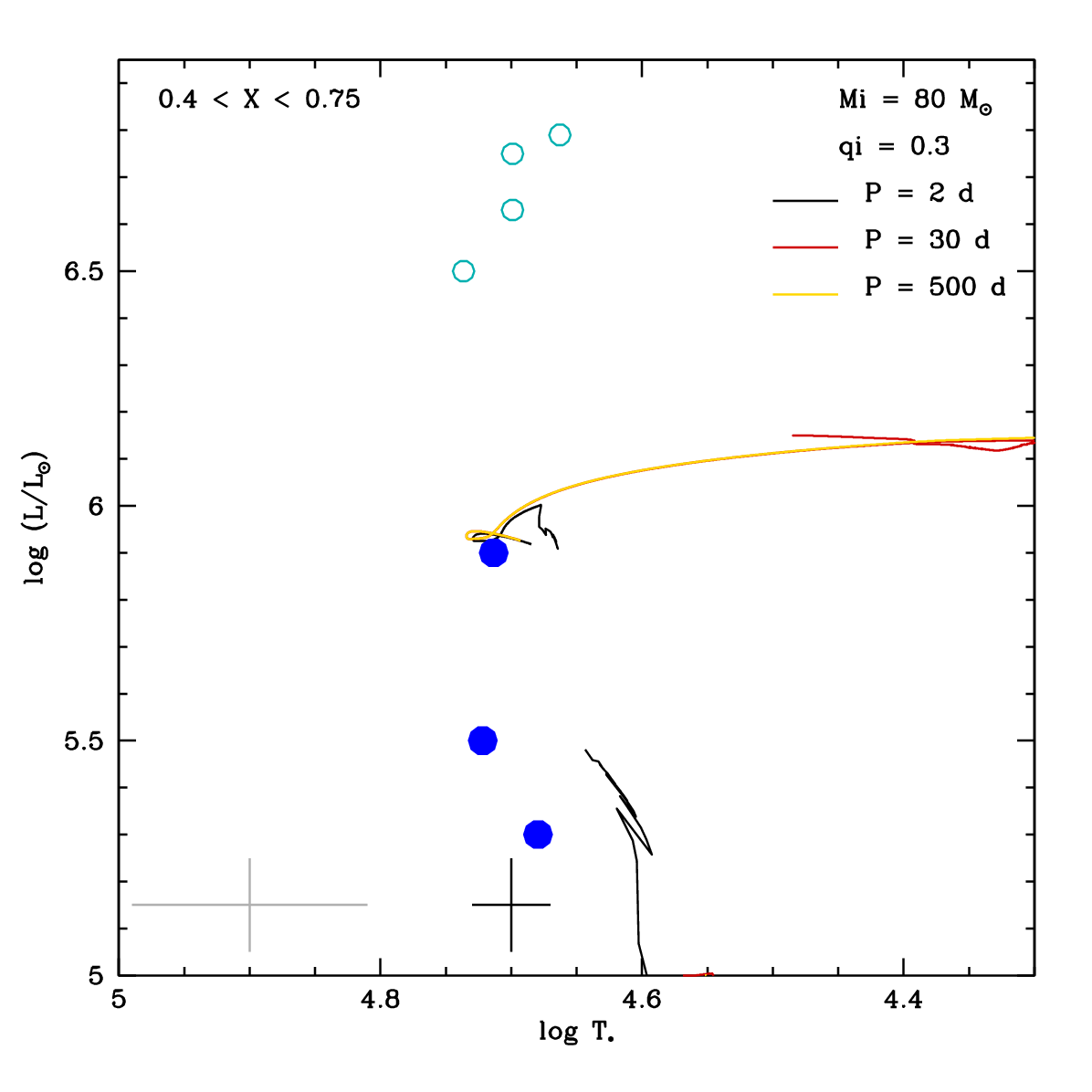}
\includegraphics[width=0.33\textwidth]{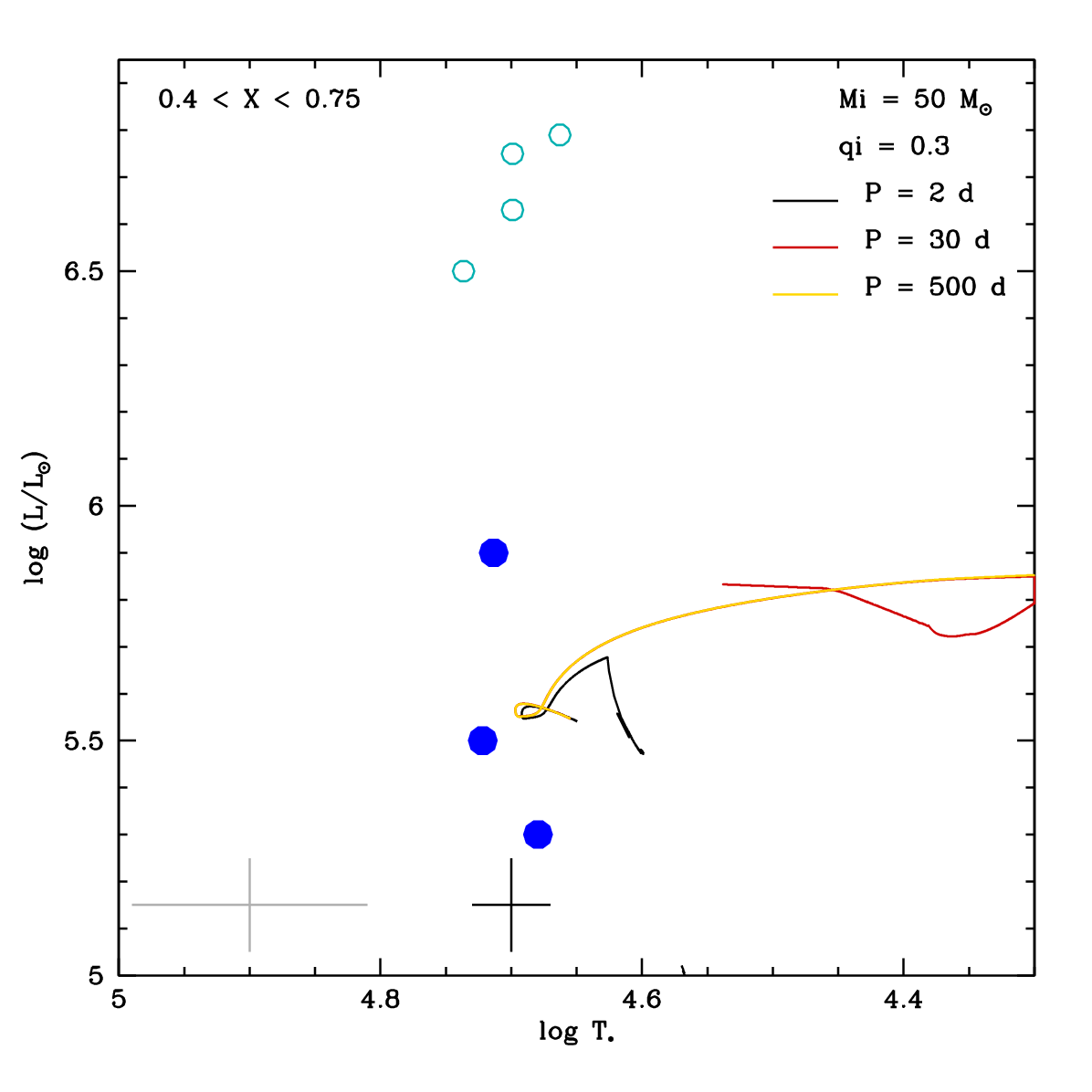}
\includegraphics[width=0.33\textwidth]{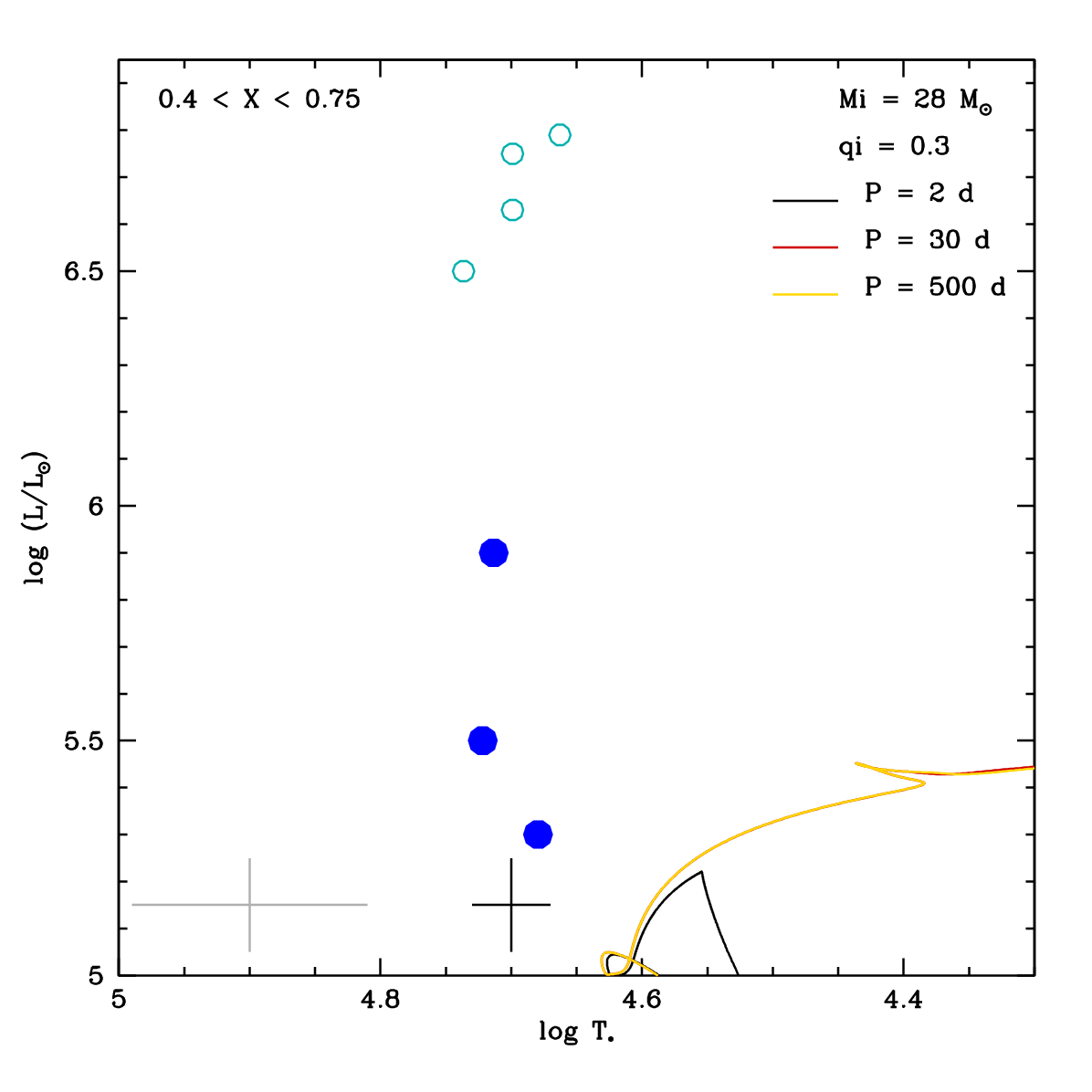}
\caption{LMC stars together with selected binary models from \citet{pauli22}. The hydrogen mass fraction is larger than 0.4 in both stars and models. The left, middle, and right panels are for a donor mass of 80, 50, and 28 \msun{, respectively}. Top and bottom panels correspond to mass ratios of 0.8 and 0.3. In each panel, the black, red, and orange tracks are for systems with periods of 2, 30, and 500 days{, respectively}. }
\label{hrd_lmc_bin_Xhigh}
\end{figure*}

\begin{figure*}[t]
\centering
\includegraphics[width=0.33\textwidth]{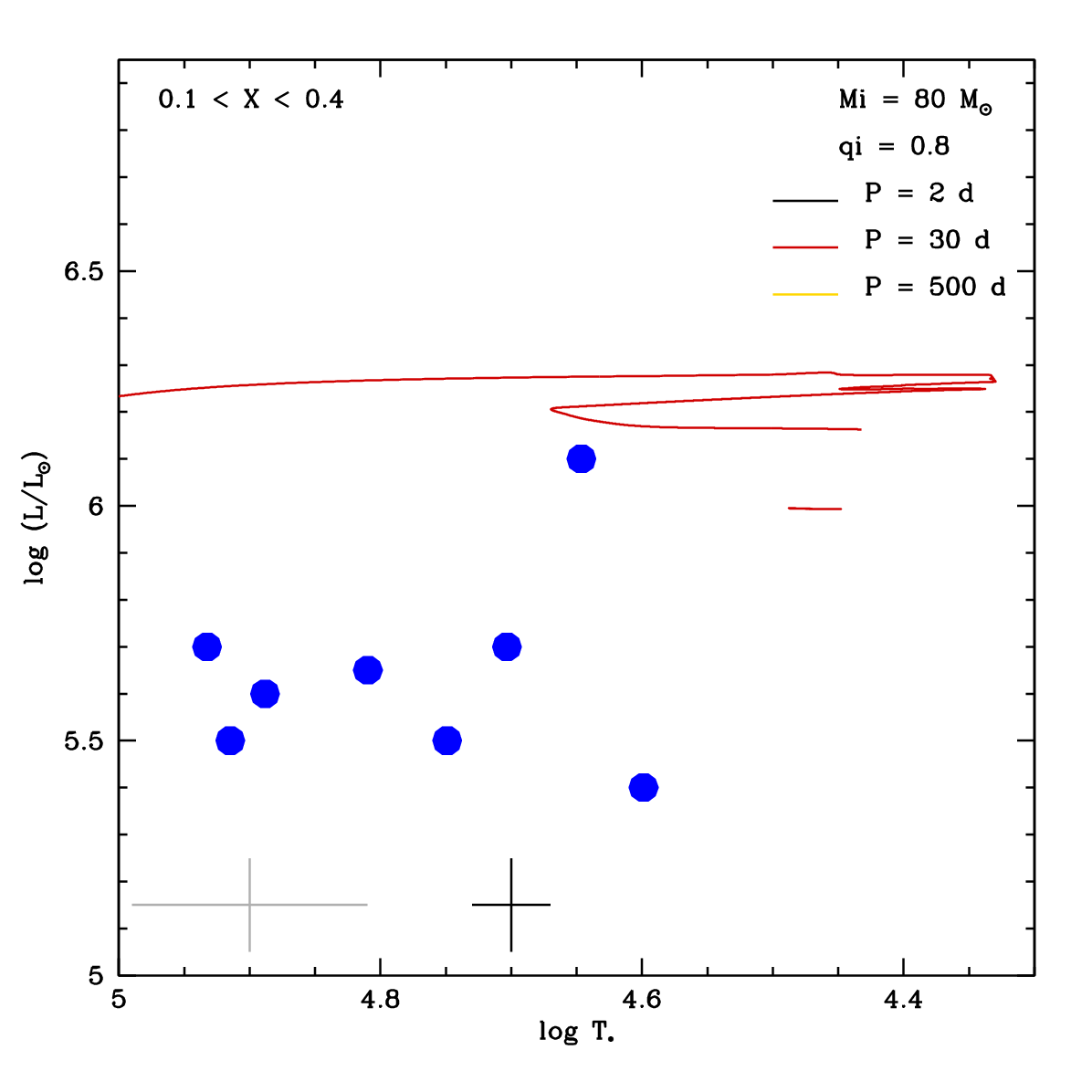}
\includegraphics[width=0.33\textwidth]{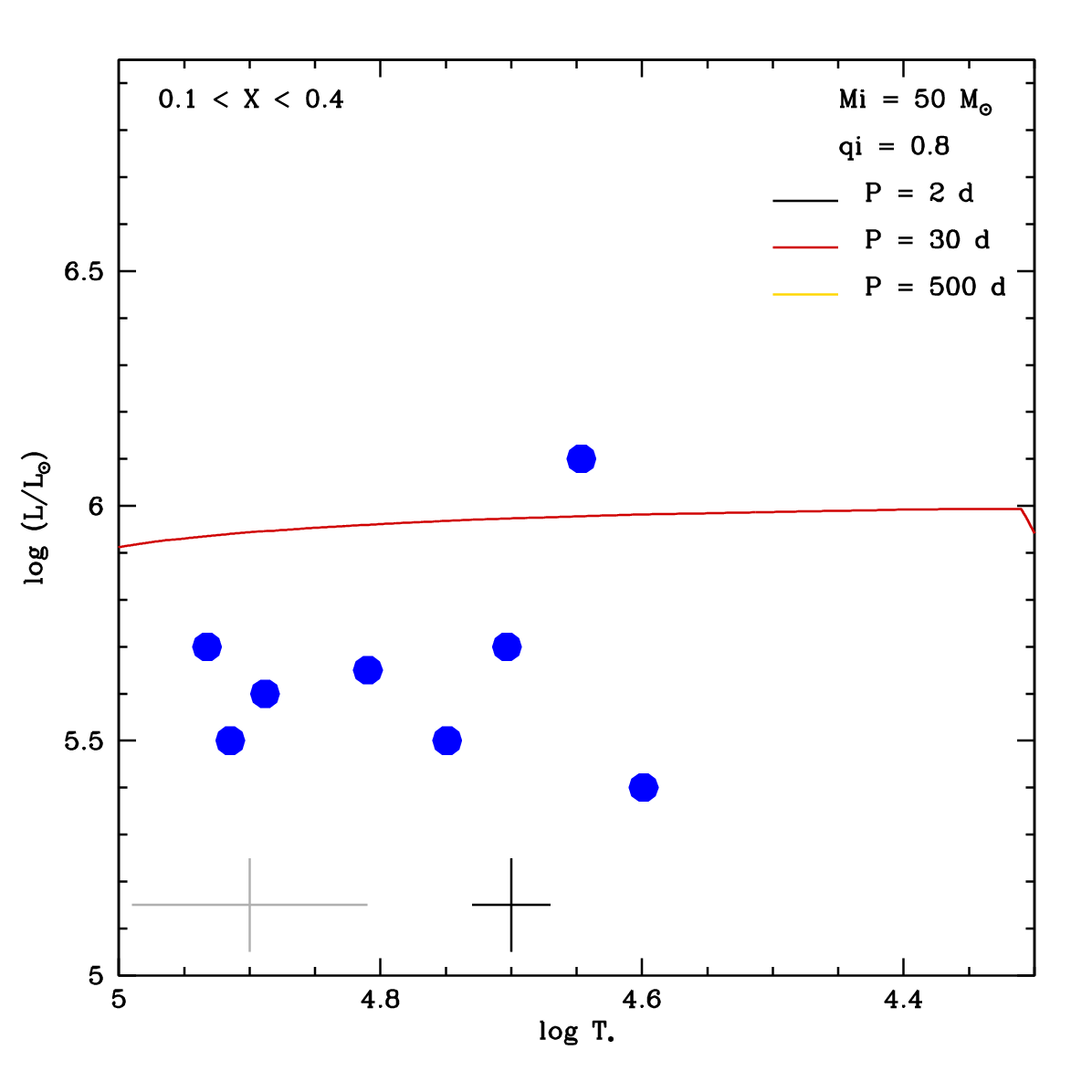}
\includegraphics[width=0.33\textwidth]{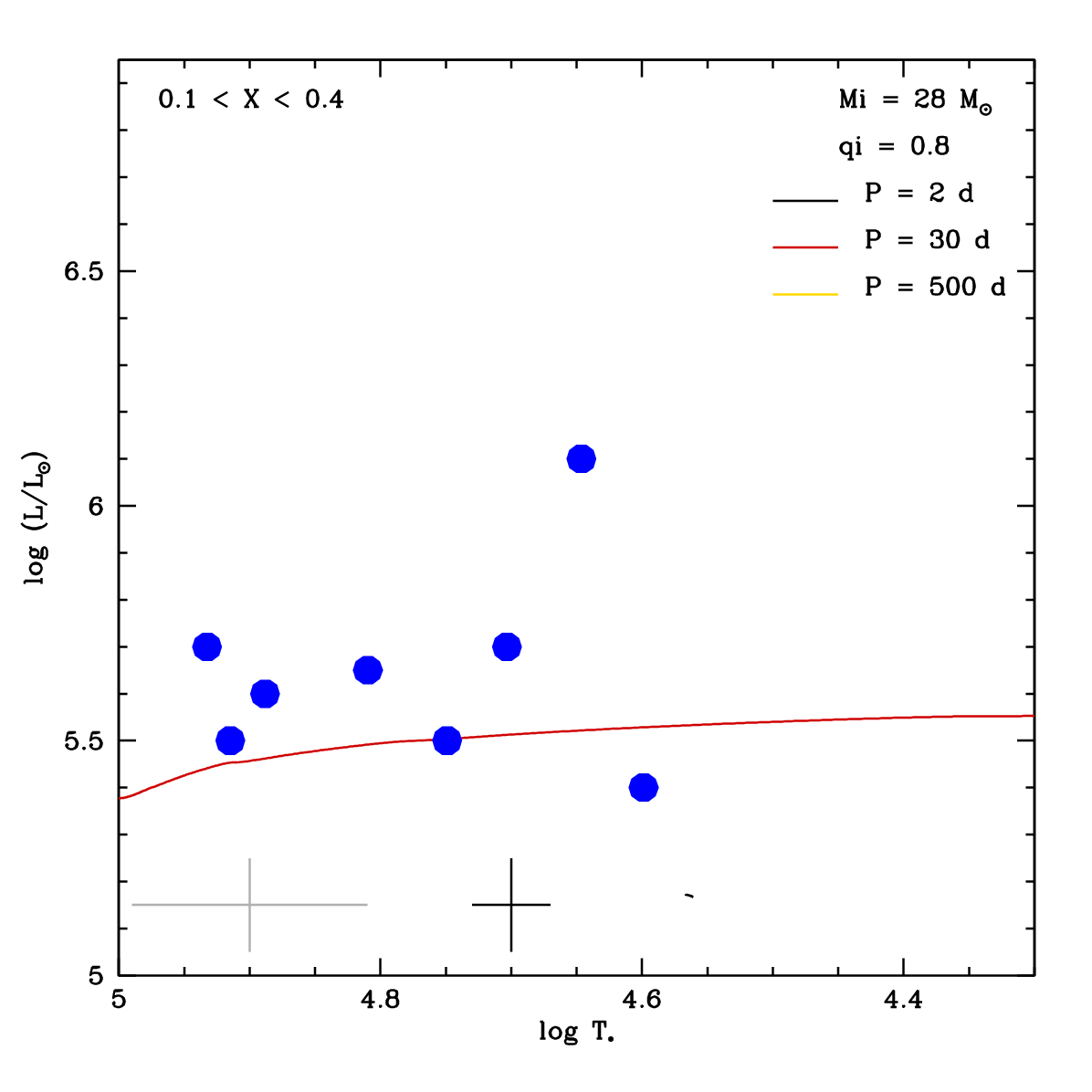}

\includegraphics[width=0.33\textwidth]{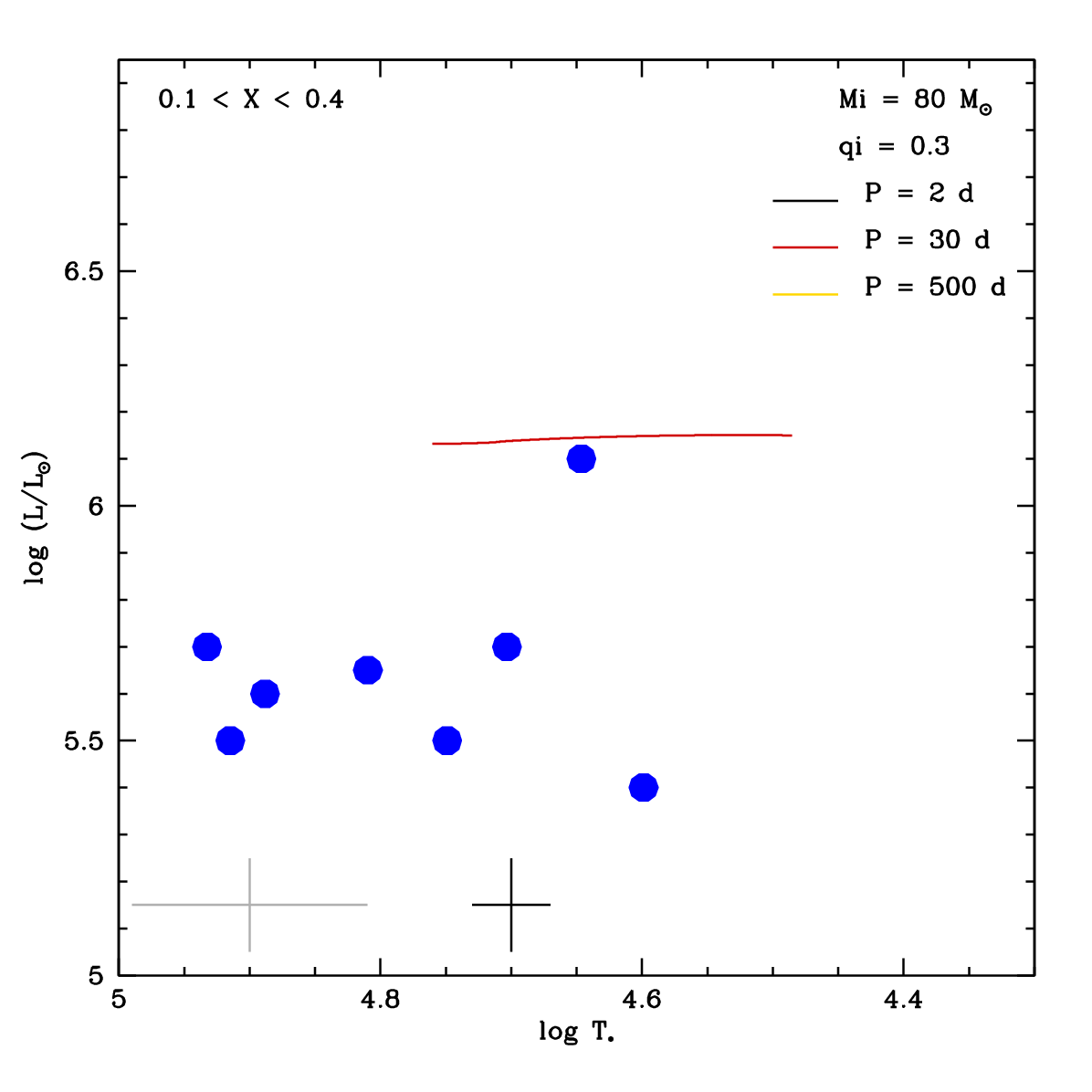}
\includegraphics[width=0.33\textwidth]{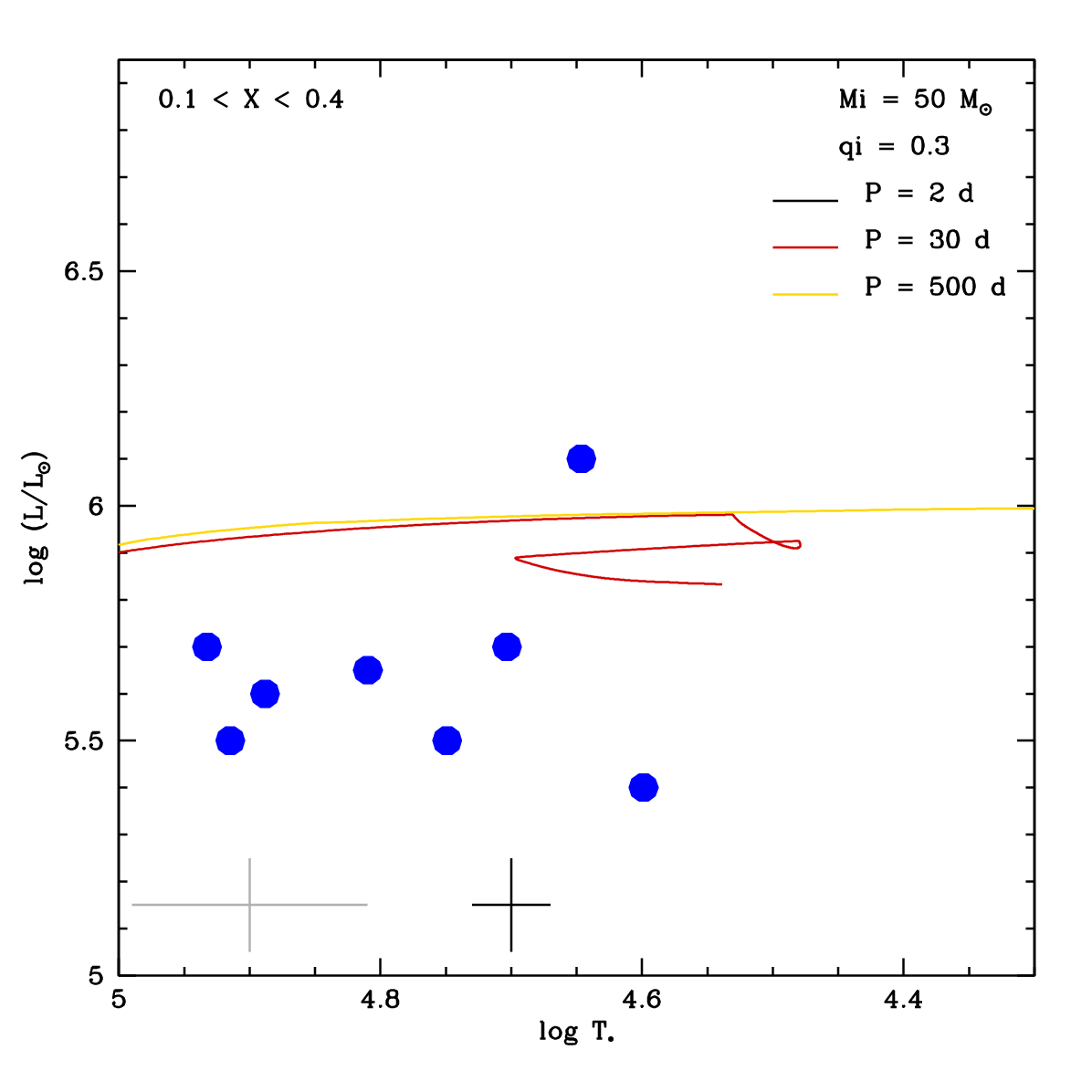}
\includegraphics[width=0.33\textwidth]{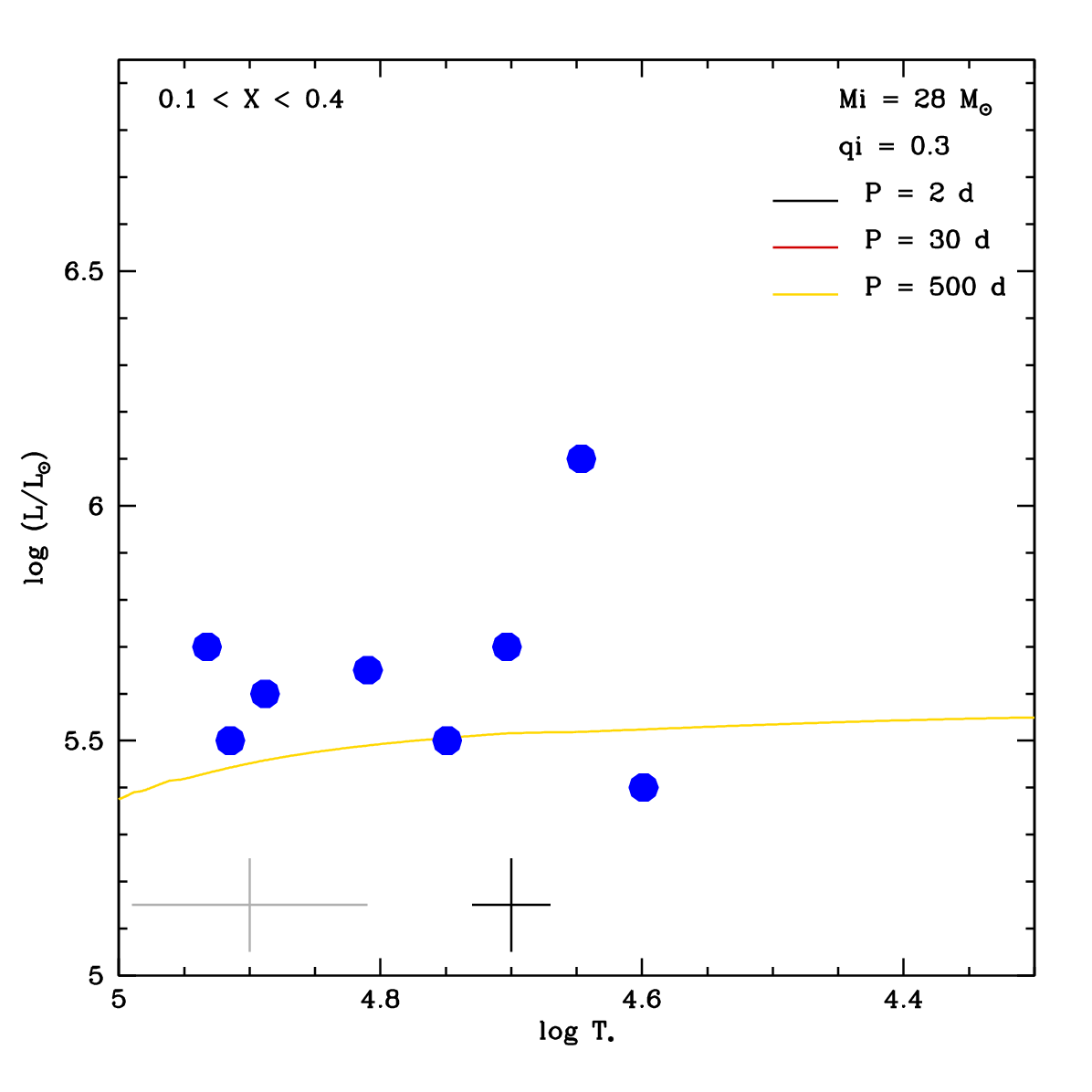}
\caption{Same as Fig.~\ref{hrd_lmc_bin_Xhigh} but for hydrogen mass fractions of between 0.1 and 0.4.}
\label{hrd_lmc_bin_Xlow}
\end{figure*}

Figures~\ref{hrd_lmc_bin_Xhigh} and \ref{hrd_lmc_bin_Xlow} show a selection of these binary models; they roughly sample the parameter space of the larger grid computed by Pauli et al. The primary mass is 28, 50, or 80~\msun, the mass ratio 0.3 or 0.8, and the period 2, 30, or 500 days. Figure~\ref{hrd_lmc_bin_Xhigh} compares models with stars that have X$>$0.4, while Fig.~\ref{hrd_lmc_bin_Xlow} is for the case where 0.1$<$X$<$0.4.
For high hydrogen mass fractions, several models cross the area occupied by the sample stars. In most cases, mass transfer has not happened yet and models behave essentially as single star models. In the case of low hydrogen mass fraction (Fig.~\ref{hrd_lmc_bin_Xlow}), the selected binary models appear to be better suited to reproducing the position of the sample stars. Contrary to Fig.~\ref{hrd_lmc} for single stars, several binary models with periods of between 30 and 500 days encompass the region where the LMC low-X stars are located.

At first sight, binary models are therefore an attractive option to explain the hydrogen-poor LMC stars. However, there is a limitation to the current set of models. As is the case  for very-fast rotating single-star models, when the binary tracks reach the position of the observed stars, they have the correct range of hydrogen mass fraction but  are much more N-poor than the LMC objects. Indeed Figs.~\ref{hrd_lmc_bin_Xhigh} and \ref{hrd_lmc_bin_Xlow} use the same coding as in Fig.~\ref{hrd_lmc}, that is, bold tracks should represent the part of the tracks with X(N) larger than 0.0032 (i.e. the strict lower limit on X(N) discussed above). No bold tracks are seen in Figs.~\ref{hrd_lmc_bin_Xhigh} and \ref{hrd_lmc_bin_Xlow} because X(N) remains below the 0.0032 threshold. The largest value reached by any of the binary models is 0.0029. In spite of the speculative binary nature of the LMC WNh stars in our sample, the models of Pauli et al. do not produce enough surface nitrogen to explain their properties.

\subsection{Mass-loss rates}

\begin{figure}[t]
\centering
\includegraphics[width=0.49\textwidth]{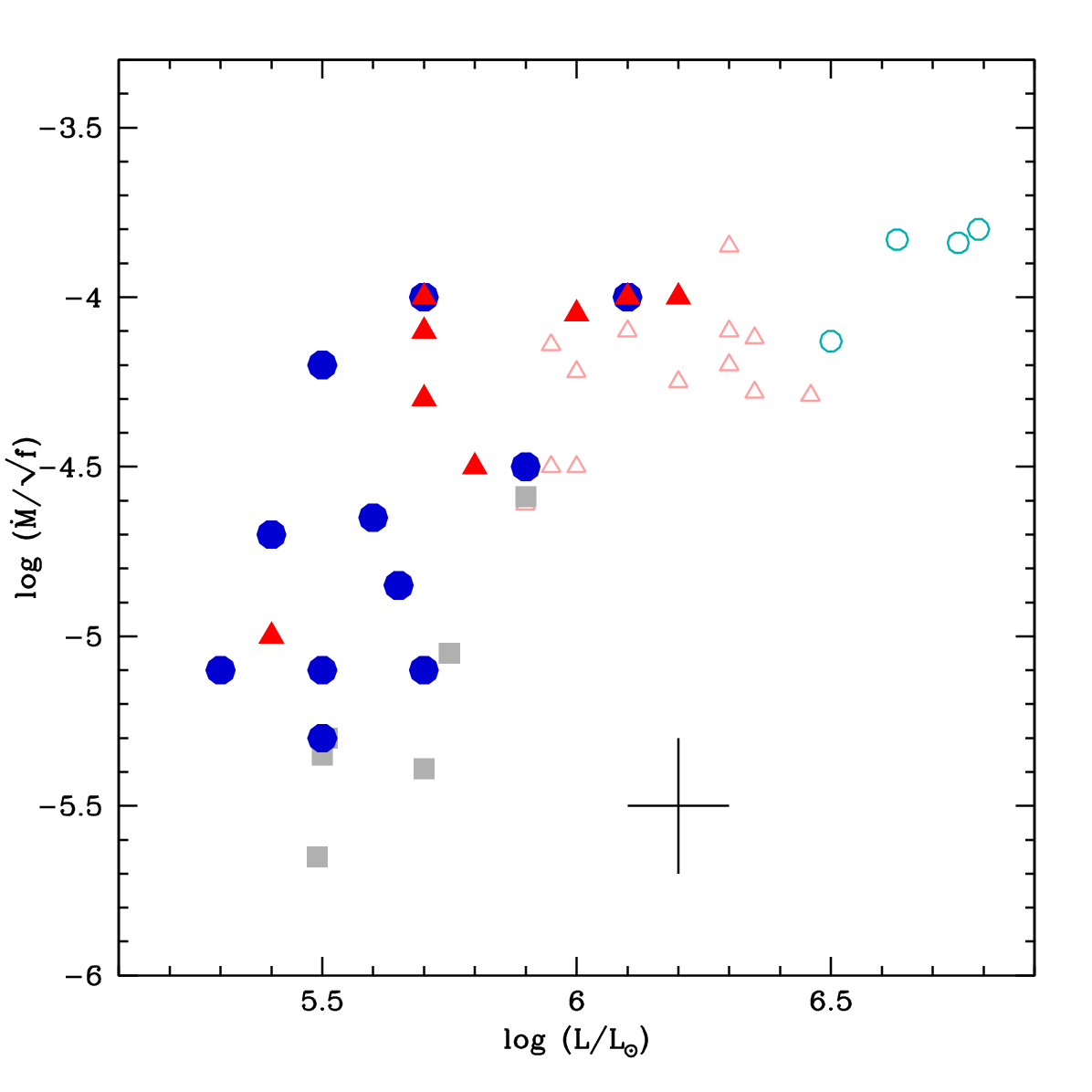}
\caption{Clumping-corrected mass-loss rate as a function of luminosity for the sample stars. Filled symbols have the same meaning as in Fig.~\ref{X_L}. Open symbols are literature data, and triangles and circles are Galactic and LMC stars{, respectively}.}
\label{lmdotf}
\end{figure}

Before discussing shortcomings and potential improvements of evolutionary models and spectroscopic analysis, we briefly summarise the results on the mass-loss rates of the sample WNh stars.
Radiatively driven winds are known to scale with luminosity \citep{cak,kud89,puls96,mokiem07} as a results of the interaction of photons with ions in the upper layers of massive stars. As a result, a metallicity dependence of mass-loss rates has been predicted \citep{vink01,bjork21,vs21} and observed \citep{mokiem07,rickard22} for OB-type stars. For Wolf-Rayet stars, the mechanisms at the origin of stellar winds are more complex. Multiple scattering and proximity to the Eddington limit imply that WR winds depend in an intricate way on luminosity, Eddington factor, and metallicity \citep{gh08,graefener11,sv20}. The amount of hydrogen in the upper layers also appears to be an important factor \citep{sander22}.
In absence of a reliable mass estimate for our sample stars, we cannot estimate the Eddington factor. The clumping-corrected mass-loss rates are therefore plotted as a function of luminosity in Fig.~\ref{lmdotf}. Additional single WNh stars from the literature are added. Two main qualitative conclusions can be drawn. First, above $10^6$~\lsun, mass-loss rates are quite homogeneous with a dispersion of about a factor of 2 around $10^{-4}$~\msun. Below $10^6$~\lsun, a global decrease in mass-loss rates is observed as \lL\ decreases. At the same time, the scatter increases considerably. For instance, in the LMC, differences in \mdot\  of an order of magnitude are observed. In the low-luminosity range, another general trend is seen: the SMC stars have smaller \mdot\ than LMC stars for a given luminosity. We refrain from further quantifying this trend because of (1) the small number of objects and (2) the hidden dependencies in that figure, in particular that on the Eddington factor. We checked whether or not the hydrogen mass fraction could explain the dispersion at low luminosity, but no clear trend emerges. Unfortunately, we cannot probe the Eddington factor, because we do not have access to the mass. Surface gravity is unconstrained and mass estimates from the HR diagram are only possible for Galactic stars. Indeed, in the MCs, evolutionary models struggle to account for the position of the sample stars in the HRD, meaning that an evolutionary mass cannot be determined. Consequently, the Eddington factor is unconstrained.
Irrespective of this latter issue, the qualitative trend with metallicity we observe for WNh stars is consistent with that quantified by \citet{hainich15}. Using published values from their group, these authors showed that average mass-loss rates of WN-type stars scale with $Z^{1.2 \pm 0.1}$, $Z$ being the global metallicity. 

\section{Discussion}
\label{s_disc}

\subsection{Stellar evolution}
\label{s_disc_evol}

We show that evolutionary models at low metallicity face serious difficulties in reproducing the observed properties of WNh stars. This is not a surprise given that this has already been stressed in the past for both H-rich and H-free WN stars (see below). However, the difficulties are amplified when considering the constraints on the surface nitrogen abundance we present in this study.

\citet{hainich14} noted that the evolutionary models of \citet{mm05} were not able to reproduce the lowest-luminosity LMC WN stars. The problem was mainly due to the presence of numerous H-free WN stars below the least luminous track able to reach the left part of the ZAMS, where these WN stars are found (see their Fig.~10). At the same time, these tracks reproduced the H-rich WN stars reasonably
well. At that time, no constraint on the surface nitrogen content existed and no further test could be performed. We show in Sect.~\ref{s_sing} that more recent generations of evolutionary tracks faced severe problems in accounting for the population of LMC WNh stars. At present, none of the available models are able to explain the WNh stars with X$<$0.4 and \lL\ $<$ 6.0.

The models of \citet{brott11} and \citet{pauli22} adopt the following initial abundances: 12+log(C/H)=7.75, 12+log(N/H)=6.90, and 12+log(O/H)=8.35. The average LMC values reported by \citet{vink23} are 8.03, 7.03, and 8.40, respectively. For nitrogen, measurements based on supernova remnants \citep{dopita19} and on the gas in 30~Dor \citep{peimbert03} indicate 12+log(N/H)$\sim$7.2, which is about a factor of two larger than that adopted by Brott et al. and Pauli et al. We recall that the maximum nitrogen mass fraction observed in our sources corresponds to CNO equilibrium, at which most of the mass initially present in C, N, and O is found as nitrogen. With the above assumption, the models of Brott et al. and Pauli et al. reach at most X(N)=0.00288.

The average nitrogen mass fraction we obtain in the LMC is 0.0066$\pm$0.0028 (see Sect.~\ref{s_surfchem}). This is larger than the maximum nitrogen content of the Brott and Pauli models (by a factor of $\sim$2). However, slightly increasing the initial C/H and N/H values  of these models so that C/H and N/H stay in the range allowed by observational constraints would help to get closer to the average nitrogen mass fraction. In that case, the QHE models presented in the bottom panels of Fig.~\ref{hrd_lmc} may be able to reproduce the full set of observational constraints. The binary models of Pauli et al. may also display larger surface nitrogen contents while passing through the HRD areas occupied by WNh stars with \lL$<$6.0. Consequently, a potential solution to the failure of LMC evolutionary models could be to revise their initial chemical composition. \citet{martinet23} reached the same conclusion when studying the nitrogen content of the three most massive stars in R136.

\begin{figure*}[t]
\centering
\includegraphics[width=0.49\textwidth]{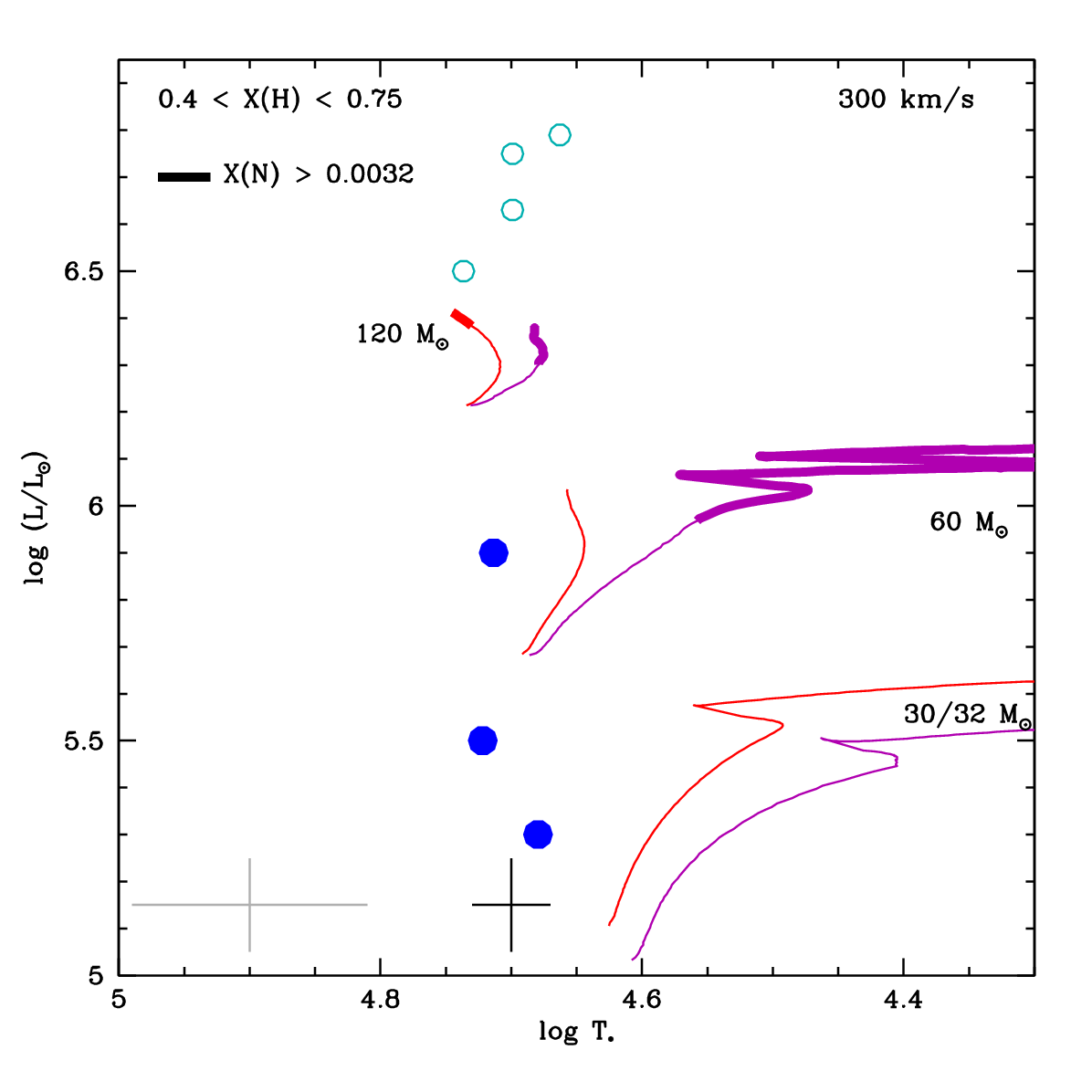}
\includegraphics[width=0.49\textwidth]{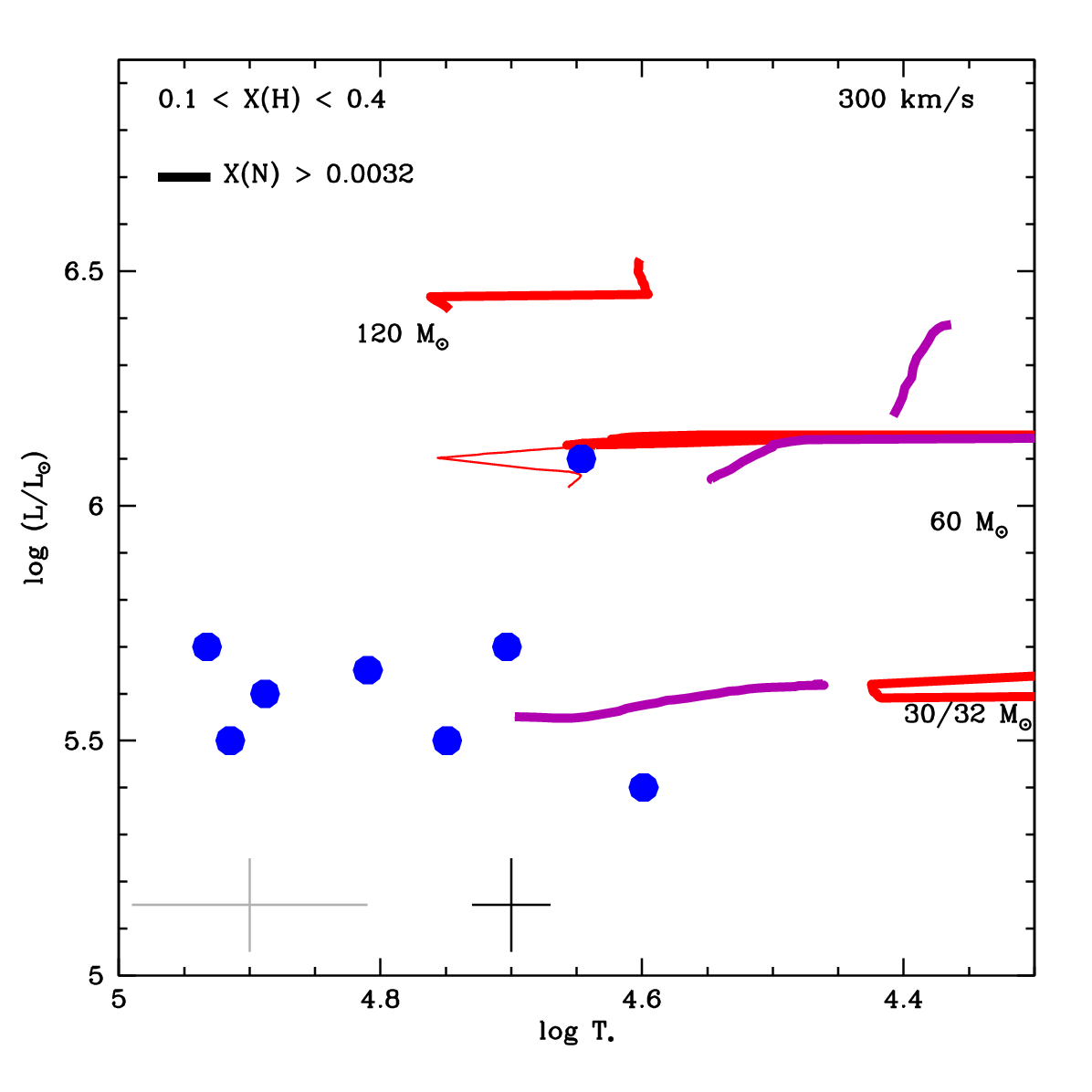}
\caption{HR diagram for LMC stars with the 2005 Geneva tracks from \citet{mm05} overplotted in magenta on top of the 2021 tracks from \citet{eggen21} shown in red. Coding is the same as in Fig.~\ref{hrd_lmc}. The 2005 (2021) tracks are for Z=0.008 (0.006). The bold part is for X(N)$>$0.0032.}
\label{hrd_lmc_mm05}
\end{figure*}

In support of this possibility, we show in Fig.~\ref{hrd_lmc_mm05} the HRD for LMC stars with the Geneva models of \citet{mm05} in addition to the more recent ones from \citet{eggen21}. The initial metal content seems to be the major difference between the two calculations: the global metallicity is 0.008 in the models of Meynet et al., while it is 0.006 in those of Eggenberger et al. For the H-rich stars, the older models of Meynet et al. do not lead to any significant improvement. On the contrary, the stars with X$<$0.4 are partly reproduced by the 30~\msun\ 2005 track, while no model of the 2013 grid populates the relevant region of the HRD. However, only the coolest of the WNh stars are explained by the 2005 track, the hottest ones remaining out of reach of any model. This exercise illustrates that fine-tuning the initial chemical composition has an impact on the exact evolutionary paths.

Returning to Fig.~\ref{xnxc}, we see that the Geneva models are able to display a large amount of nitrogen at their surface. In the LMC and the Galaxy, the surface mass fractions even exceed those observed at the surface of the sample WNh stars. We have seen that, for Galactic stars, the large nitrogen content was predicted at the correct place in the HRD, but in the LMC this is not the case: when models predict N-rich objects, they do not have the temperature and luminosity required to explain the LMC sample stars. Stellar parameters are therefore not consistently predicted.

The population of SMC WN stars was studied by \citet{hainich15}. All are of the WNh subtype and were found to contain hydrogen at their surface. Hainich et al. did not derive the carbon and nitrogen mass fractions but already from the hydrogen mass fraction they concluded that evolutionary models could not reproduce the position of most stars in the HRD and at the same time their surface hydrogen composition. Chemically homogeneous evolution was raised as a possible solution by \citet{martins09} to populate the relevant part of the HRD. However, \citet{hainich15} argued that the predictions of \citet{brott11} for these objects underestimate the surface hydrogen content. In the present study, we find that for one object ---with a high hydrogen mass fraction--- QHE is a viable solution given that (1) the HRD position, (2) the surface hydrogen content, and (3) the surface nitrogen abundance are all relatively well explained by the models of Brott et al. However, for the remainder of the sample, especially stars with the lowest surface X, QHE tracks do not consistently predict all parameters. As discussed above, an increase in the initial abundances in stellar models may help.

Evolutionary models use different mass-loss prescriptions for different phases of evolution. In the OB-star regime, the classical recipe of \citet{vink01} is widely adopted. When models enter the WR regime, a switch to different mass-loss-rate prescriptions is made (the recipe of \citealt{nl00} is often used). The definition of the different regimes relies on criteria based on the effective temperature and surface chemical composition. For instance, the models of \citet{lc18} consider WR-type mass-loss rates for \teff\ $> 10^4$ K and $X < 0.4$. Our determination of surface hydrogen mass fraction indicates that WR stars exist with $X > 0.4$. Evolutionary models may therefore use incorrect mass-loss rates for some H-rich phases. This can affect the shape of evolutionary tracks.

Shortcomings may also be due to the definition of temperatures used in evolutionary models and atmosphere models. Because of the dense winds of WR stars, including WNh objects, the classical effective temperature defined at an optical depth of two-thirds may be different from the surface temperature provided by evolutionary models. At such an optical depth, WR winds may not be in the hydrostatic part and therefore do not correspond to the stellar surface as defined in evolutionary models. This is why a temperature at an optical depth of 20, deeper in the atmosphere, is usually chosen to compare with the temperature of evolutionary models \citep[e.g.][]{hamann06}. The optical depth scale in atmosphere models depends on the wind density, and therefore on wind parameters (see discussion and references in, e.g. \citealt{lefever23}). For this reason, we must keep in mind the systematic uncertainties that are present in the determination of surface temperatures.

\subsection{WNh stars and young stellar populations}
\label{s_popsyn}

In this study, we determined the surface chemical composition of WNh stars. The hydrogen content is shown as a function of luminosity in Fig.~\ref{X_L}. We added WNh stars from various clusters to increase the size of the sample. Although it is certainly not exhaustive, this collection of data indicates that a non-negligible fraction of WNh stars have $X > 0.4$. As discussed in the previous section, this may impact the treatment of WR phases in evolutionary models; it also likely affects population synthesis models. The latter assign synthetic or empirical spectra to isochrone points in order to produce integrated spectra of clusters or galaxies (using different assumptions on the initial mass function and star formation history).
Spectra are assigned according to the predicted stellar parameters at a given position of the isochrone. To the best of our knowledge, WR spectra are considered only for isochrone points that have $X < 0.3-0.4$. This is the case in BPASS \citep{bpass}, the Bruzual \& Charlot models \citep{plat19}, STARBURST99 \citep{leitherer14}, and PYPOPSTAR \citep{hrpypopstar} to quote only a few. \citet{vms} studied the presence of VMSs in local star-forming regions and raised the issue of population synthesis models failing to predict some WR-type features. These authors discussed the above criterion to assign WR spectra and concluded that a better description of WR stars was needed. The present results confirm that WR stars are found with relatively large surface hydrogen mass fractions. These objects have various luminosities and can be present in populations dominated by VMSs or more classical, less luminous WR objects. 

We also demonstrate that evolutionary models are not able to simultaneously reproduce  the HRD position and the surface chemistry of most low-metallicity WNh stars. In particular, the surface nitrogen content is underestimated. If this is confirmed for a larger number of WR stars, this could potentially impact stellar yields of young stellar populations. In turn, this could affect the interpretation of chemical abundances in unresolved populations. A specific case has attracted attention since its discovery by \citet{oesch16}: the star-forming galaxy GN-z11 at a redshift of 10.6 and metallicity of 12+log(O/H)$\sim$7.9 \citep{bunker23} displays a peculiar nitrogen-to-oxygen ratio \citep{cameron23} that no standard chemical evolution can explain. Several solutions have been put forward to explain the super-solar value \citep{senchyna23,charbonnel23,marques23}. The leading scenario is that GN-z11 shows the same chemistry as that of globular clusters and that the mechanism at work is similar in both types of objects. \citet{marques23} studied alternative possibilities and argued that,
 when WR stars dominate, a young cluster at a well-defined age can also reproduce the observed abundances. However, these authors do not favour this solution because the predicted yields from such a stellar population only match the observed constraints  for a narrow age range.

Our results indicate that the evolution of WR stars at low metallicity is not well understood and that most current models do not produce enough surface nitrogen at the surface of the WNh stars we analysed, at least at the position of the HRD where the stars are found. One might therefore anticipate that the stellar yields of the current generation of models also incorrectly predict the amount of nitrogen released by WR stars. Of course, this may not change the global picture, the WR phase lasting only a small fraction of the life of a young cluster. \citet{higgins23} reported that VMSs, which are all WNh stars, contribute five to ten times more H-processed elements on the main sequence compared to normal massive stars. This includes nitrogen. It is therefore required that we revisit the nucleosynthesis, mixing processes, and yields of WR stars at low metallicity in order to better understand the nature of intensely star-forming regions in the early Universe.

\section{Concluding remarks}
\label{s_conc}

We performed a spectroscopic analysis of 22 single WNh stars in the Galaxy and the Magellanic Clouds. We determined the stellar parameters (temperatures, luminosity, mass-loss rates) and the surface abundances of hydrogen, helium, carbon, and nitrogen. We relied on synthetic spectra calculated with the atmosphere code CMFGEN. A $\chi^2$ analysis was used to infer the best-fit adundances and their uncertainties.

We find that the WNh stars display a wide range of surface hydrogen mass fraction, from almost no depletion to almost full exhaustion. Values of X of greater than 0.4 ---a typical threshold value to include WR-type winds in evolutionary calculation--- are not exceptional. All stars are carbon-poor and nitrogen-rich, consistent with products of CNO-cycle nucleosynthesis. The nitrogen content of WNh stars is higher for stars in higher metallicity environments. Most of the initial C, N, and O is found in N in these objects, and so this trend is expected.

At solar metallicity, WNh stars can be reasonably reproduced by evolutionary models, provided parameters controlling the amount of rotation and the implementation of its mixing effects are fine-tuned. At lower metallicity, only a minority of stars can be reproduced. For the vast majority of objects, the position in the HRD coupled to constraints on the surface chemistry, especially nitrogen, is not matched by any models. Even QHE or binary interaction (unlikely for these single objects), which offer the most promising evolutionary pathways according to past studies, face problems. We argue that increasing the initial CNO content in some models may help to solve part of this issue.

In agreement with previous studies \citep{hamann06,hainich14,hainich15}, we find that WNh stars with large amounts of hydrogen exist. Such objects are not currently taken into account properly in population synthesis models. This may affect the predicted integrated spectra of young massive clusters. In addition, WR-type mass-loss rates are only implemented for phases with a hydrogen mass fraction of lower than 0.4 in most evolutionary models. This is inconsistent with the existence of relatively H-rich WN stars.

This study focused on the surface chemistry of WNh stars. Studies of a large sample of single WN and WC stars, as well as binary systems, indicate that evolutionary models lack physical ingredients to reproduce their position in the HRD \citep{sander12,hamann19}. Future analysis of the surface chemical composition of these H-free objects will place more constraints on these models and will be the subject of subsequent publications.

\begin{acknowledgements}
I thank an anonymous referee for a timely and positive report. I thank Rainer Hainich for sharing the data of Foellmi et al. (2003, 2003b), Daniel Pauli and Koushik Sen for providing binary models from their grid, and John Hillier for making CMFGEN available to the community. I thank Ana Palacios for comments on an early version of the manuscript. This research has made use of the SIMBAD database, operated at CDS, Strasbourg, France. This research is based on observations made with the International Ultraviolet Explorer, obtained from the MAST data archive at the Space Telescope Science Institute, which is operated by the Association of Universities for Research in Astronomy, Inc., under NASA contract NAS 5–26555. Based on data obtained from the ESO Science Archive Facility. 
\end{acknowledgements}

\bibliographystyle{aa}
\bibliography{ab_wnh}


\begin{appendix}

\FloatBarrier

\onecolumn

\section{Observational data}
\label{ap_obsdata}

Table~\ref{tab_obs} gathers the information on the spectroscopic data used in the present analysis.

\begin{table}[!h]
\caption{Observational data} \label{tab_obs}
\begin{tabular}{lc|c|c}
\hline
Star           &    ST   & UV        &  optical  \\
\hline
WR16           & WN8h             &  IUE swp13893 + lwr18488      & ESO/FEROS 073.D-0609 \\
WR24           & WN6ha-w          &  IUE swp27275 + lwr03825      & ESO/FEROS 073.D-0609 \\
WR40           & WN8h             &  IUE swp27968 + lwp07344      & ESO/FEROS 073.D-0609 \\
WR78           & WN7h             &  IUE swp33450 + lwp13166      & ESO/FEROS 095.D-0929 \\
WR89           & WN8h             &  --                           & ESO/FEROS 075.D-0532 \\
WR108          & WN9h             &  IUE swp38433 + lwr10529      & ESO/FEROS 089.D-0730 \\
WR124          & WN8h             &  IUE swp28945 + lwr07624      & ESO/FEROS 091.D-0622 \\
WR128          & WN4(h)-w         &  IUE swp15101 + lwr03873      & ESO/FEROS 089.D-0730 \\
\hline
BAT99-35       & WN3(h)           &  IUE swp23387 + lwr03702       & \citet{foellmi03}    \\
BAT99-44       & WN8ha            &  IUE swp06187 + lwr05352       & ESO/GIRAFFE 074.D-0518 \\
BAT99-50       & WN5h             &  --                           & \citet{foellmi03}    \\
BAT99-63       & WN4ha:           &  IUE swp09157 + lwr07906       & \citet{foellmi03}    \\
BAT99-66       & WN3h             &  --                           & \citet{foellmi03}    \\
BAT99-67       & WN5ha            &  --                           & \citet{foellmi03}    \\
BAT99-73       & WN5ha            &  --                           & \citet{foellmi03}    \\
BAT99-74       & WN3(h)a          &  --                           & \citet{foellmi03}    \\
BAT99-81       & WN5h             &  IUE swp40900 + lwp20463      & \citet{foellmi03}    \\
BAT99-89       & WN7h             &  IUE swp48106 + lwp2506       & ESO/GIRAFFE 182.D-0222 \\
BAT99-122      & WN5h             &  IUE swp04139 + lwr03670      & \citet{foellmi03}    \\
\hline
AB9            & WN3ha            &  --                           & \citet{foellmi03b}    \\
AB10           & WN3ha            &  --                           & \citet{foellmi03b}    \\
AB11           & WN4ha            &  --                           & \citet{foellmi03b}    \\
\hline
\end{tabular}
\tablefoot{Columns are source ID, spectral type, source of UV data and source of optical data. }
\end{table}

\newpage

\FloatBarrier

\section{Comparison to literature parameters}
\label{ap_complit}

Table~\ref{tab_comp} gathers the stellar parameters available in the literature for some our sample stars.

\begin{table}[!h]  
\begin{center}
  \caption{Comparison with literature parameters} \label{tab_comp}
  \footnotesize
\begin{tabular}{llllllllllllllllllllll}
\hline
Star        &   T* & R*     & \lL    & \lmdot & \vinf & $f_{\infty}$ & X  & Y     & X(C) & X(N) & reference  \\
            & [kK] &[\rsun] &        & [\myr] & [\kms] &           & & & & \\
\hline
WR16        & 42.3 &  13.2       & 5.7    & -5.3        &  800  & 0.01     & 0.28$\pm$0.03 & 0.71$\pm$0.03  & 5.0$\pm$3.0 10$^{-5}$ & 11.2$\pm$3.0 10$^{-3}$ & 1 \\
            & 44.7 &  11.56      & 5.72   & -4.6        &  650  & 0.1       & 0.25 &-- &-- &-- & 2 \\
WR24        & 43.1 &  20.2       & 6.1    & -5.3        &  2100 & 0.01     & 0.53$\pm$0.09 & 0.46$\pm$0.08  & 4.6$\pm$2 10$^{-5}$ & 11.5$\pm$4 10$^{-3}$ & 1  \\
            & 50.1 &  21.73      & 6.47   & -4.3        &  2160 & 0.1      & 0.44 &-- &-- &-- & 2 \\
WR40        & 41.7 &  13.6       & 5.7    & -4.6        &  800  & 0.1      & 0.19$\pm$0.03 & 0.80$\pm$0.03  & 7.0$\pm$5.0 10$^{-5}$ & 5.1$\pm$1.1 10$^{-3}$ & 1 \\
            & 44.7 &  14.51      & 5.91   & -4.2        &  650  & 0.1      & 0.23 &-- &-- &-- & 2 \\
WR78        & 41.5 &  19.5       & 6.0    & -4.7        &  1600 & 0.05     & 0.17$\pm$0.05 & 0.82$\pm$0.05 & 4.7$\pm$3.0 10$^{-5}$ & 7.4$\pm$2.0 10$^{-3}$ & 1 \\
            & 50.1 &  10.14      & 5.80   & -4.5        &  1385 & 0.1      & 0.11 &-- &-- &-- & 2 \\
WR89        & 35.1 &  34.3       & 6.2    & -4.5        &  1500 & 0.1      & 0.26$\pm$0.04 & 0.73$\pm$0.04 & 1.0$\pm$0.7 10$^{-4}$ & 12.3$\pm$10.0 10$^{-3}$ & 1 \\
            & 39.8 &  30.04      & 6.33   & -4.4        &  1600 & 0.1      & 0.20 &-- &-- &-- & 2 \\
WR108       & 35.2 &  21.5       & 5.8    & -5.0        &  900  & 0.1      & 0.17$\pm$0.04 & 0.82$\pm$0.03 & 5.9$\pm$5.4 10$^{-4}$ & 9.3$\pm$4.3 10$^{-3}$ & 1 \\
            & 39.8 &  16.07      & 5.77   & -4.9        &  1170 & 0.1      & 0.27 &-- &-- &-- & 2 \\
WR124       & 41.5 &  13.8       & 5.7    & -4.5        &  800  & 0.1      & 0.19$\pm$0.08 & 0.80$\pm$0.03 &  --                   & 2.5$\pm$1.0 10$^{-3}$ & 1 \\
            & 44.7 &  11.93      & 5.75   & -4.3        &  710  & 0.1      & 0.13 &-- &-- &-- & 2 \\
WR128       & 65.4 &  3.9        & 5.4    & -5.5        &  1800 & 0.1      & 0.23$\pm$0.18 & 0.75$\pm$0.19 & 6.2$\pm$2.5 10$^{-5}$ & 21.6$\pm$6 10$^{-3}$ & 1  \\
            & 70.8 &  2.69       & 5.22   & -5.4        &  2050 & 0.1      & 0.16 &-- &-- &-- & 2 \\
            & 59.9 &  5.43       & 5.40   & -5.3        &  1800 & 0.1      & 0.26$\pm$0.07  & 0.73$\pm$0.20 & 6.1$\pm$1.6 10$^{-5}$ & 11.0$\pm$3 10$^{-3}$ & 3 \\
\hline
BAT99-35    & 77.3 & 3.54        & 5.6    & -5.15       &  1600 & 0.1      & 0.11$\pm$0.10 & 0.88$\pm$0.10 & 3.7$\pm$0.4 10$^{-5}$ & 10.0$\pm$2.6 10$^{-3}$ & 1 \\
            & 71.0 & 4.2         & 5.60   & -5.11       &  1600 & 0.1      & 0.1 &-- &-- &-- & 4 \\
BAT99-44    & 39.7 & 10.64       & 5.4    & -5.2        &  600  & 0.1      & 0.31$\pm$0.05 & 0.69$\pm$0.04 & --                    & 4.2$\pm$1.7 10$^{-3}$ & 1 \\
            & 45.0 & 11.3        & 5.66   & -5.12       &  700  & 0.1      & 0.4 &-- &-- &-- & 4 \\
BAT99-50    & 56.1 & 5.97        & 5.5    & -5.6        &  1600 & 0.1      & 0.39$\pm$0.16 & 0.60$\pm$0.16 & 5.4$\pm$3.0 10$^{-5}$ & 9.0$\pm$5.0 10$^{-3}$ & 1 \\
            & 56.0 & 7.1         & 5.65   & -5.52       &  1600 & 0.1      & 0.4 &-- &-- &-- & 4 \\
BAT99-63    & 64.5 & 5.38        & 5.65   & -5.35       &  2000 & 0.1      & 0.31$\pm$0.02 & 0.68$\pm$0.03 & 1.3$\pm$0.2 10$^{-4}$ & 10.6$\pm$9.0 10$^{-3}$ & 1  \\
            & 63.0 & 5.2         & 5.58   & -5.42       &  1600 & 0.1      & 0.4 &-- &-- &-- & 4 \\
            & 68.9 & 3.73        & 5.45   & -5.45       &  2000 & 0.1      & 0.40$\pm$0.20  & 0.56$\pm$0.14 & 2.0$\pm$1.5 10$^{-4}$ & 2.3$\pm$1.6 10$^{-3}$ & 3 \\
BAT99-66    & 85.6 & 3.23        & 5.7    & -5.6        &  1600 & 0.1      & 0.27$\pm$0.10 & 0.72$\pm$0.11 & $<$3.0 10$^{-4}$      & 7.8$\pm$1.0 10$^{-3}$ & 1 \\
            & 89.0 & 3.3         & 5.78   & -5.42       &  1600 & 0.1      & 0.2 &-- &-- &-- & 4 \\
BAT99-67    & 51.7 & 11.1        & 5.90   & -5.5        &  1800 & 0.01     & 0.45$\pm$0.10 & 0.53$\pm$0.10 & 2.7$\pm$1.1 10$^{-5}$ & 8.8$\pm$7.6 10$^{-3}$ & 1  \\
            & 47.0 & 14.3        & 5.96   & -4.91       &  1600 & 0.1      & 0.3 &-- &-- &-- & 4 \\
BAT99-73    & 52.7 & 6.78        & 5.5    & -5.2        &  1300 & 0.01     & 0.42$\pm$0.08 & 0.57$\pm$0.07 & 3.3$\pm$1.1 10$^{-5}$ & 6.6$\pm$2.9 10$^{-3}$ & 1 \\
            & 60.0 & 6.8         & 5.72   & -5.54       &  1600 & 0.1      & 0.4 &-- &-- &-- & 4 \\
BAT99-74    & 82.2 & 2.79        & 5.5    & -5.8        &  2200 & 0.1      & 0.30$\pm$0.05 & 0.68$\pm$0.04 & $<$1.0 10$^{-4}$      & 8.1$\pm$0.6 10$^{-3}$ & 1 \\
            & 79.0 & 3.7         & 5.69   & -5.62       &  2000 & 0.1      & 0.2 &-- &-- &-- & 4 \\ 
BAT99-81    & 47.8 & 6.5         & 5.3    & -5.6        &  1250 & 0.1      & 0.61$\pm$0.13 & 0.38$\pm$0.23 & 2.3$\pm$0.5 10$^{-4}$ & 7.3$\pm$3.8 10$^{-3}$ & 1 \\
            & 47.0 & 8.2         & 5.48   & -5.55       &  1000 & 0.1      & 0.4 &-- &-- &-- & 4 \\
BAT99-89    & 50.5 & 9.3         & 5.7    & -4.5        &  1000 & 0.1      & 0.22$\pm$0.04 & 0.77$\pm$0.06 & --                    & 3.2$\pm$0.7 10$^{-3}$ \\
            & 50.0 & 10.3        & 5.78   & -4.73       &  1000 & 0.1      & 0.2 &-- &-- &-- & 4 \\
BAT99-122   & 44.3 & 19.12       & 6.1    & -4.5        &  1800 & 0.1      & 0.17$\pm$0.10 & 0.82$\pm$0.10 & 4.2$\pm$0.9 10$^{-5}$ & 3.7$\pm$1.3 10$^{-3}$ & 1 \\
            & 50.0 & 17.3        & 6.23   & -4.56       &  1600 & 0.1      & 0.2 &-- &-- &-- & 4 \\
\hline
AB9         & 81.6 &  3.6        & 5.7    & -5.9        &  2000 & 0.1      & 0.28$\pm$0.08  & 0.71$\pm$0.10      &$<$1.0 10$^{-4}$  & 3.8$\pm$0.8 10$^{-3}$ & 1 \\
            & 100.0&  3.5        & 6.05   & -5.65       &  1800 & 0.1      & 0.35 &-- &-- & -- & 5 \\
AB10        & 82.4 &  2.8        & 5.5    & -5.8        &  1600 & 0.1      & 0.35$\pm$0.08  & 0.65$\pm$0.11     &$<$2.0 10$^{-5}$  & 3.5$\pm$1.3 10$^{-3}$ & 1 \\
            & 100.0&  2.2        & 5.65   & -5.64       &  2000 & 0.1      & 0.35 &-- &-- & -- & 5 \\
AB11        & 81.9 &  2.8        & 5.5    & -6.15       &  2000 & 0.1      & 0.26$\pm$0.05  & 0.73$\pm$0.06      &$<$7.0 10$^{-5}$ & 7.6$\pm$1.6 10$^{-3}$ & 1 \\
            & 89.0 &  3.5        & 5.85   & -5.56       &  2200 & 0.1      & 0.4 &-- &-- &-- & 5 \\
\hline
\end{tabular}
\normalsize
\tablefoot{Columns are source ID, temperature, radius, mass-loss rate, terminal velocity, clumping factor, H, He, C and N surface mass fractions. References: 1- this study; 2- \citet{hamann19}; 3- \citet{martins13}; 4- \citet{hainich14}; 5- \citet{hainich15} }
\end{center}
\end{table}

\FloatBarrier
\newpage

\section{Effect of the choice of X(N) on HRDs}
\label{ap_hrdxn}

\begin{figure}[h]
\centering
\includegraphics[width=0.49\textwidth]{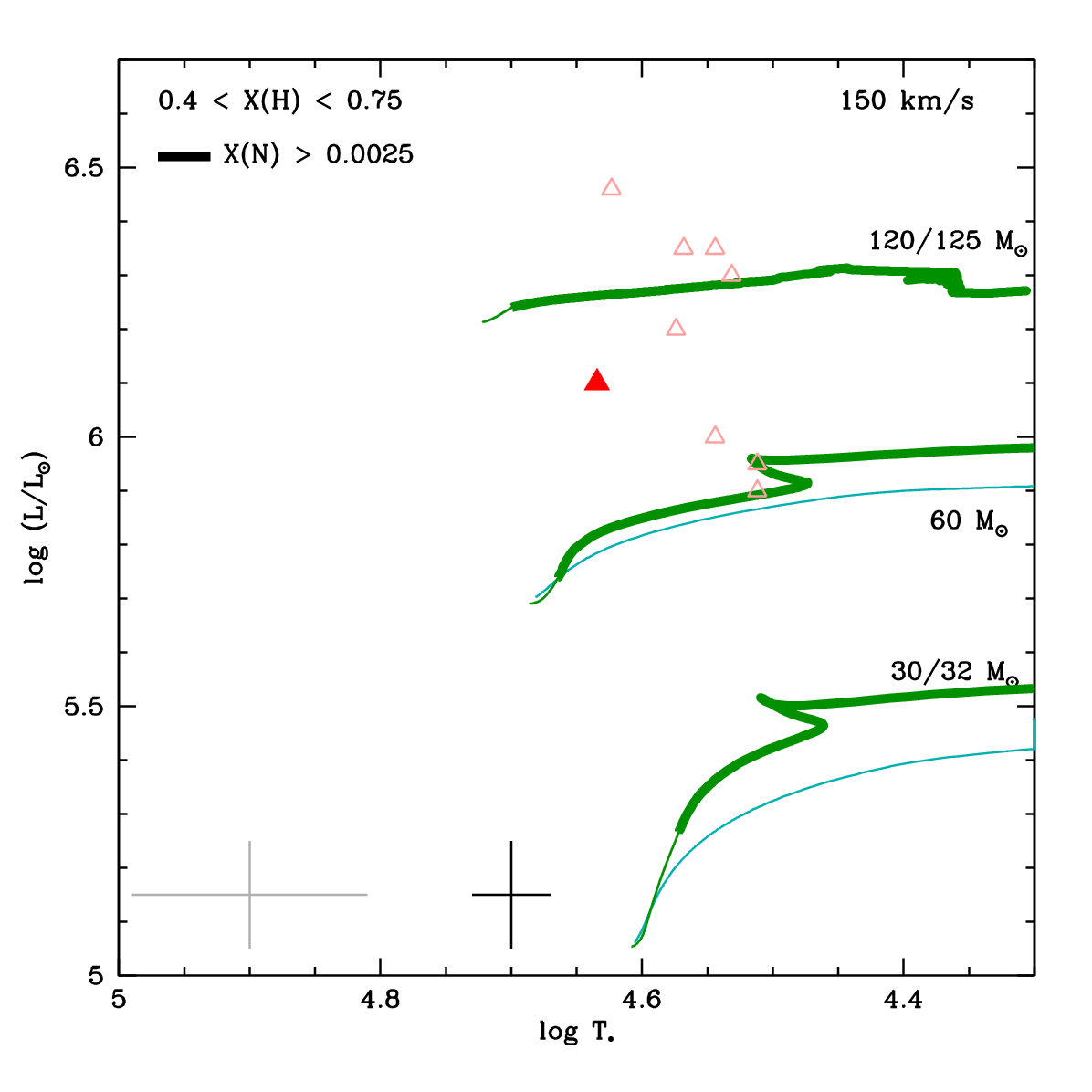}
\includegraphics[width=0.49\textwidth]{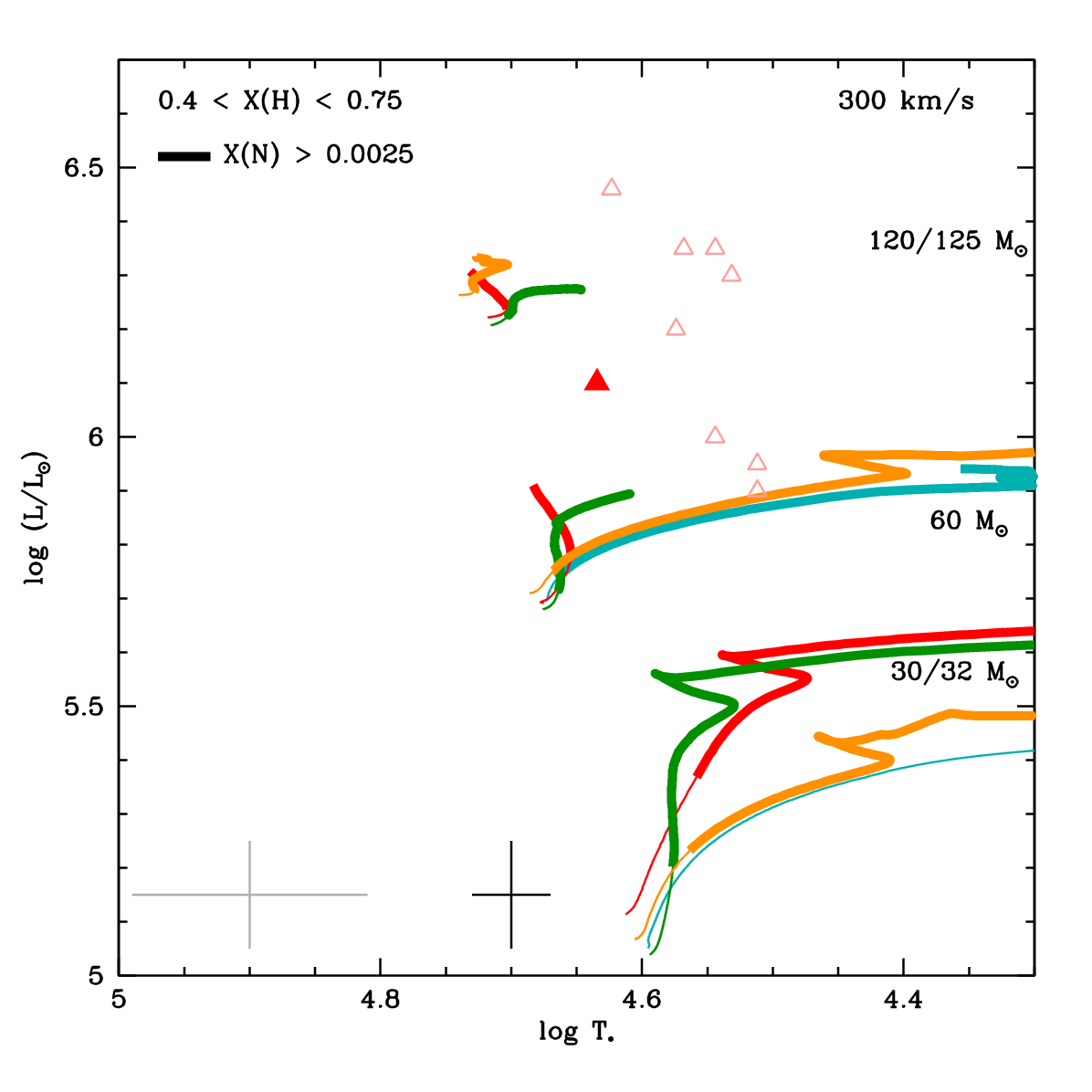}

\includegraphics[width=0.49\textwidth]{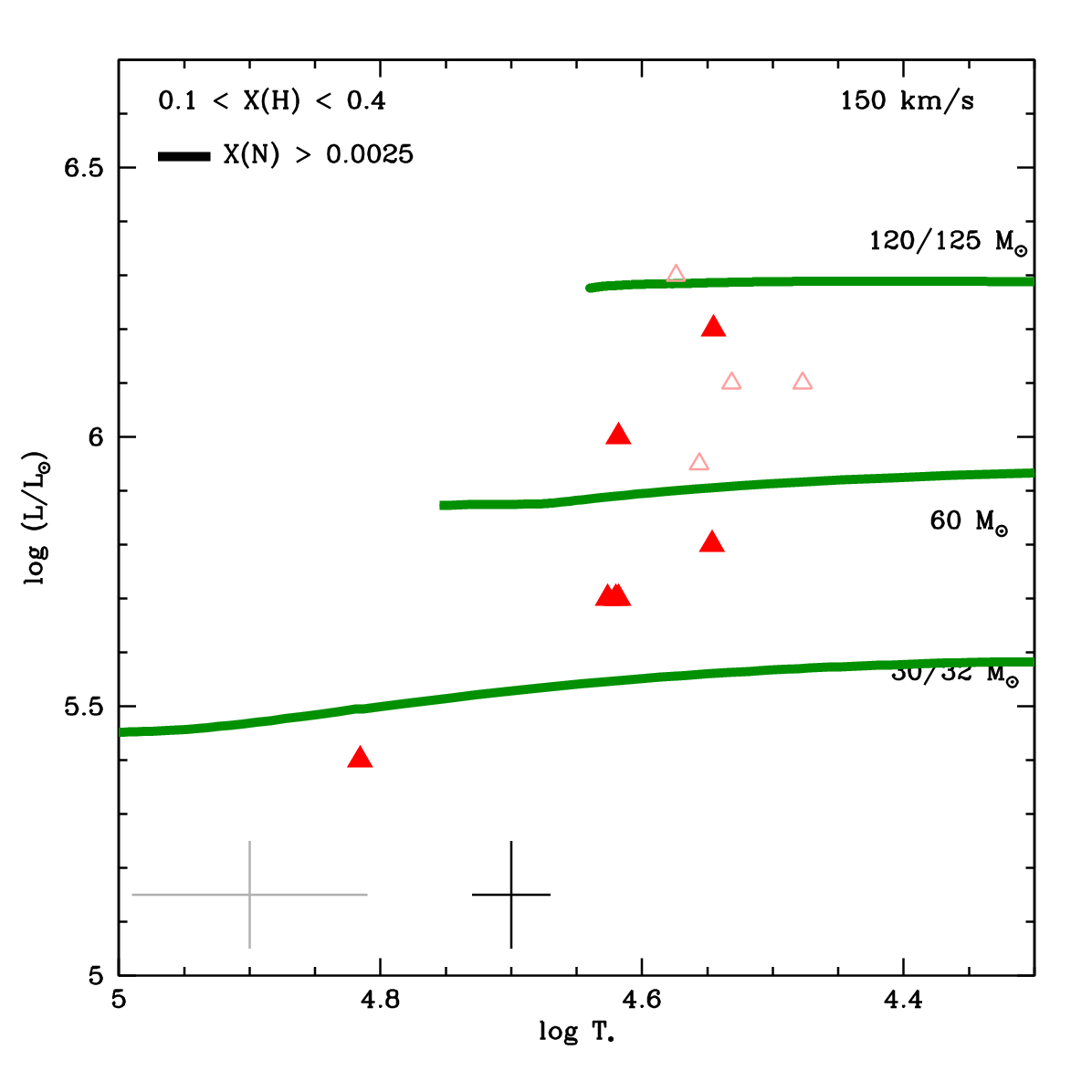}
\includegraphics[width=0.49\textwidth]{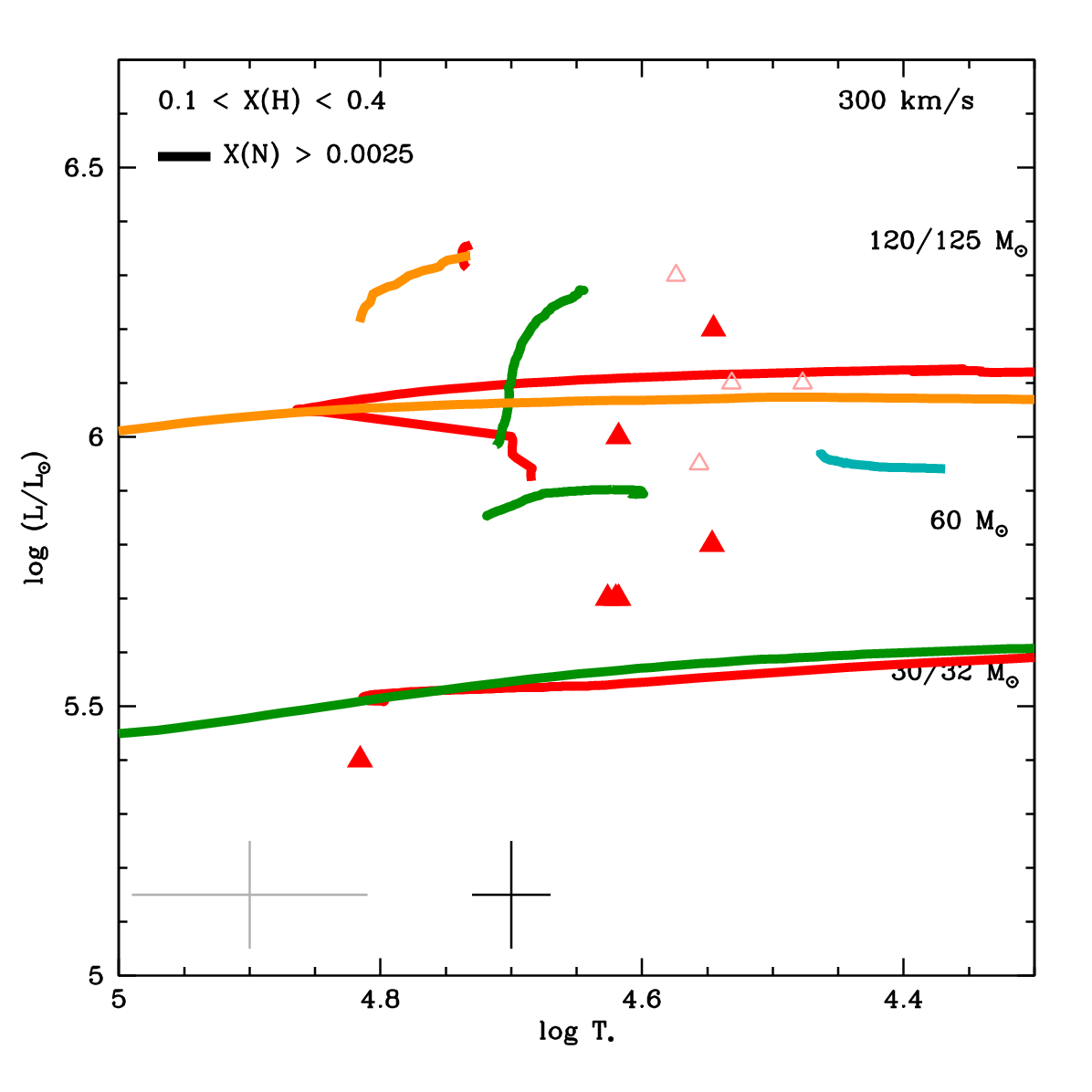}
\caption{Same as Fig.~\ref{hrd_gal} except that the bold part of the tracks is for a nitrogen mass fraction of greater than 0.0025.}
\label{hrd_gal2}
\end{figure}

\begin{figure}[h]
\centering
\includegraphics[width=0.3\textwidth]{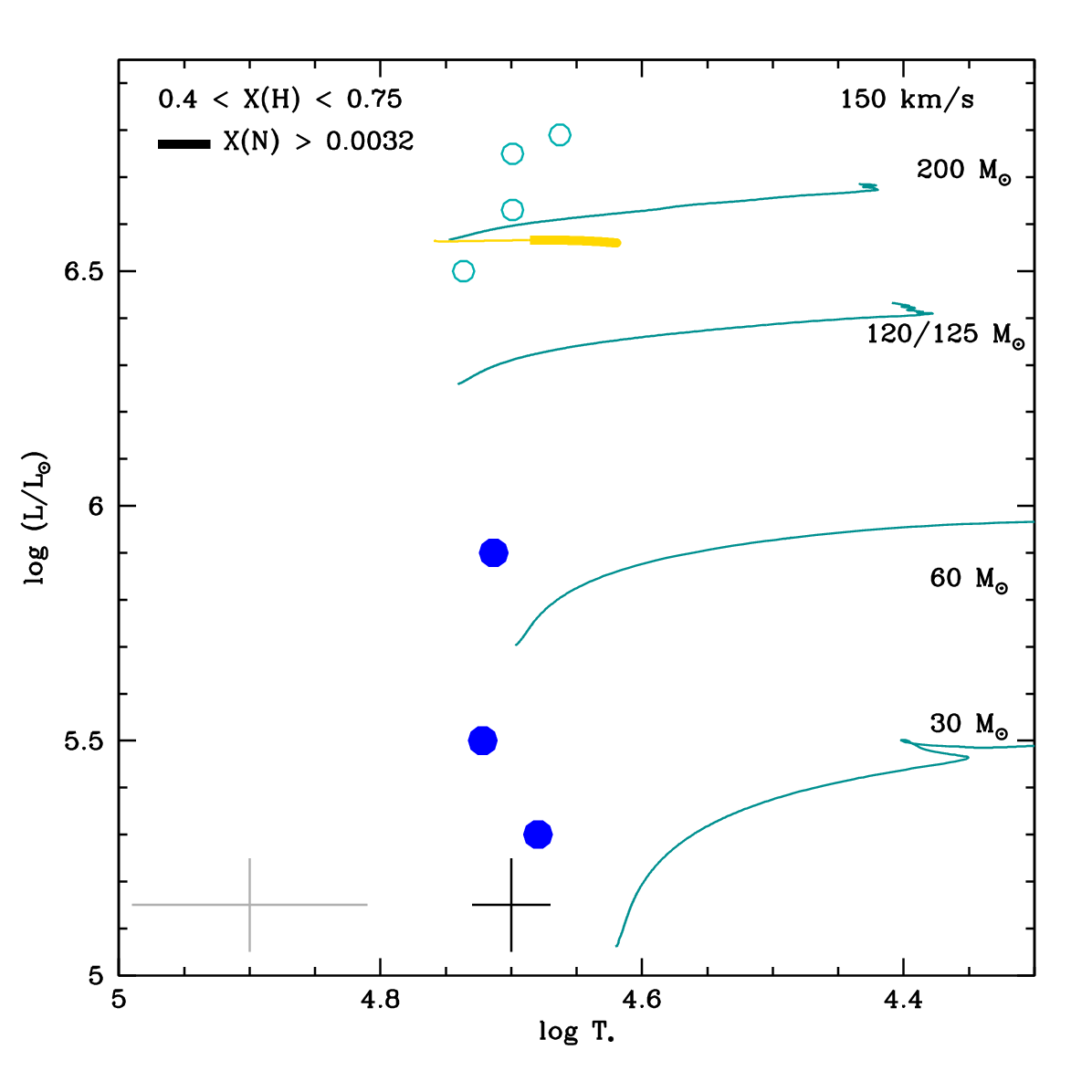}
\includegraphics[width=0.3\textwidth]{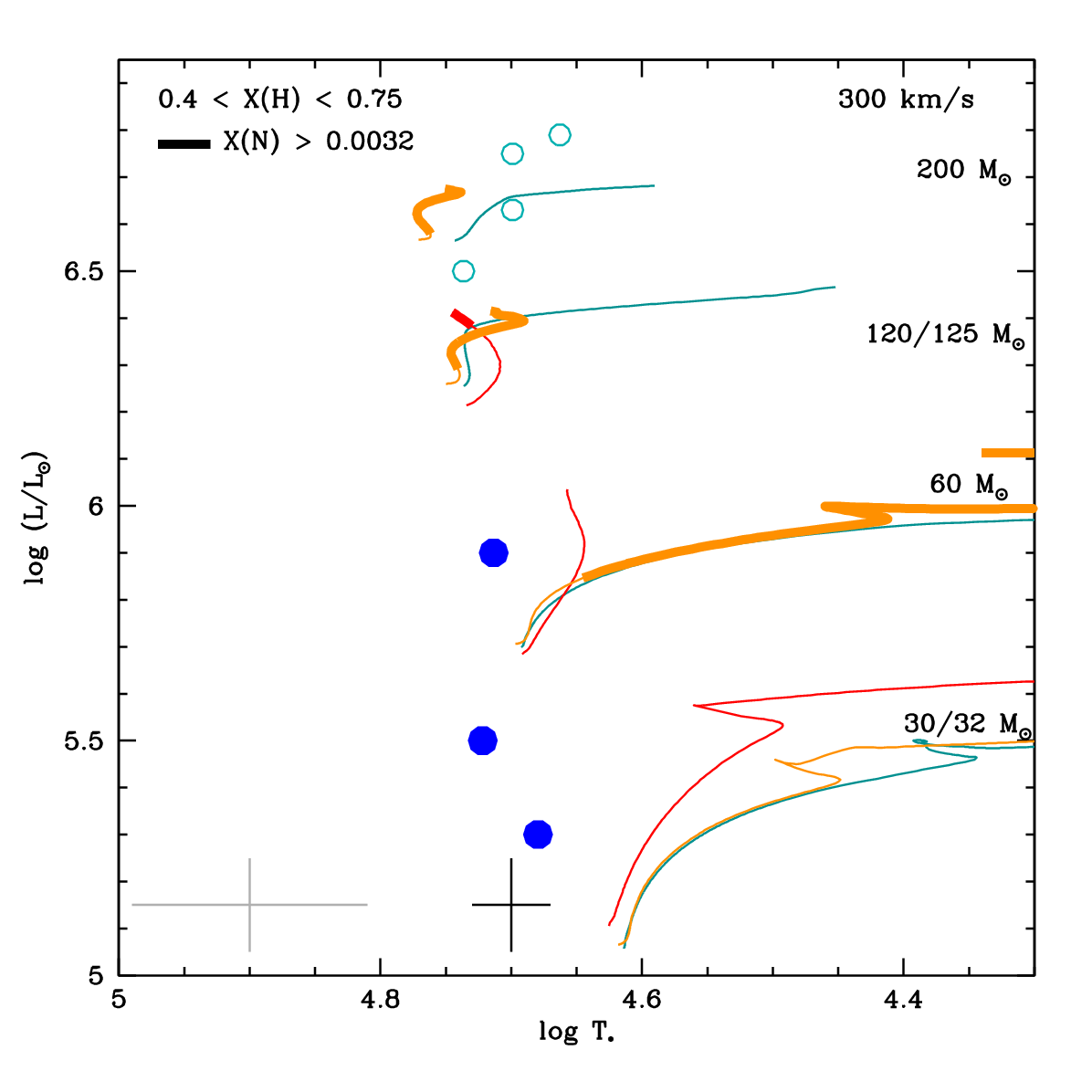}
\includegraphics[width=0.3\textwidth]{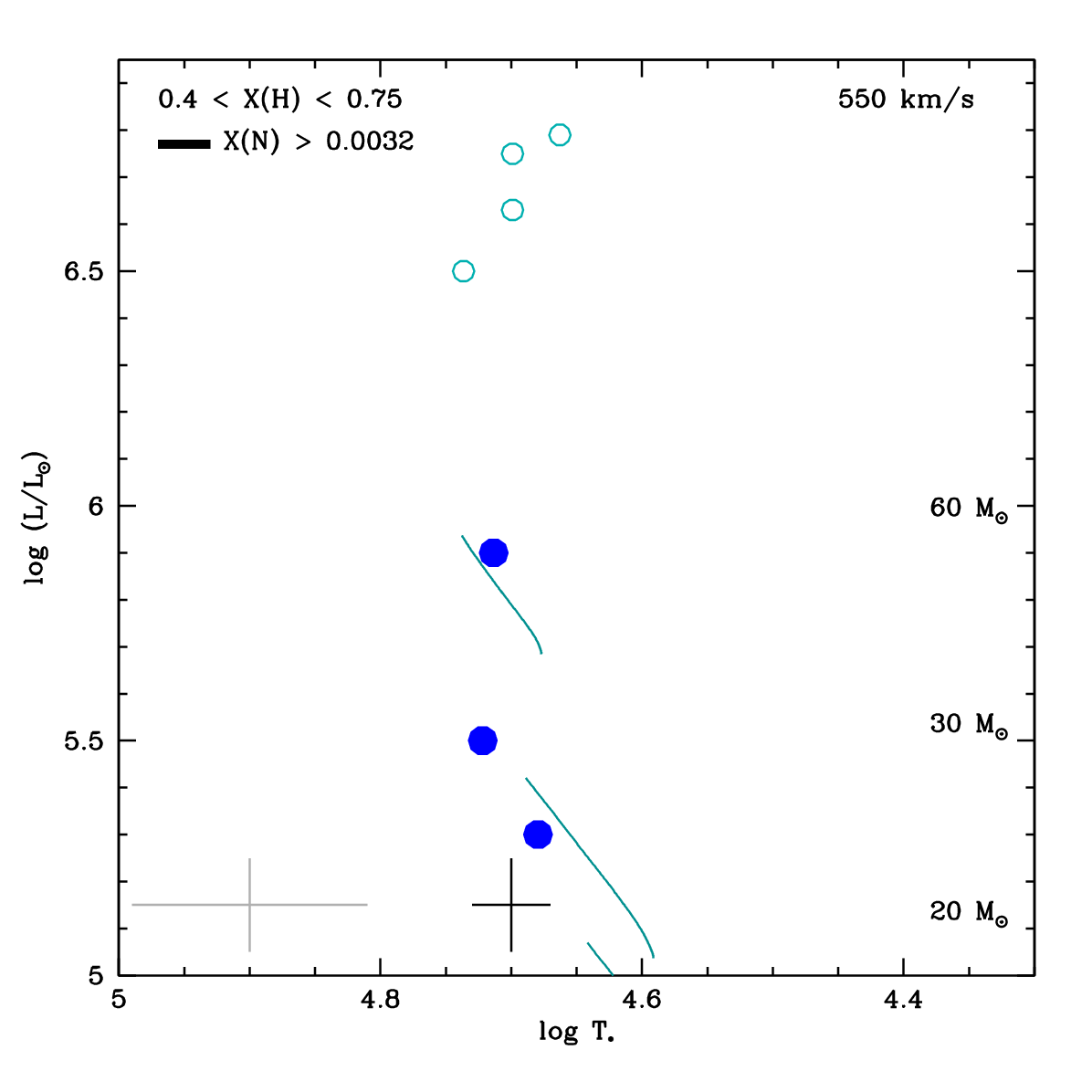}

\includegraphics[width=0.3\textwidth]{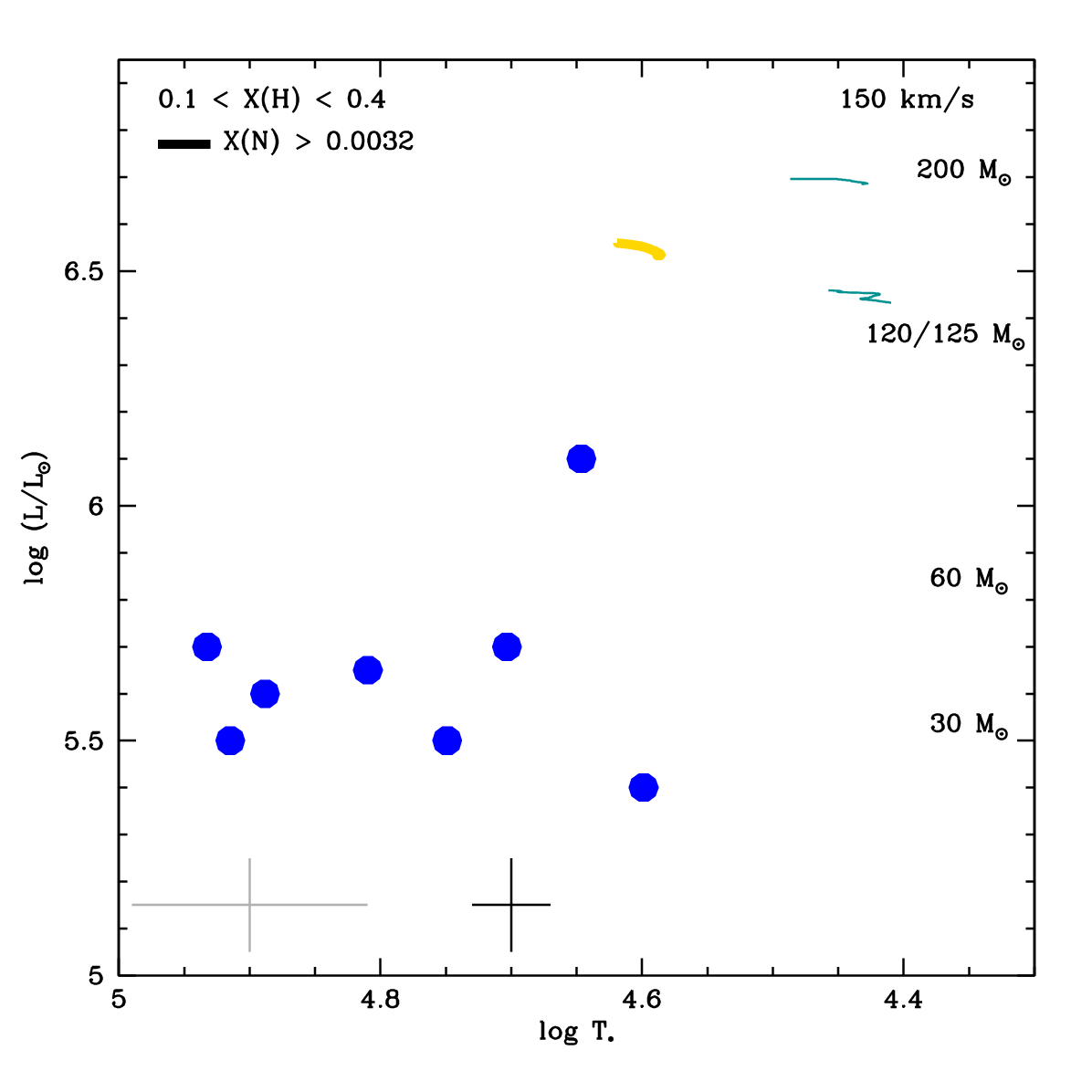}
\includegraphics[width=0.3\textwidth]{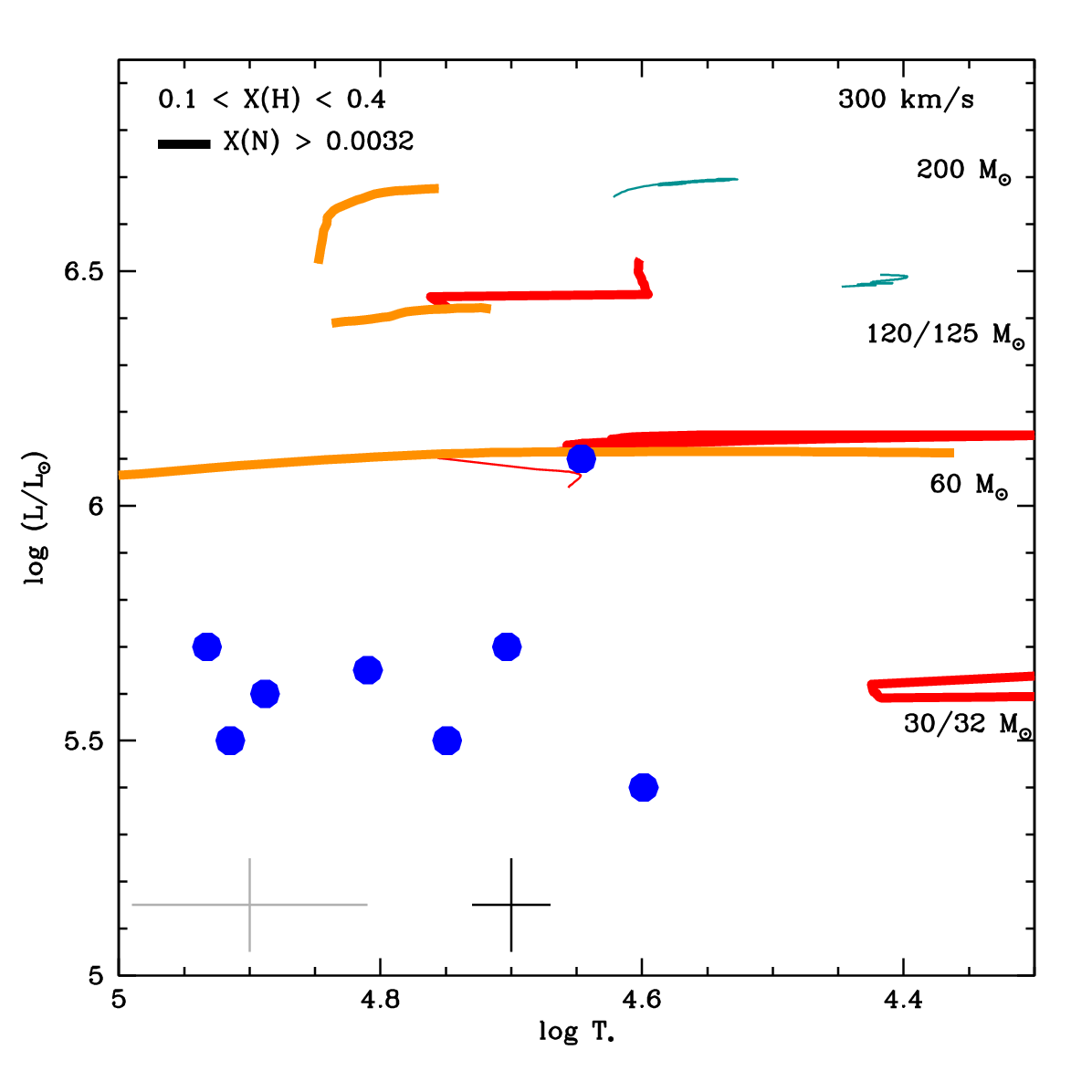}
\includegraphics[width=0.3\textwidth]{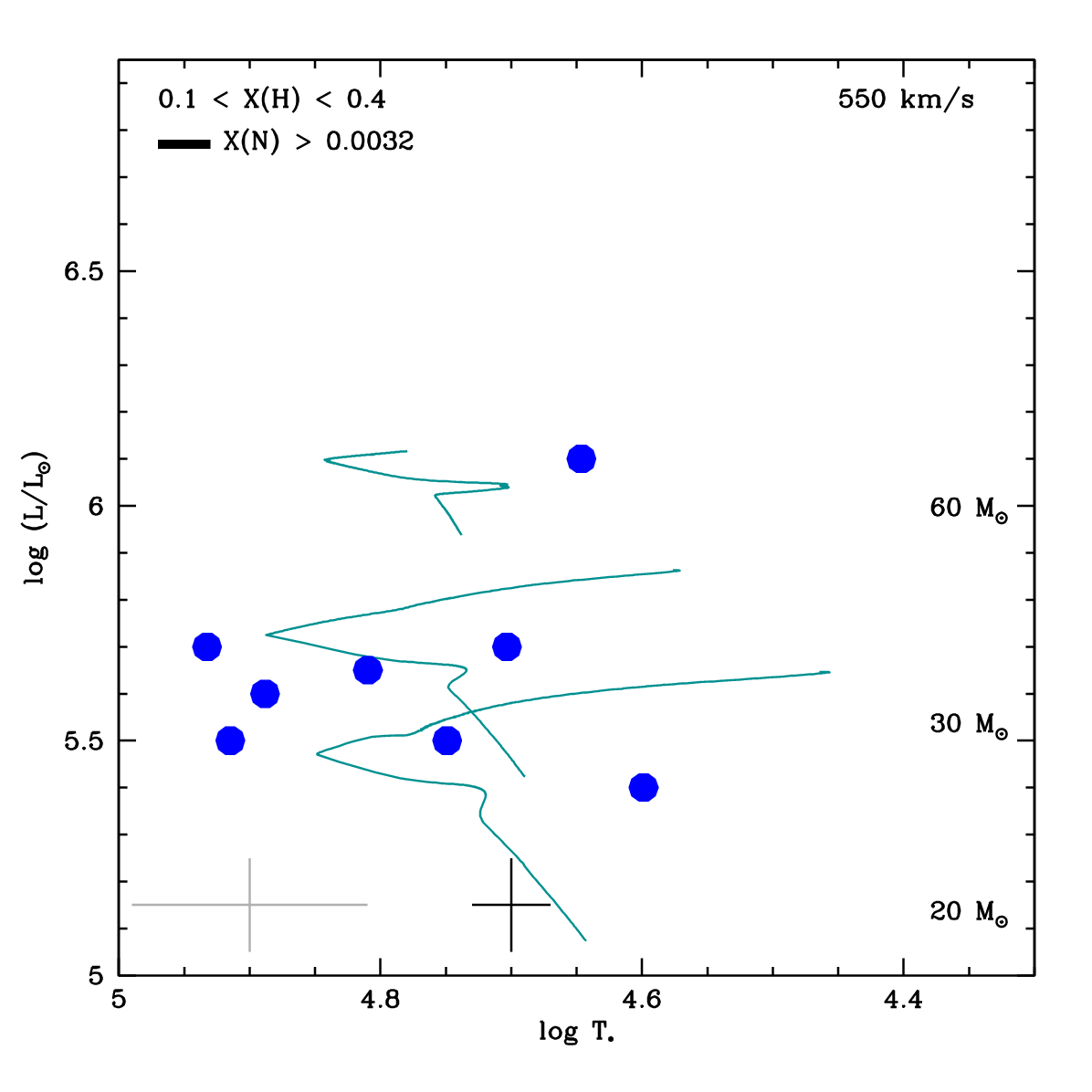}
\caption{Same as Fig.~\ref{hrd_lmc} except that the bold part of tracks is for a nitrogen mass fraction of greater than 0.0032.}
\label{hrd_lmc2}
\end{figure}

\begin{figure}[h]
\centering
\includegraphics[width=0.3\textwidth]{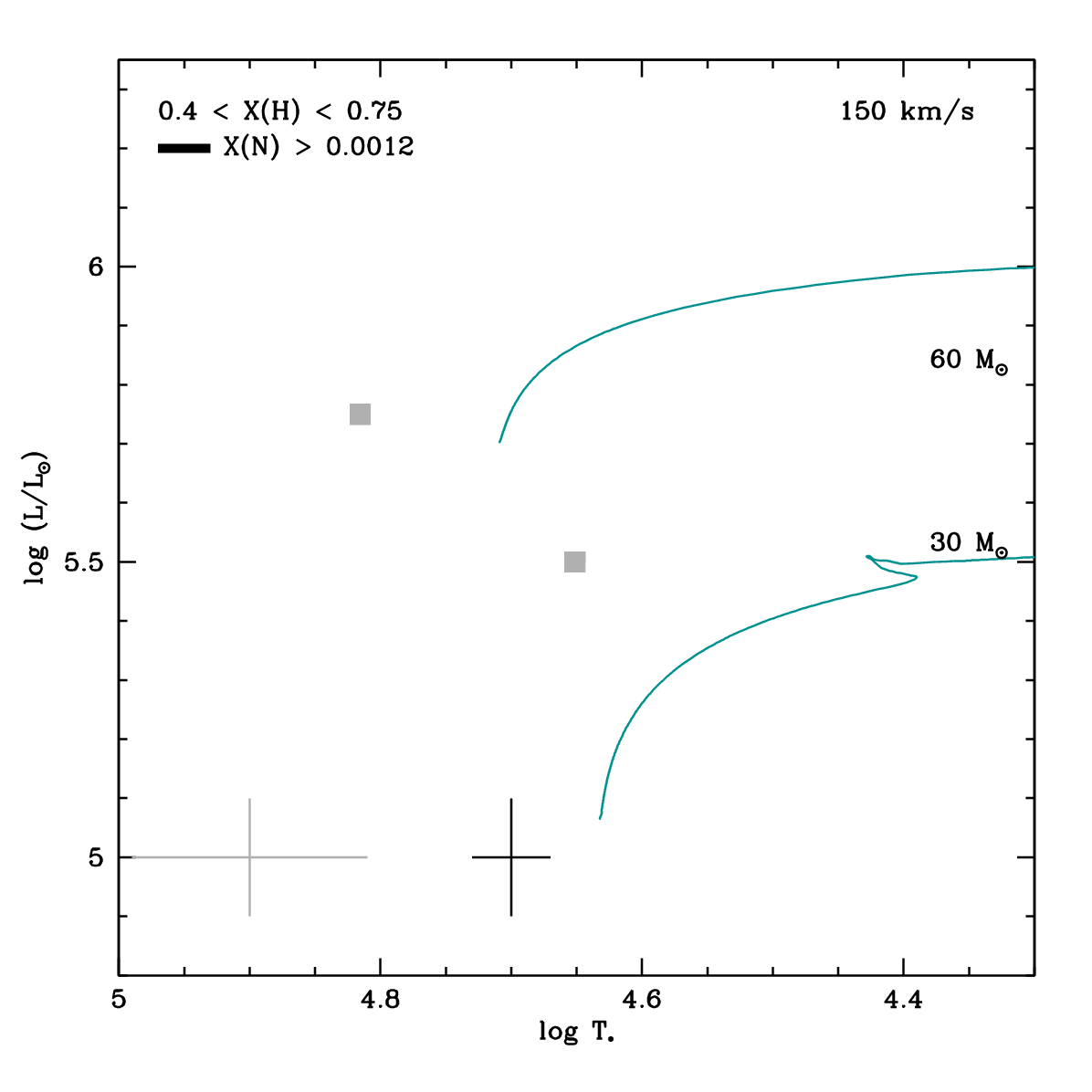}
\includegraphics[width=0.3\textwidth]{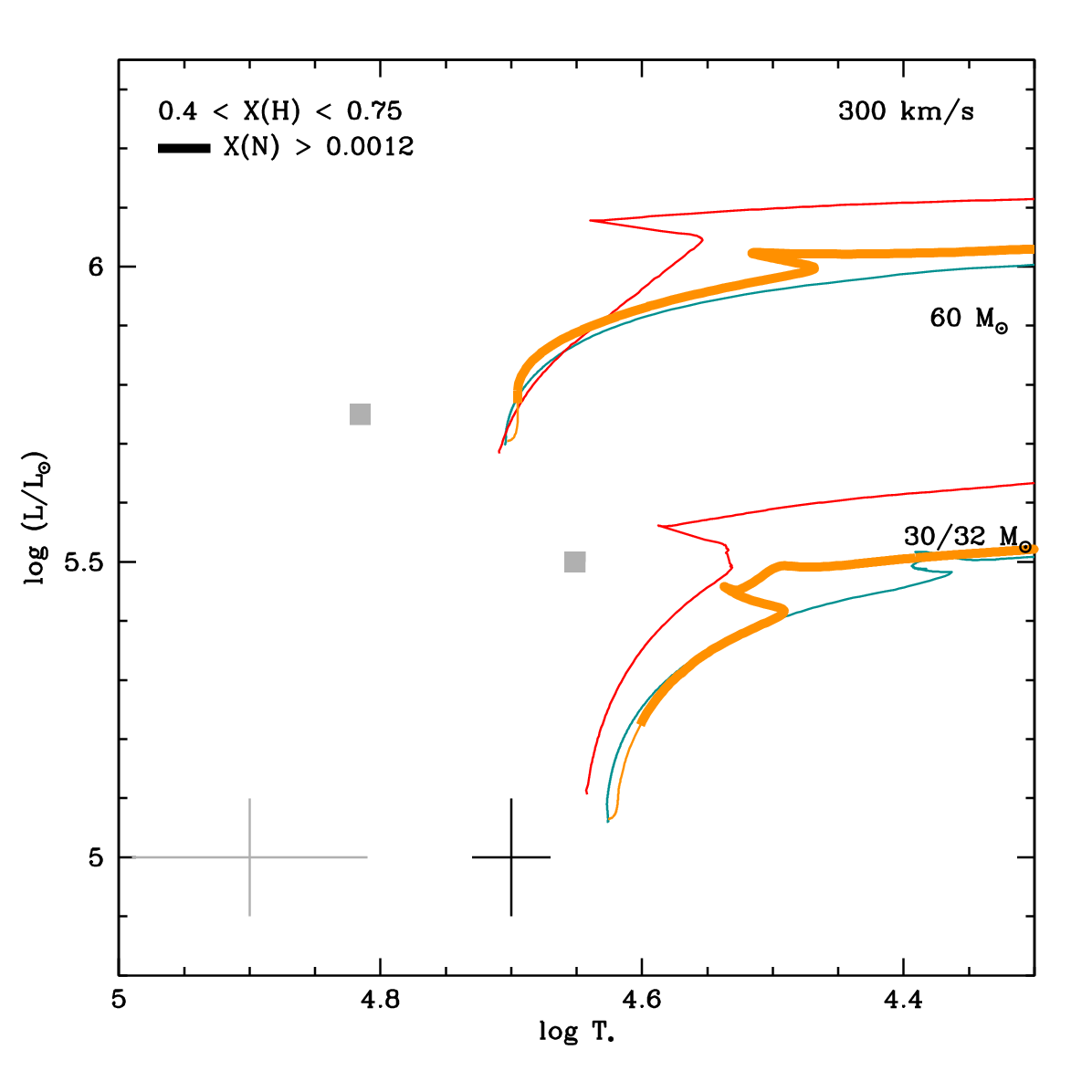}
\includegraphics[width=0.3\textwidth]{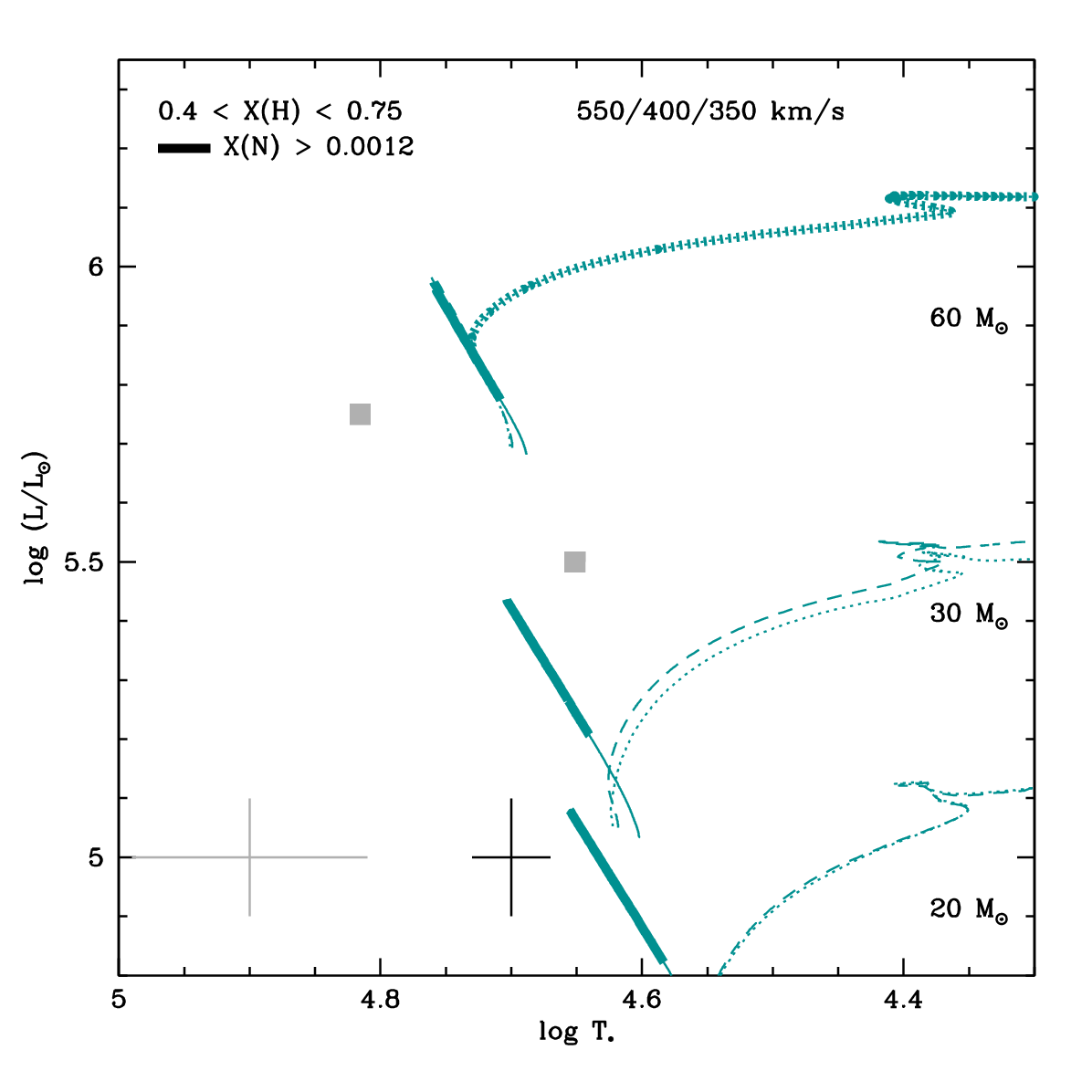}

\includegraphics[width=0.3\textwidth]{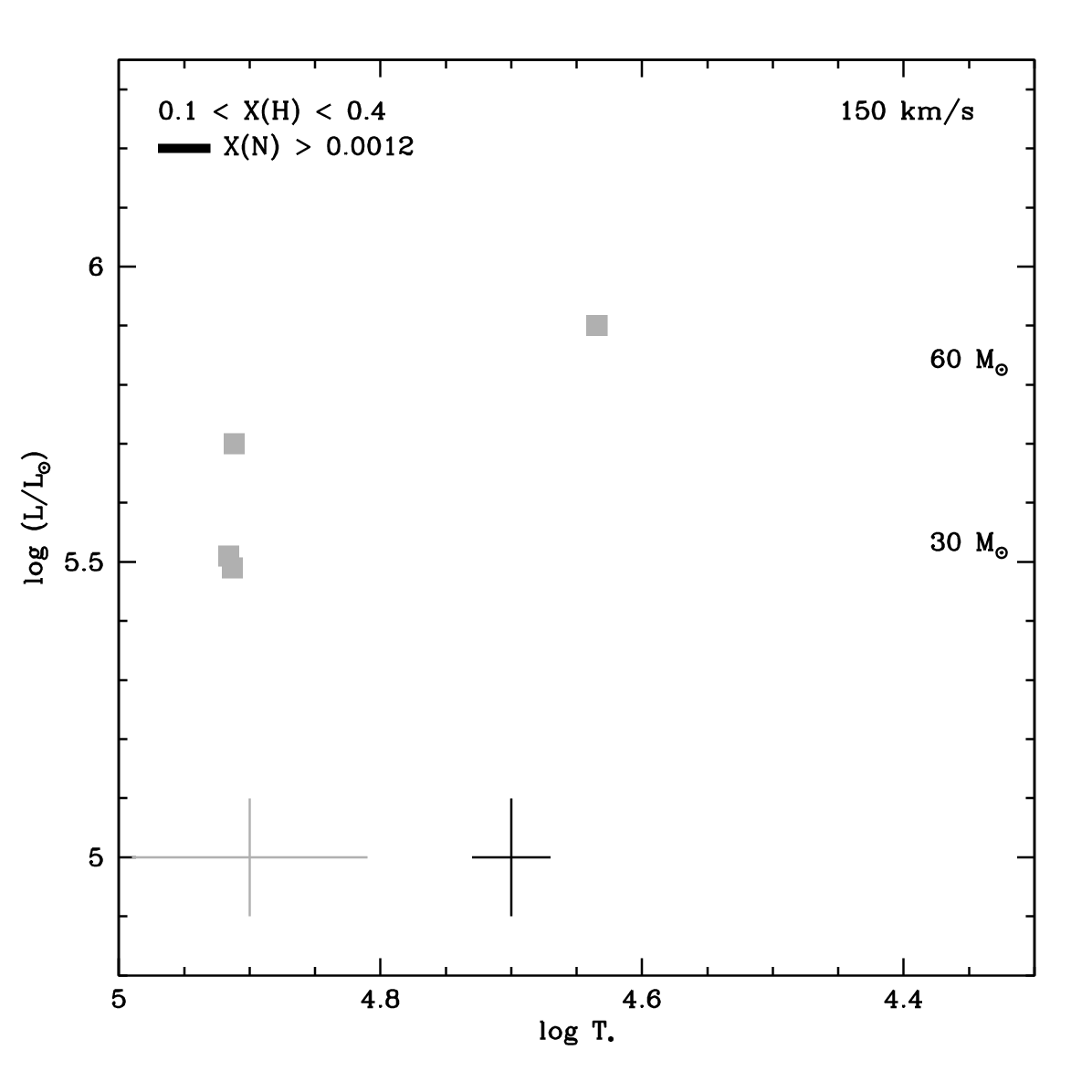}
\includegraphics[width=0.3\textwidth]{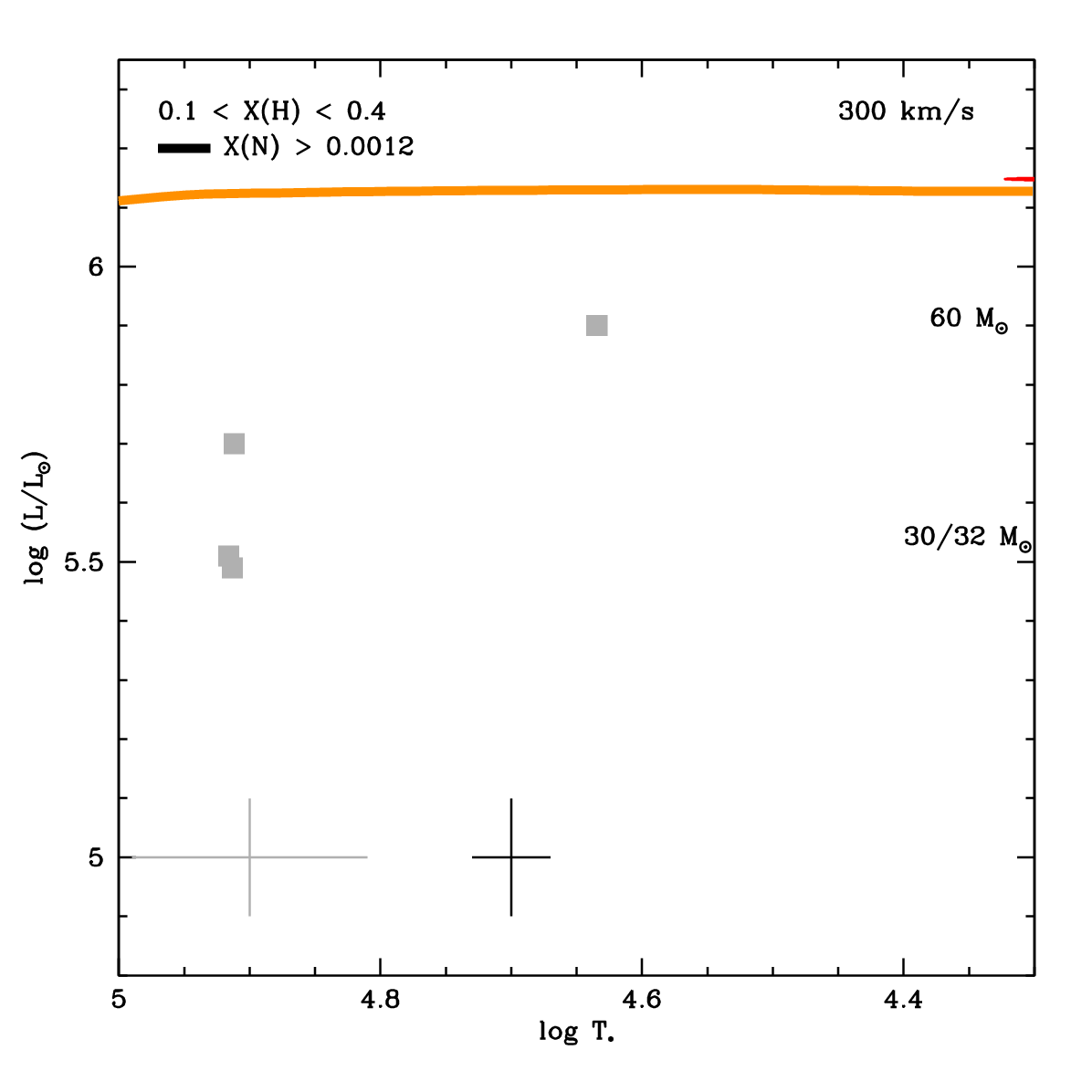}
\includegraphics[width=0.3\textwidth]{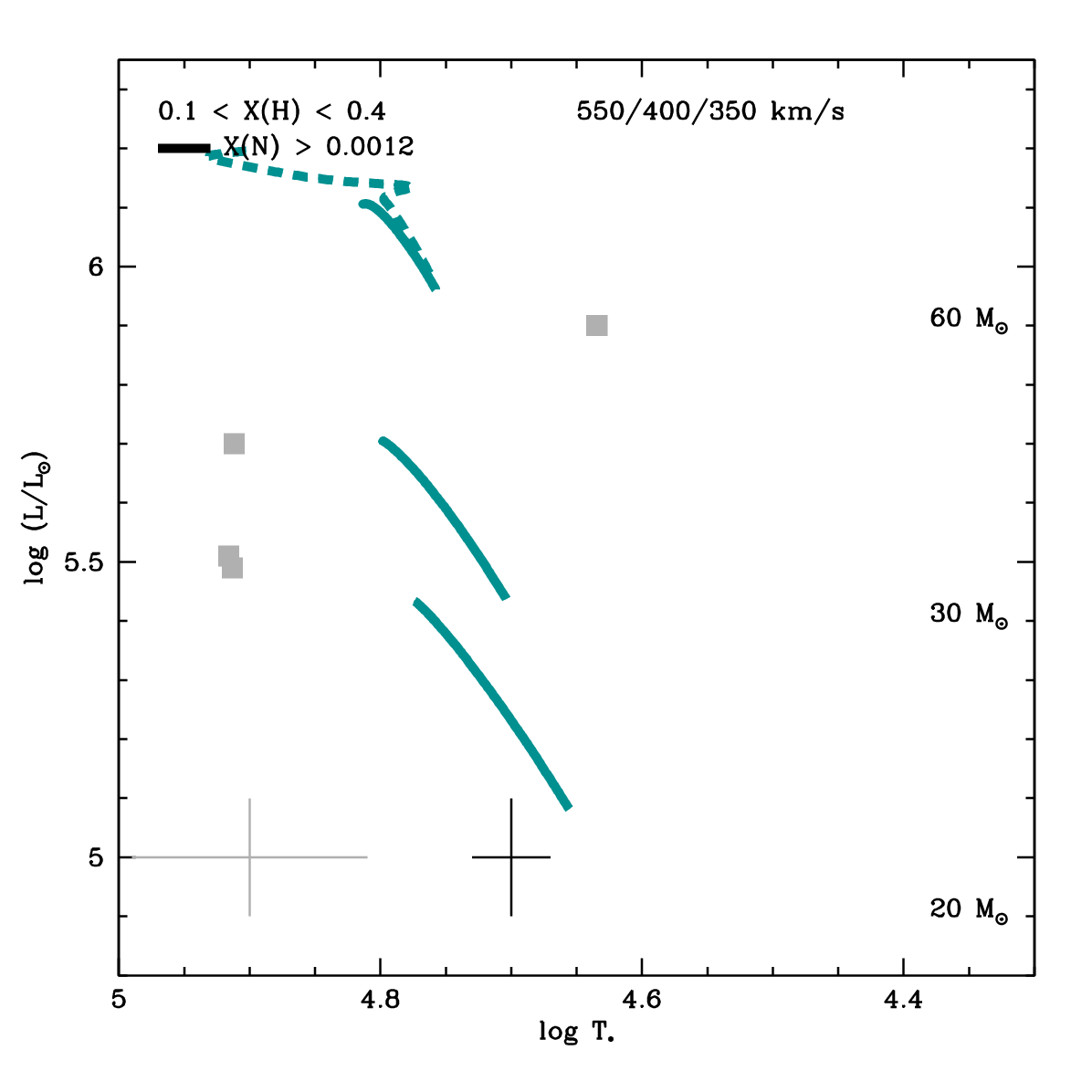}
\caption{Same as Fig.~\ref{hrd_smc} except that the bold part of tracks is for a nitrogen mass fraction  of greater than 0.0012.}
\label{hrd_smc2}
\end{figure}

Figures~\ref{hrd_gal2} to \ref{hrd_smc2} show the HRD for the sample stars with evolutionary tracks in which the minimum nitrogen mass fraction for bold coding has been changed from 0.0043 to 0.0025 in the Galaxy, from 0.0047 to 0.0032 in the LMC and from 0.0018 to 0.0012 in the SMC. These new values correspond to the minimum values of all stars as seen in Table~\ref{tab_param} and Fig.~\ref{histoxn}. See section~\ref{s_sing} for more details and discussion.

\FloatBarrier
\newpage

\twocolumn

\section{Best fit}
\label{ap_bf}

In this section, we gather the figures showing the best-fit models (in red) compared to the observational data (in black).

\begin{figure}[!h]
\centering
\includegraphics[width=0.49\textwidth]{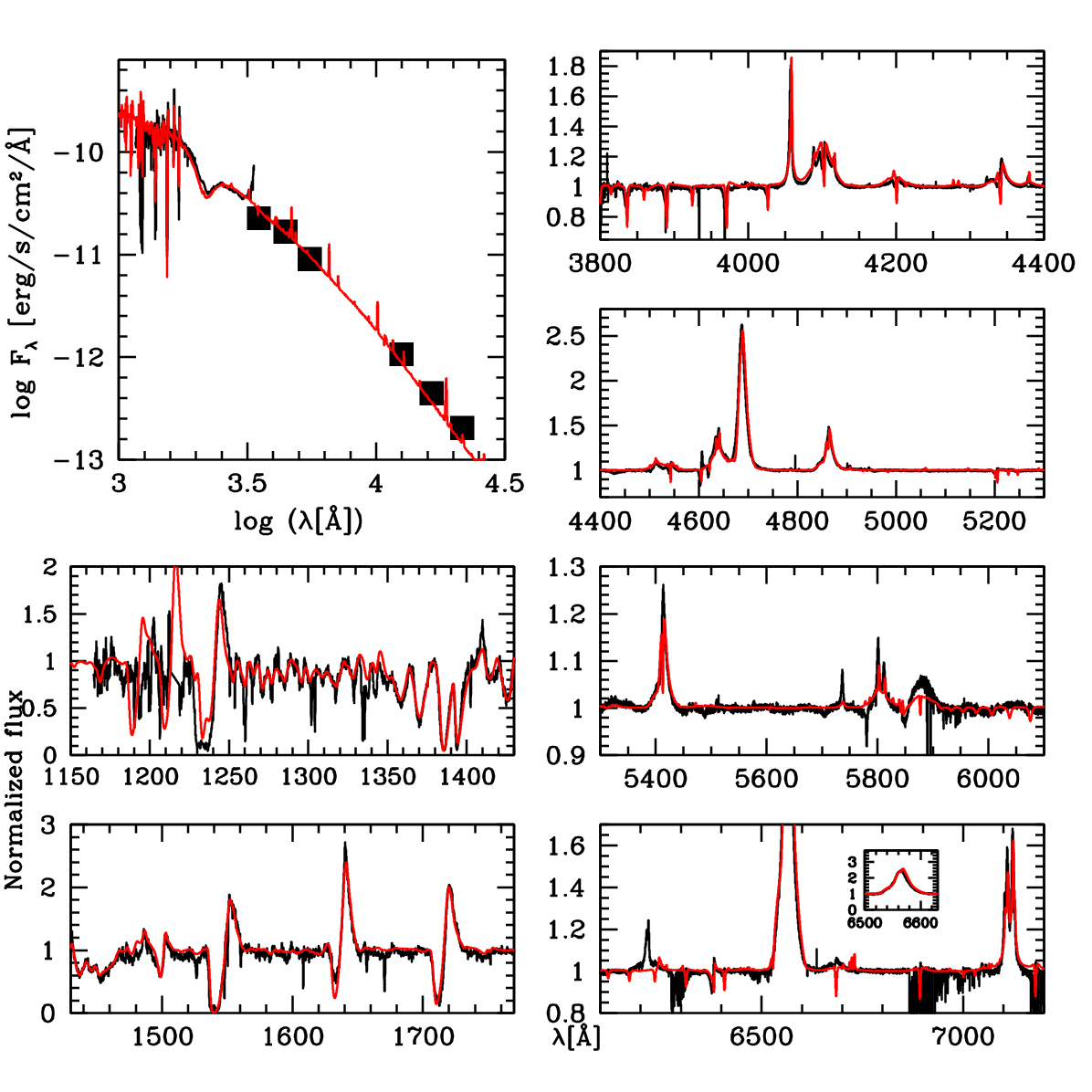}
\caption{Comparison of the observed spectrum of WR24 (black) with the best-fit model (red).}
\label{fig_wr24}
\end{figure}

\begin{figure}[!h]
\centering
\includegraphics[width=0.49\textwidth]{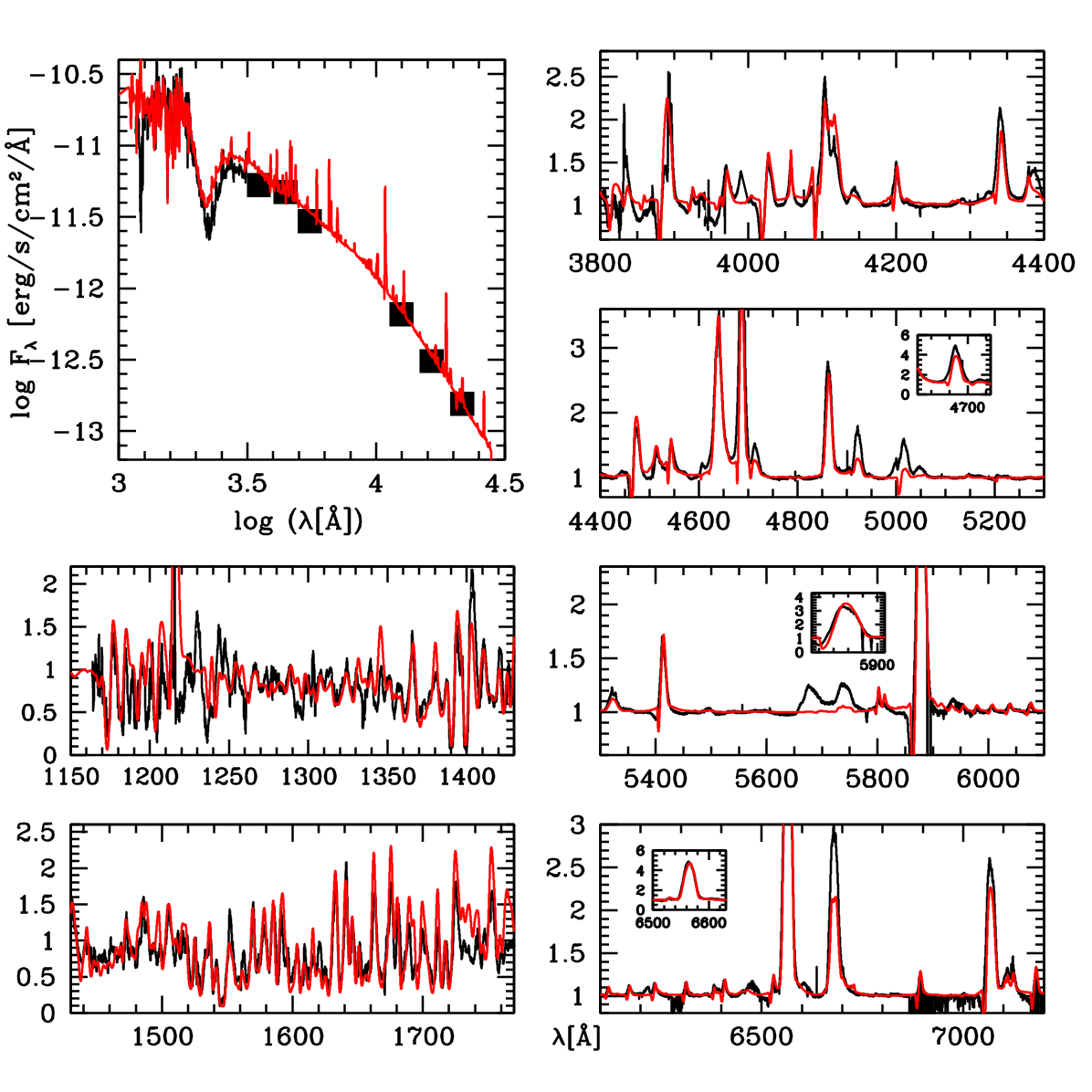}
\caption{Comparison of the observed spectrum of WR40 (black) with the best-fit model (red).}
\label{fig_wr40}
\end{figure}

\begin{figure}[!h]
\centering
\includegraphics[width=0.49\textwidth]{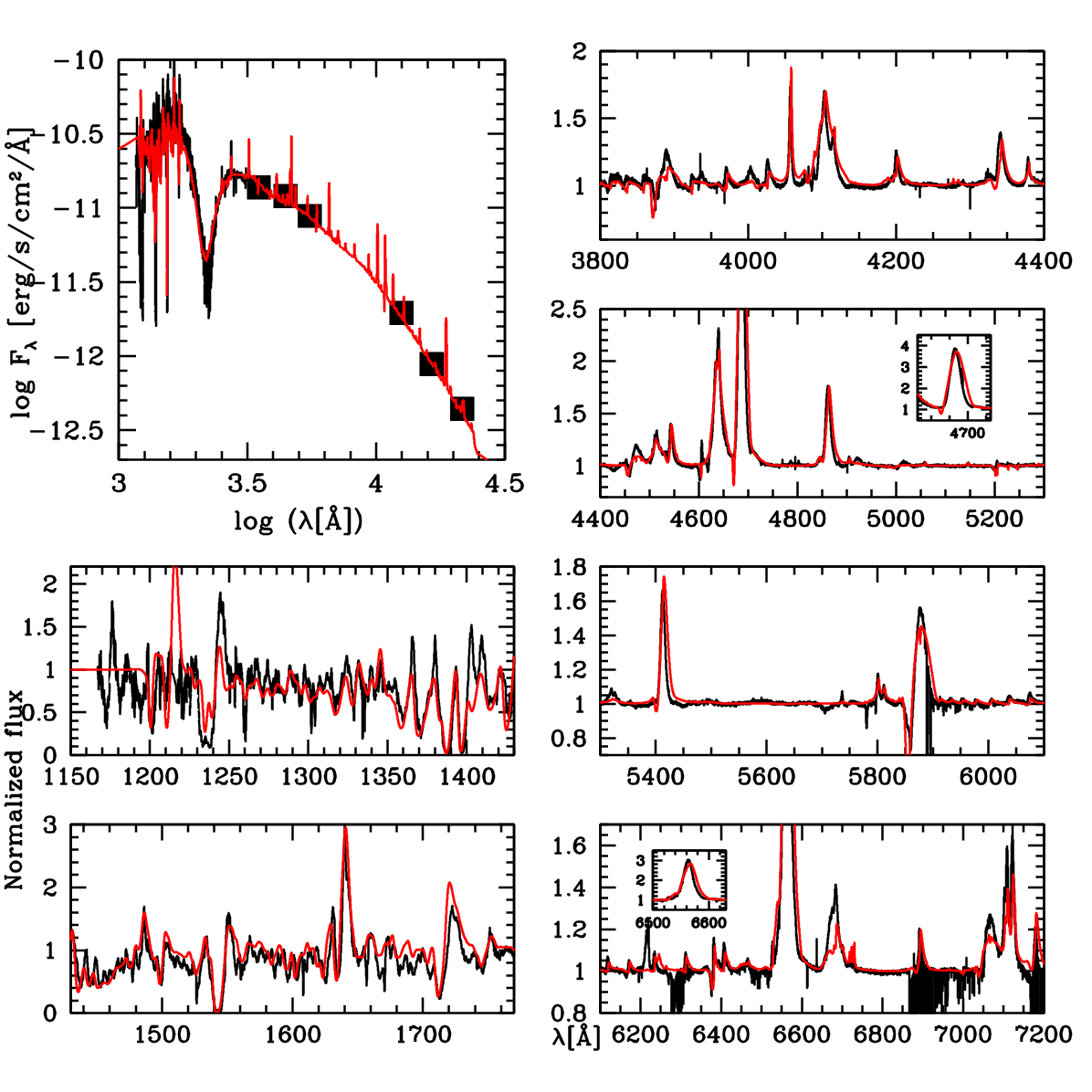}
\caption{Comparison of the observed spectrum of WR78 (black) with the best-fit model (red).}
\label{fig_wr78}
\end{figure}

\begin{figure}[!h]
\centering
\includegraphics[width=0.49\textwidth]{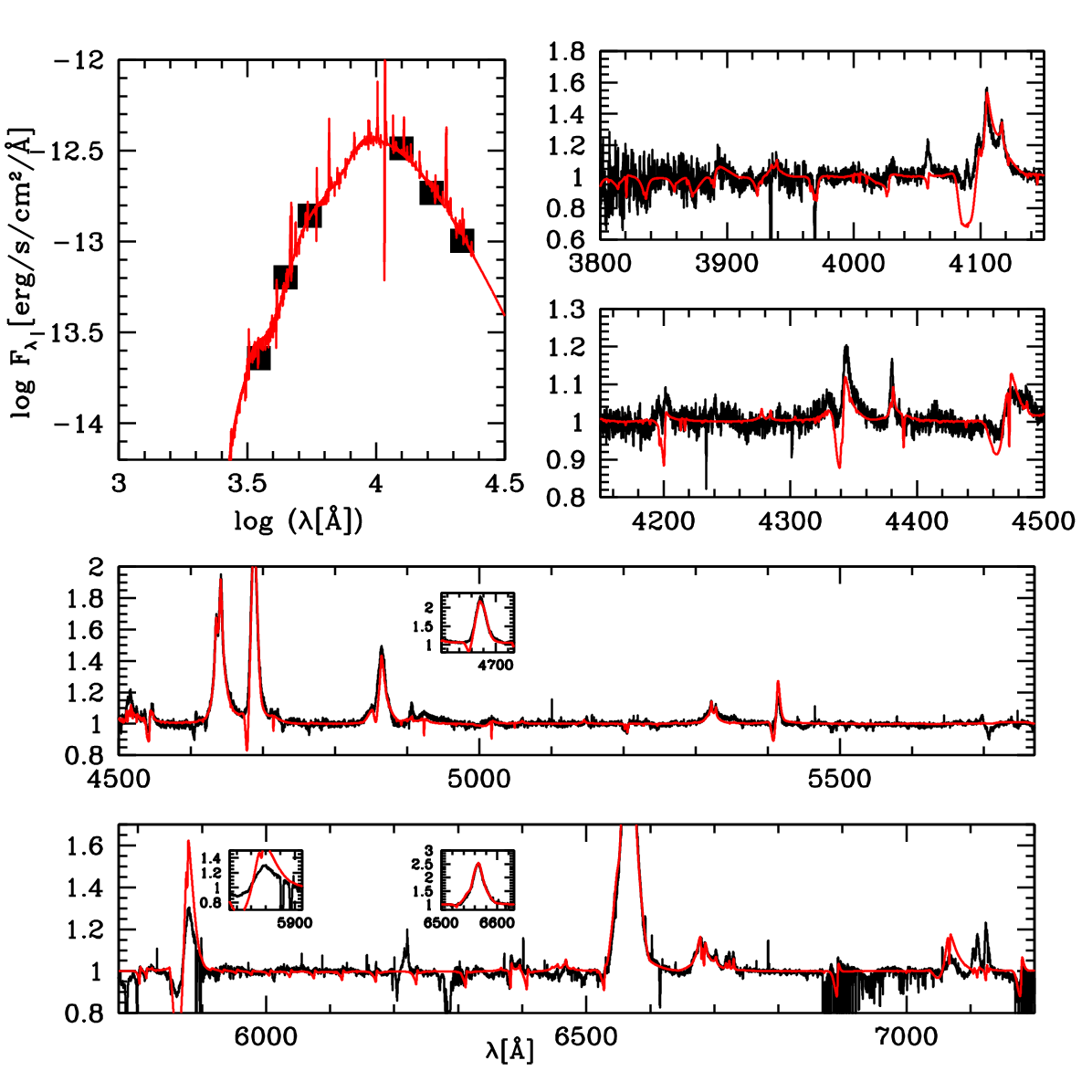}
\caption{Comparison of the observed spectrum of WR89 (black) with the best-fit model (red).}
\label{fig_wr89}
\end{figure}

\FloatBarrier
\newpage

\begin{figure}[]
\centering
\includegraphics[width=0.45\textwidth]{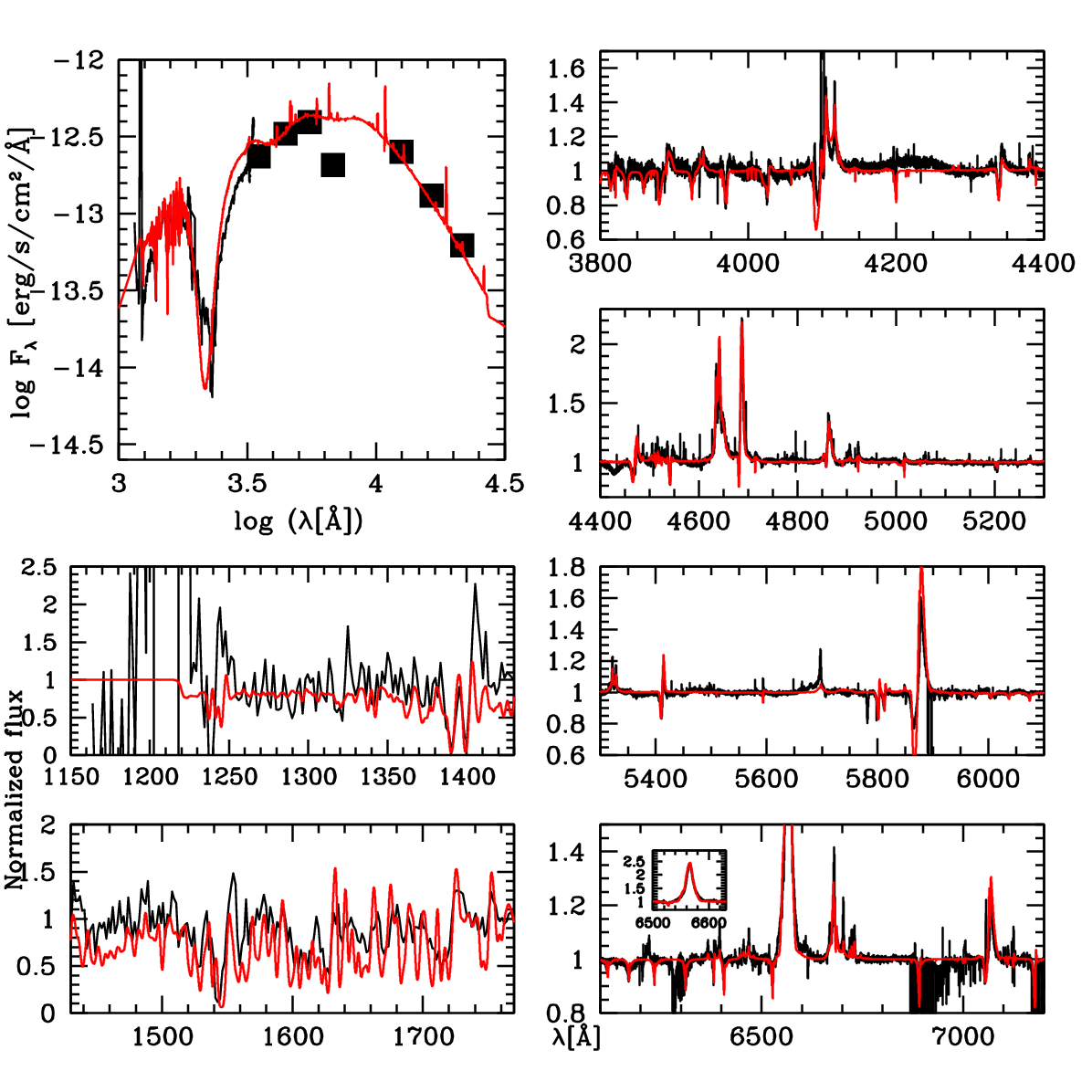}
\caption{Comparison of the observed spectrum of WR108 (black) with the best-fit model (red).}
\label{fig_wr108}
\end{figure}

\begin{figure}[]
\centering
\includegraphics[width=0.45\textwidth]{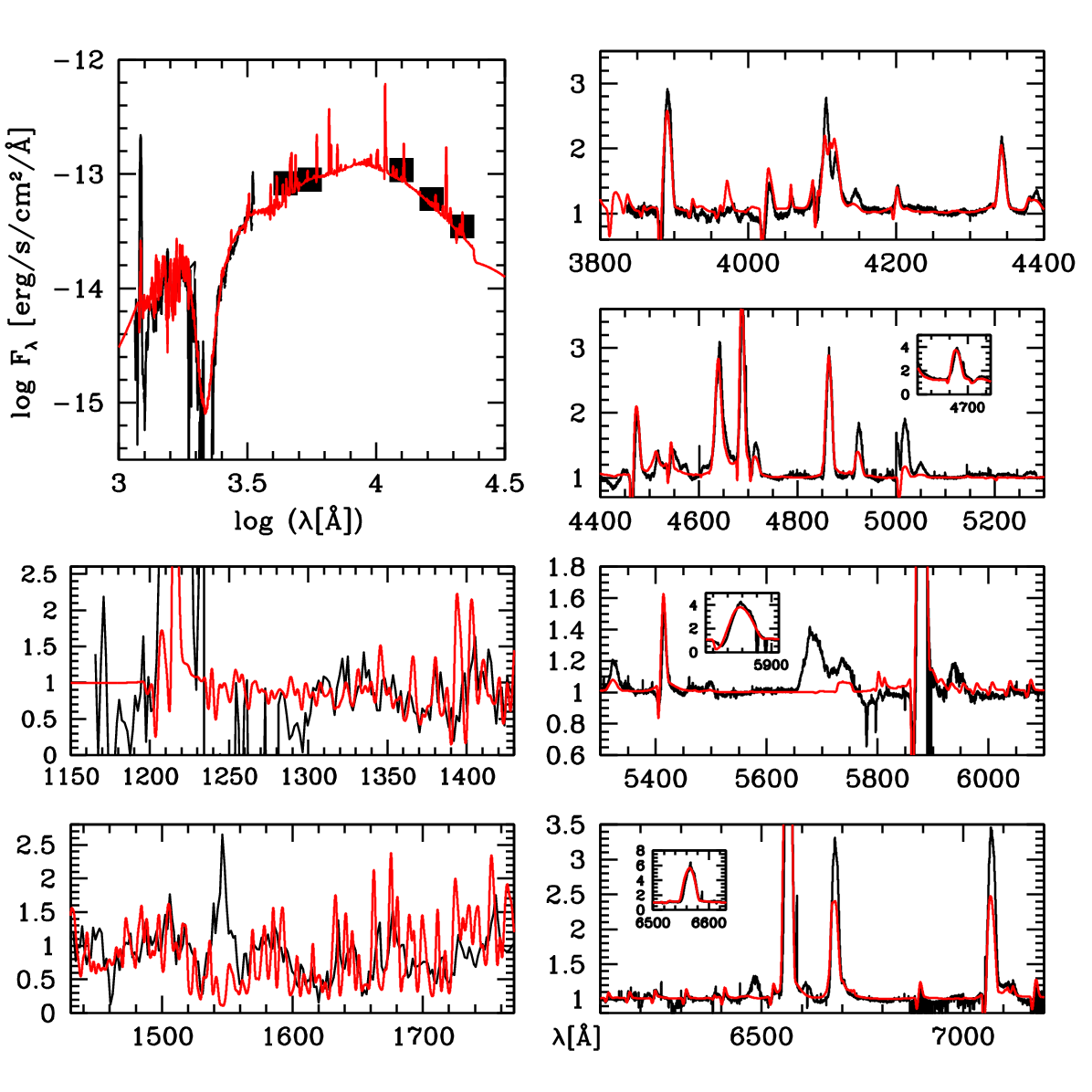}
\caption{Comparison of the observed spectrum of WR124 (black) with the best-fit model (red).}
\label{fig_wr124}
\end{figure}

\begin{figure}[]
\centering
\includegraphics[width=0.45\textwidth]{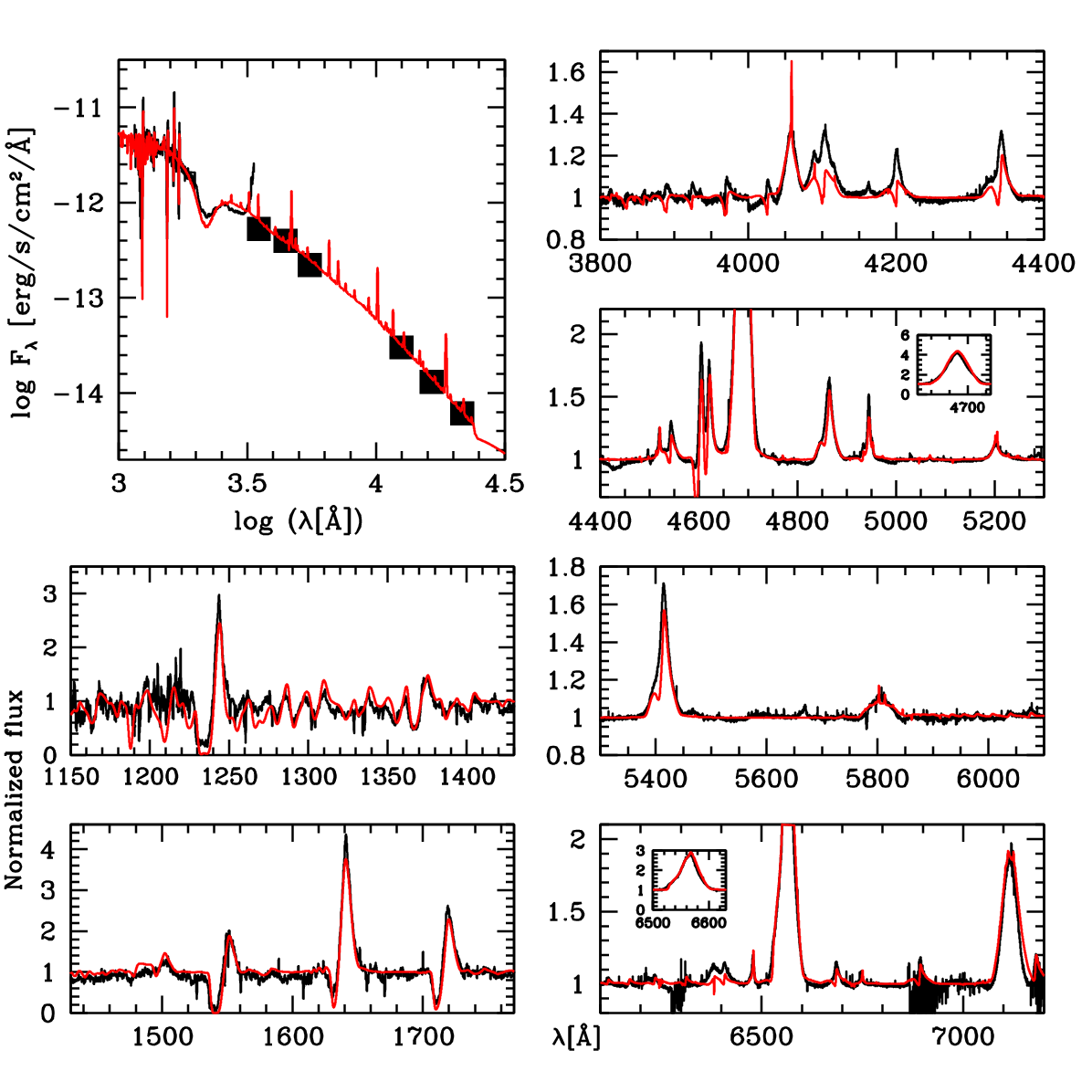}
\caption{Comparison of the observed spectrum of WR128 (black) with the best-fit model (red).}
\label{fig_wr128}
\end{figure}

\begin{figure}[]
\centering
\includegraphics[width=0.45\textwidth]{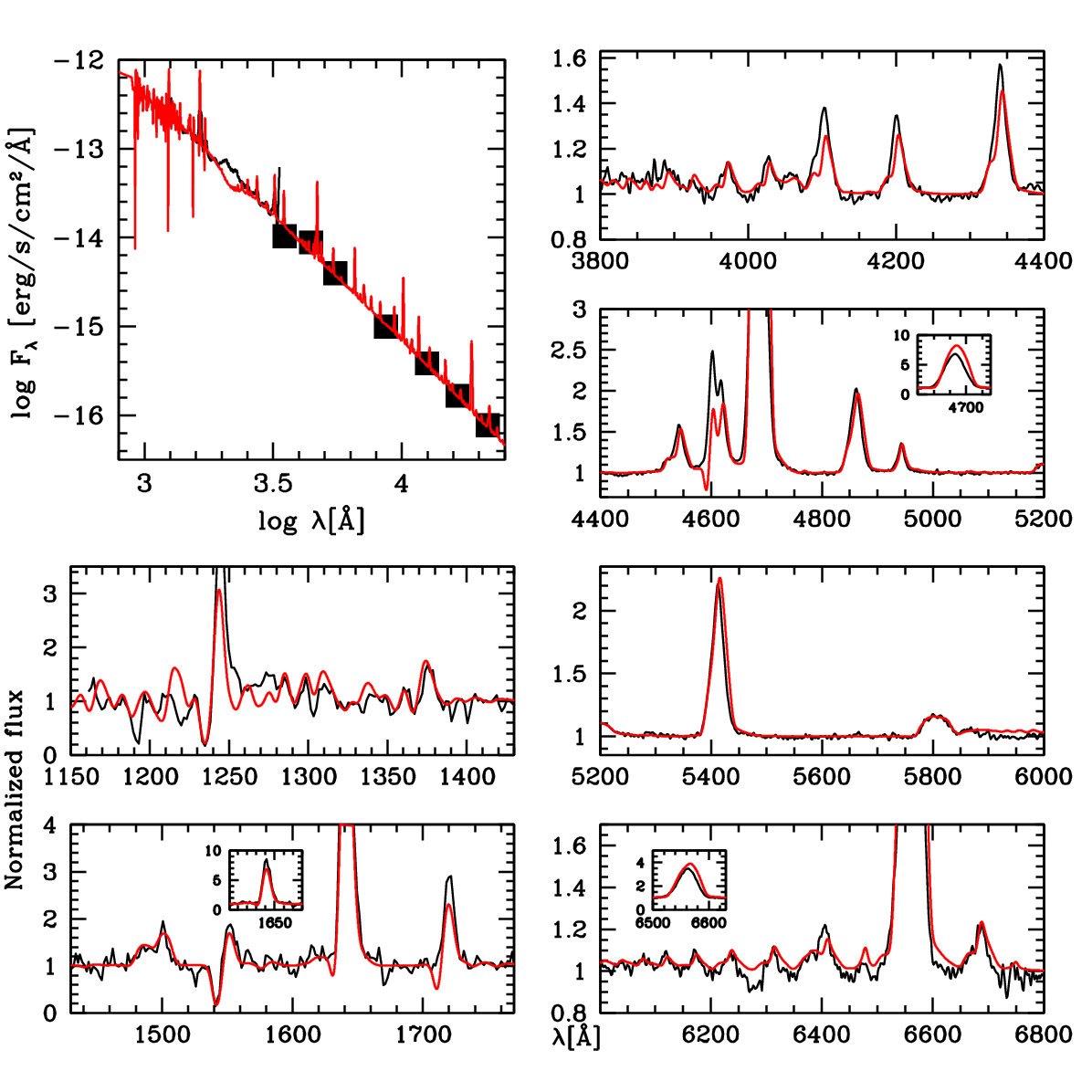}
\caption{Comparison of the observed spectrum of BAT99-35 (black) with the best-fit model (red).}
\label{fig_bat99_35}
\end{figure}

\begin{figure}[]
\centering
\includegraphics[width=0.45\textwidth]{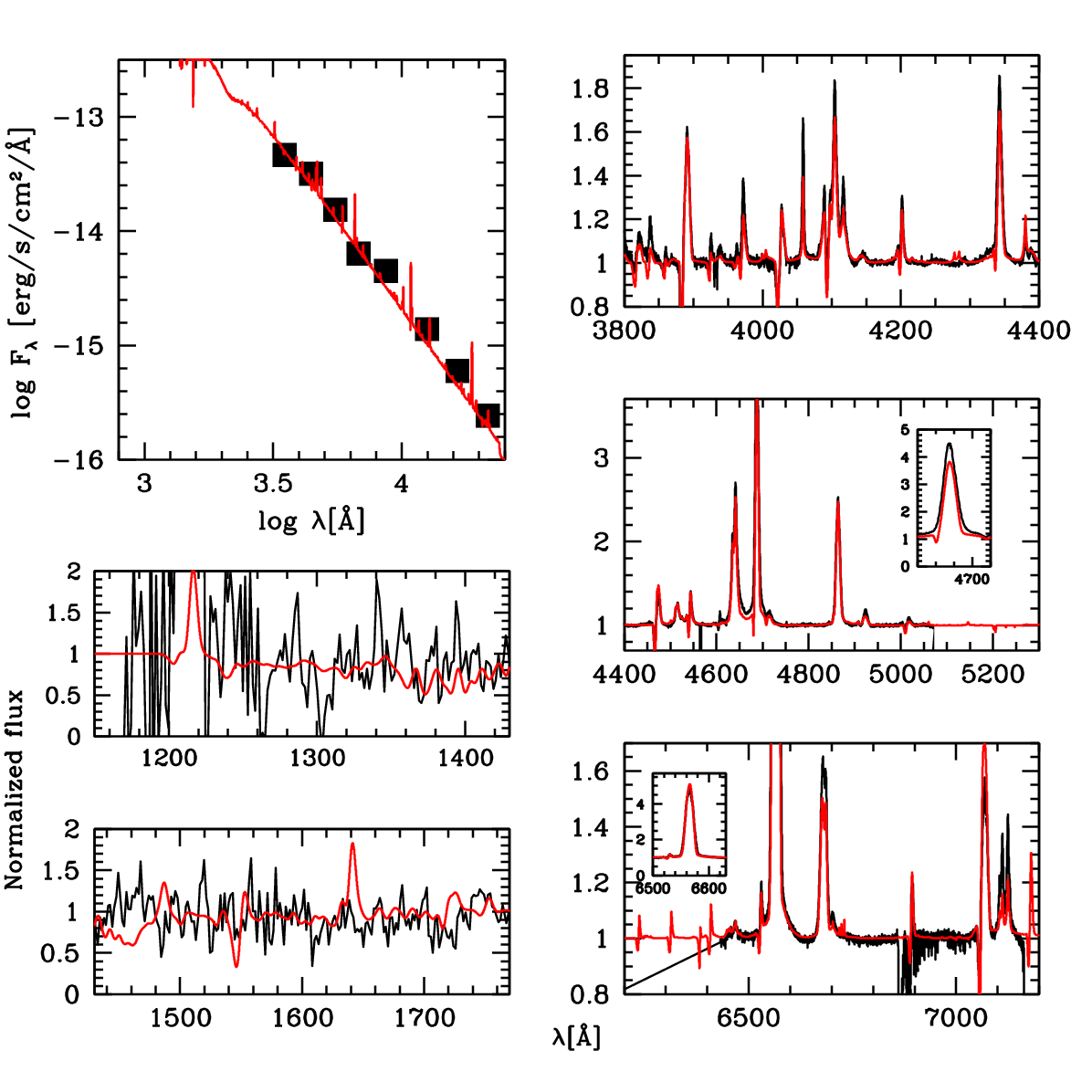}
\caption{Comparison of the observed spectrum of BAT99-44 (black) with the best-fit model (red).}
\label{fig_bat99_44}
\end{figure}

\begin{figure}[]
\centering
\includegraphics[width=0.45\textwidth]{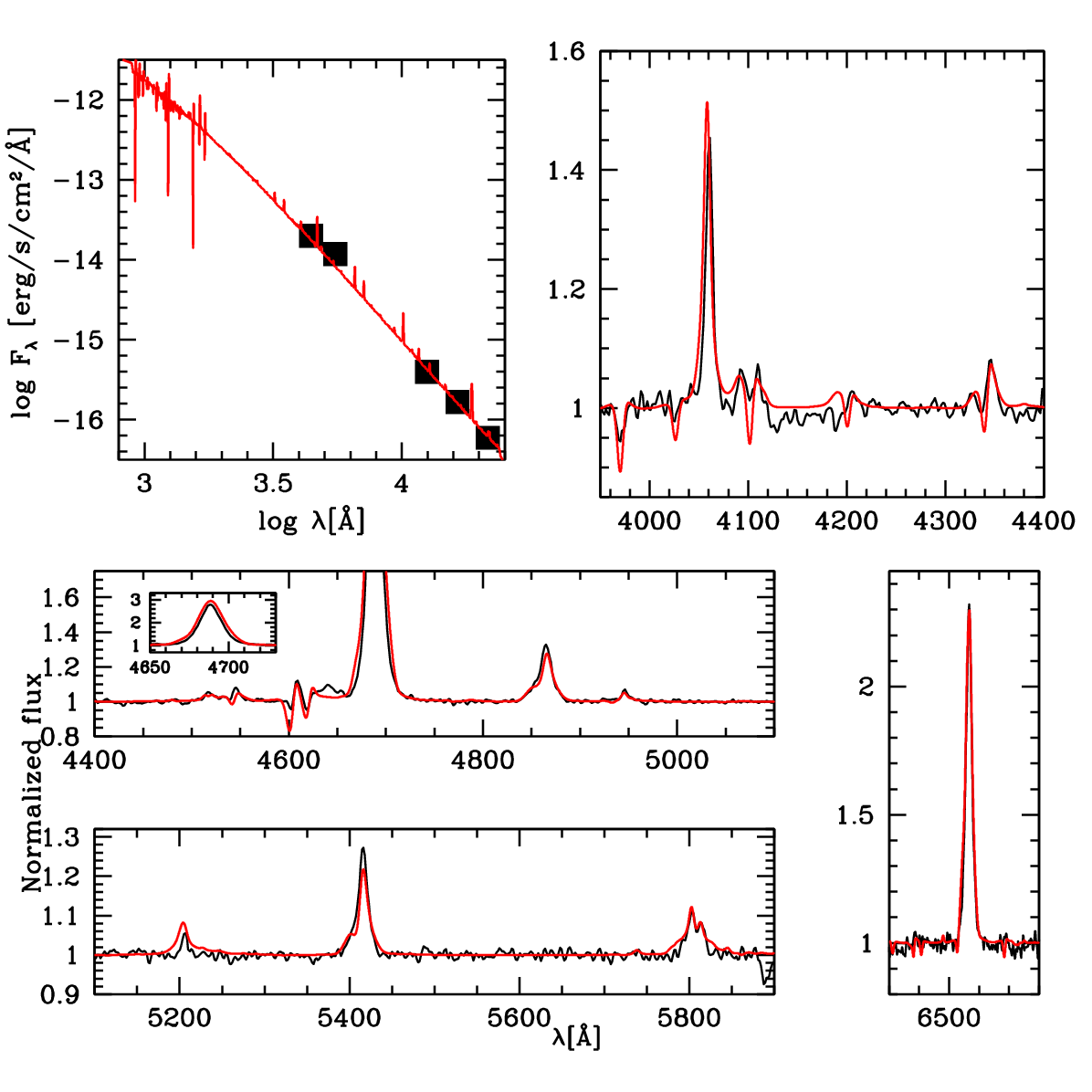}
\caption{Comparison of the observed spectrum of BAT99-50 (black) with the best-fit model (red).}
\label{fig_bat99_50}
\end{figure}

\begin{figure}[]
\centering
\includegraphics[width=0.45\textwidth]{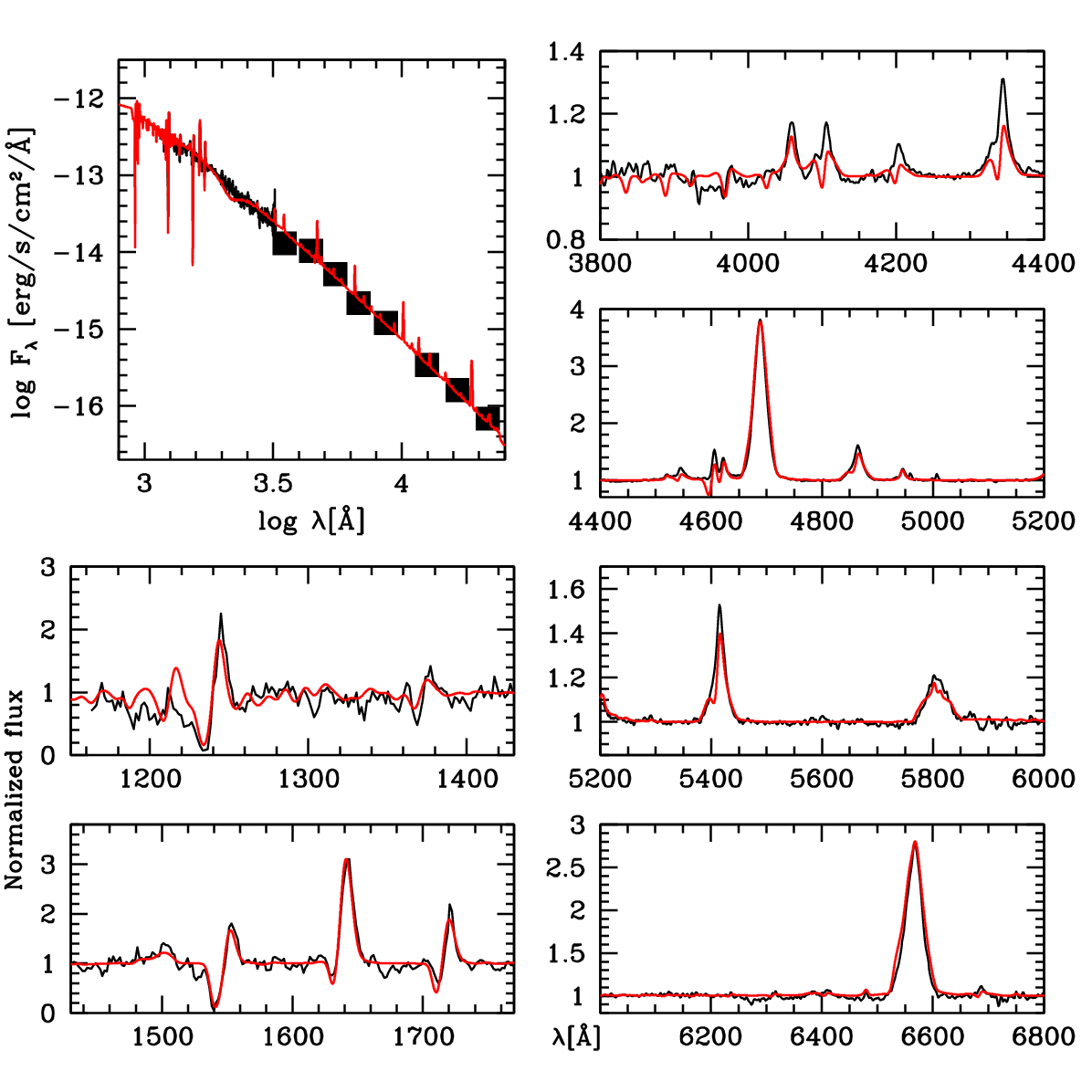}
\caption{Comparison of the observed spectrum of BAT99-63 (black) with the best-fit model (red).}
\label{fig_bat99_63}
\end{figure}

\begin{figure}[]
\centering
\includegraphics[width=0.45\textwidth]{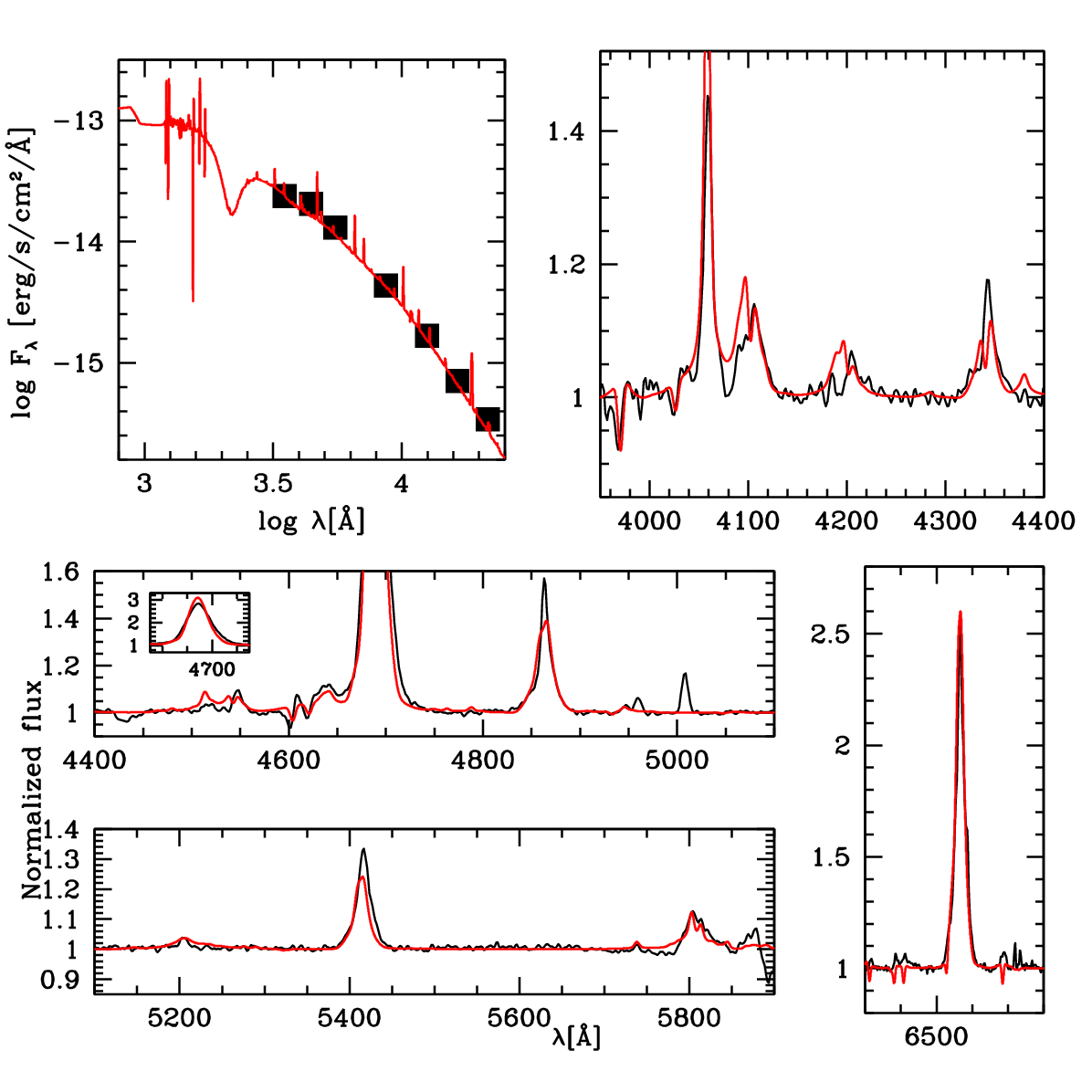}
\caption{Comparison of the observed spectrum of BAT99-67 (black) with the best-fit model (red).}
\label{fig_bat99_67}
\end{figure}

\begin{figure}[]
\centering
\includegraphics[width=0.45\textwidth]{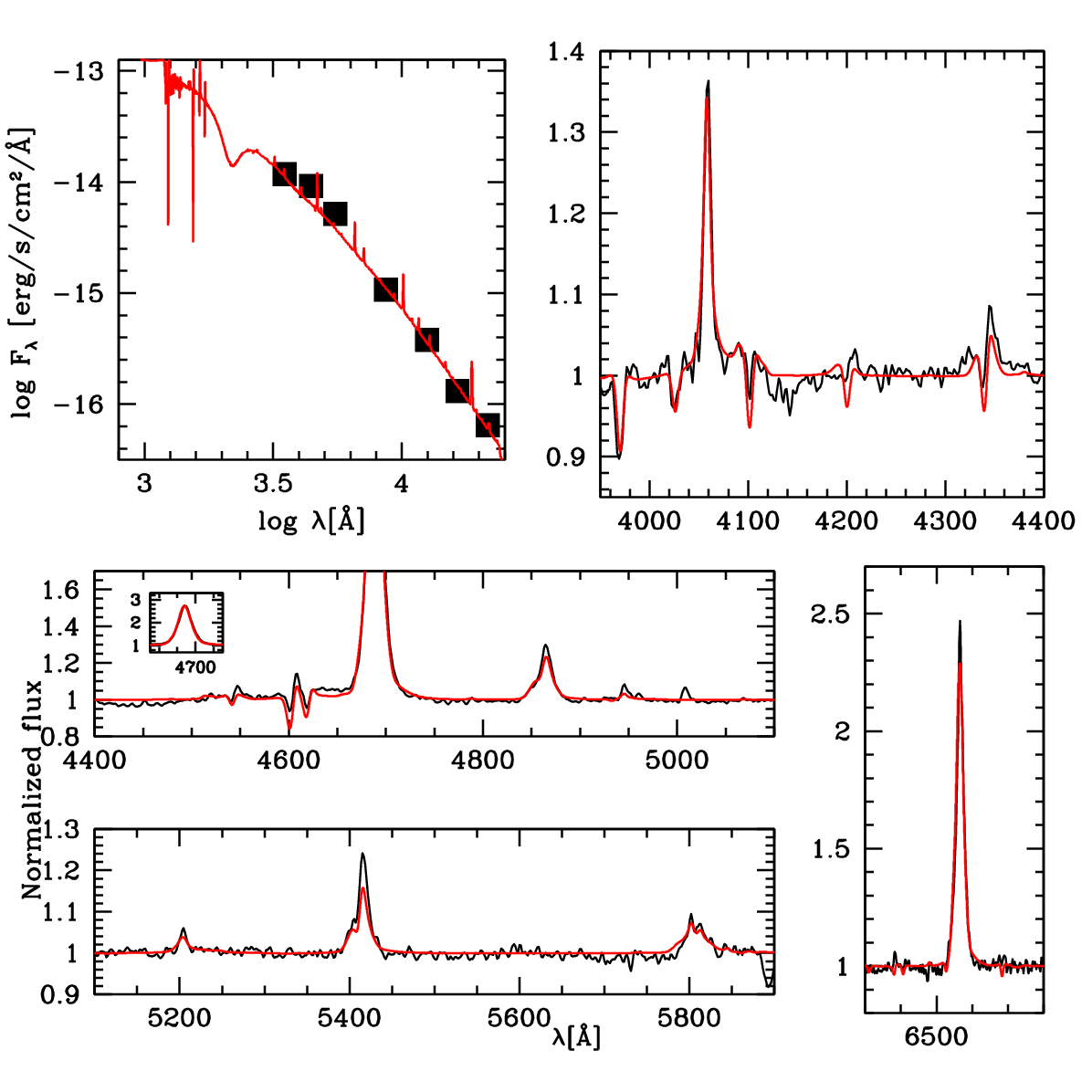}
\caption{Comparison of the observed spectrum of BAT99-73 (black) with the best-fit model (red).}
\label{fig_bat99_73}
\end{figure}

\begin{figure}[]
\centering
\includegraphics[width=0.45\textwidth]{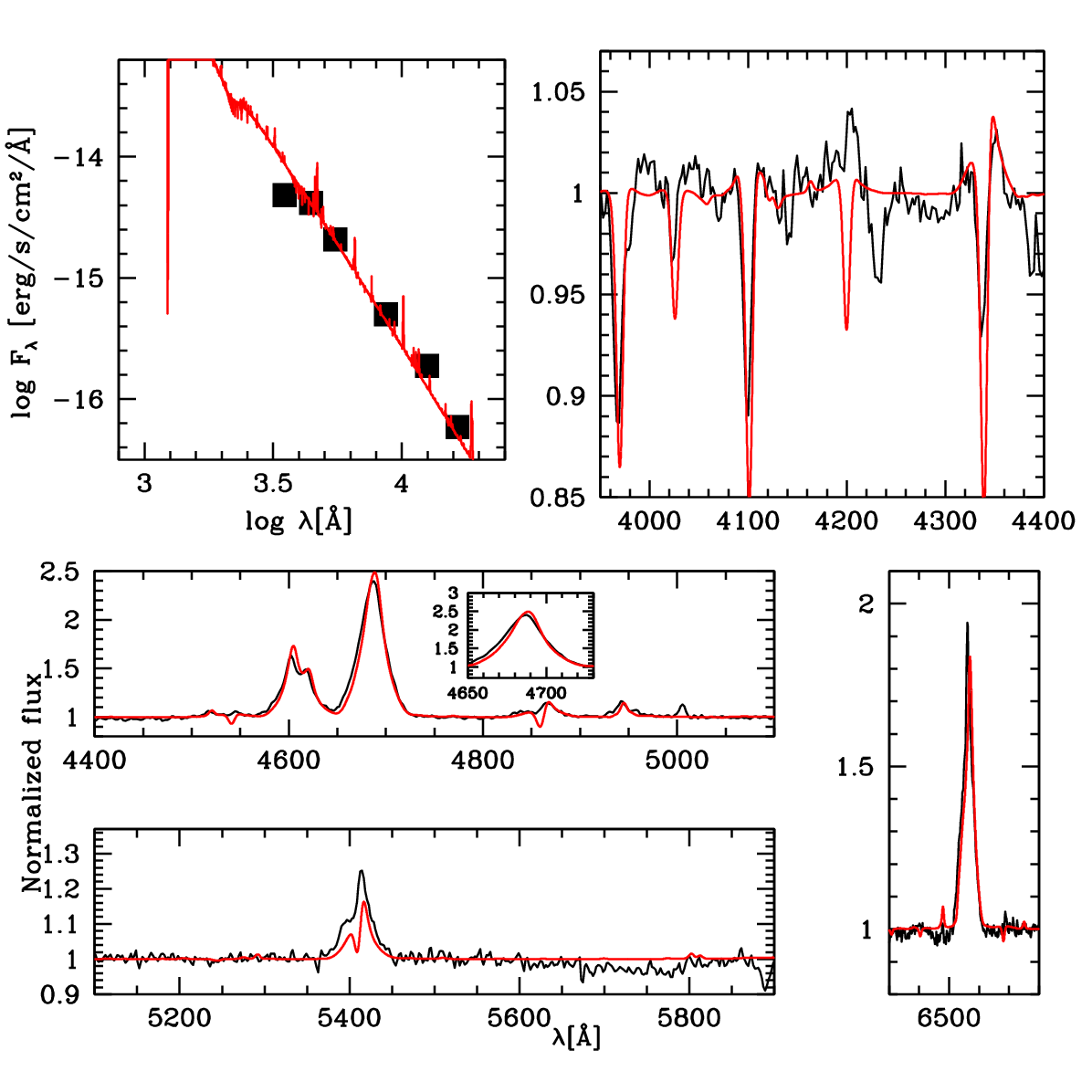}
\caption{Comparison of the observed spectrum of BAT99-74 (black) with the best-fit model (red).}
\label{fig_bat99_74}
\end{figure}

\begin{figure}[]
\centering
\includegraphics[width=0.45\textwidth]{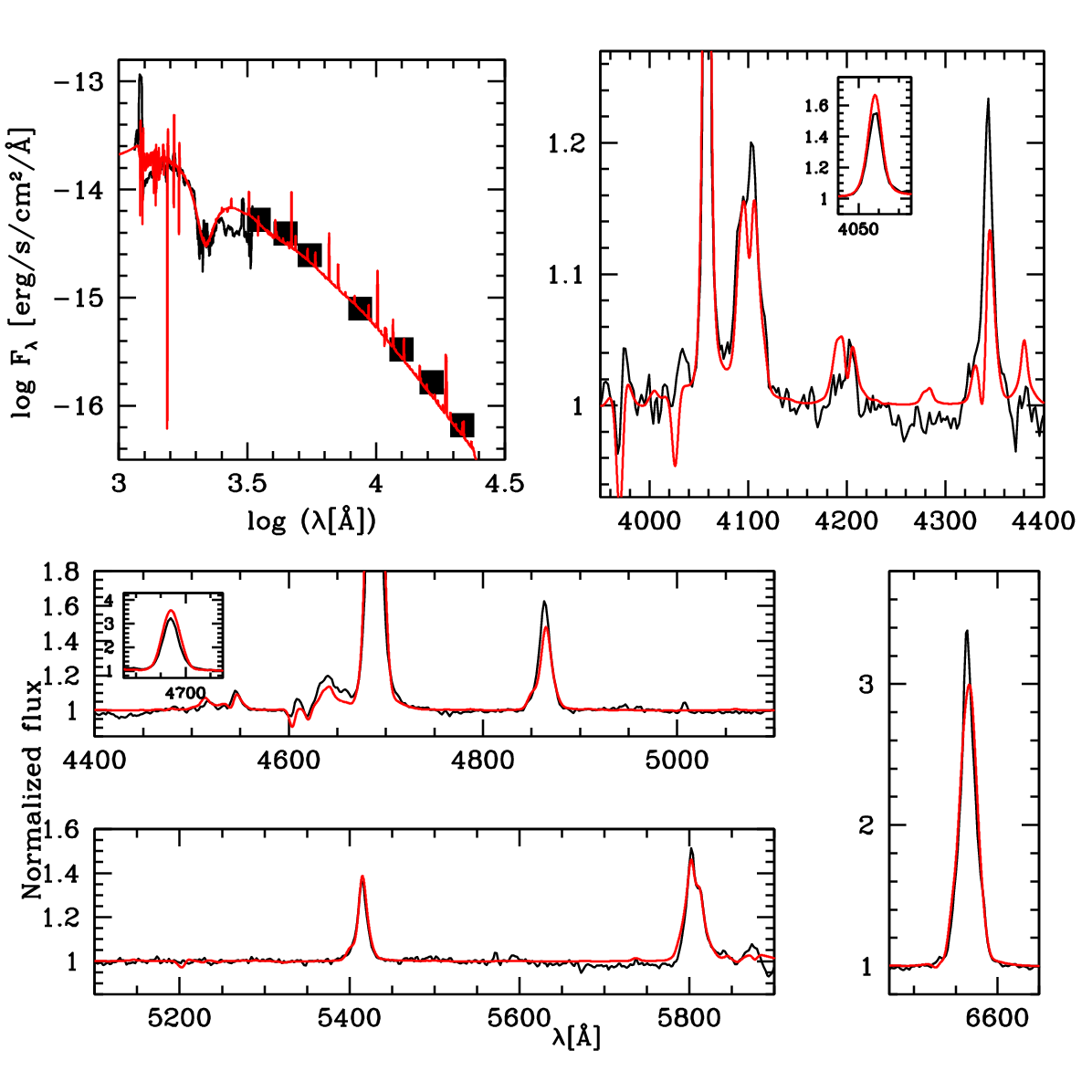}
\caption{Comparison of the observed spectrum of BAT99-81 (black) with the best-fit model (red).}
\label{fig_bat99_81}
\end{figure}

\begin{figure}[]
\centering
\includegraphics[width=0.45\textwidth]{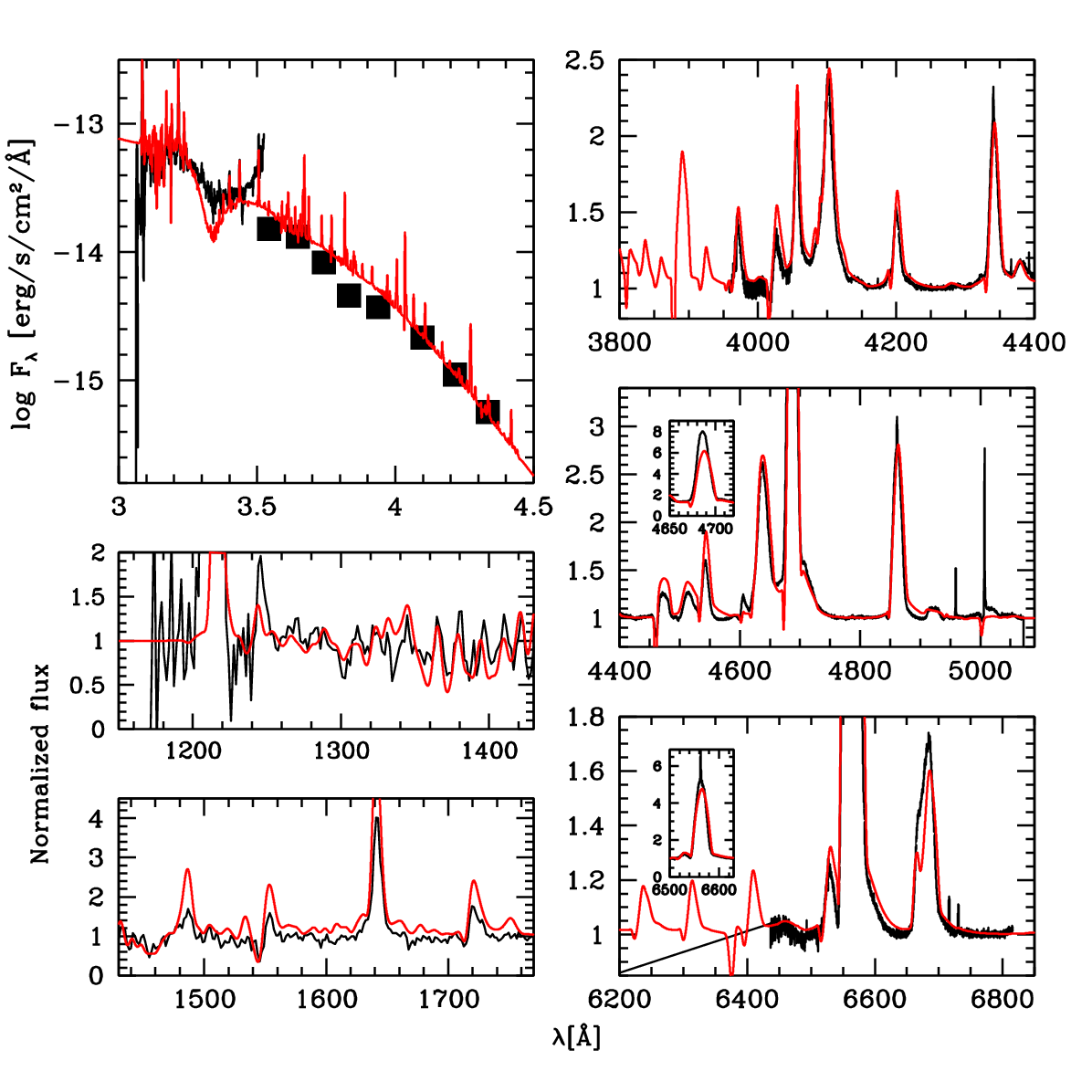}
\caption{Comparison of the observed spectrum of BAT99-89 (black) with the best-fit model (red).}
\label{fig_bat99_89}
\end{figure}

\begin{figure}[]
\centering
\includegraphics[width=0.45\textwidth]{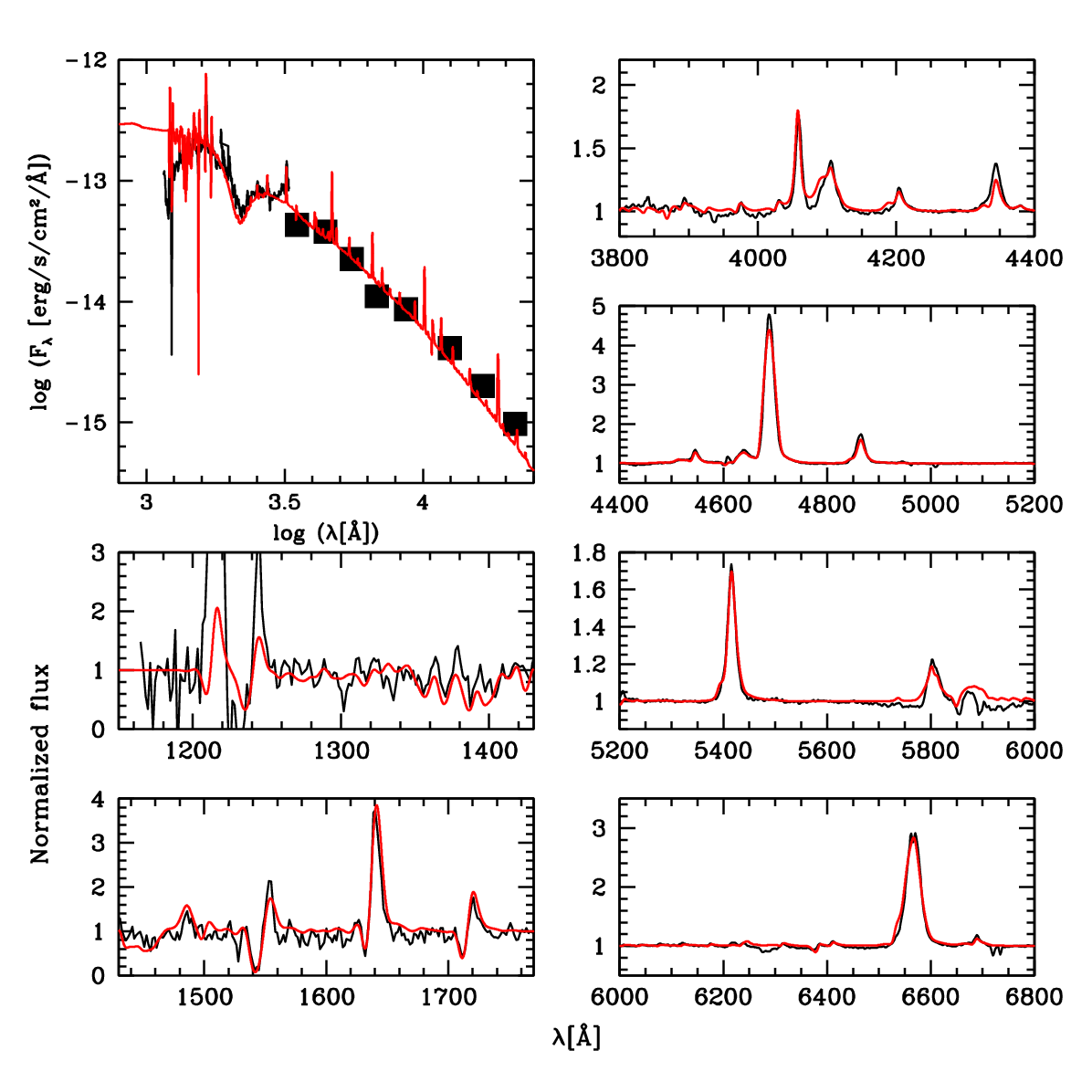}
\caption{Comparison of the observed spectrum of BAT99-122 (black) with the best-fit model (red).}
\label{fig_bat99_122}
\end{figure}

\begin{figure}[]
\centering
\includegraphics[width=0.45\textwidth]{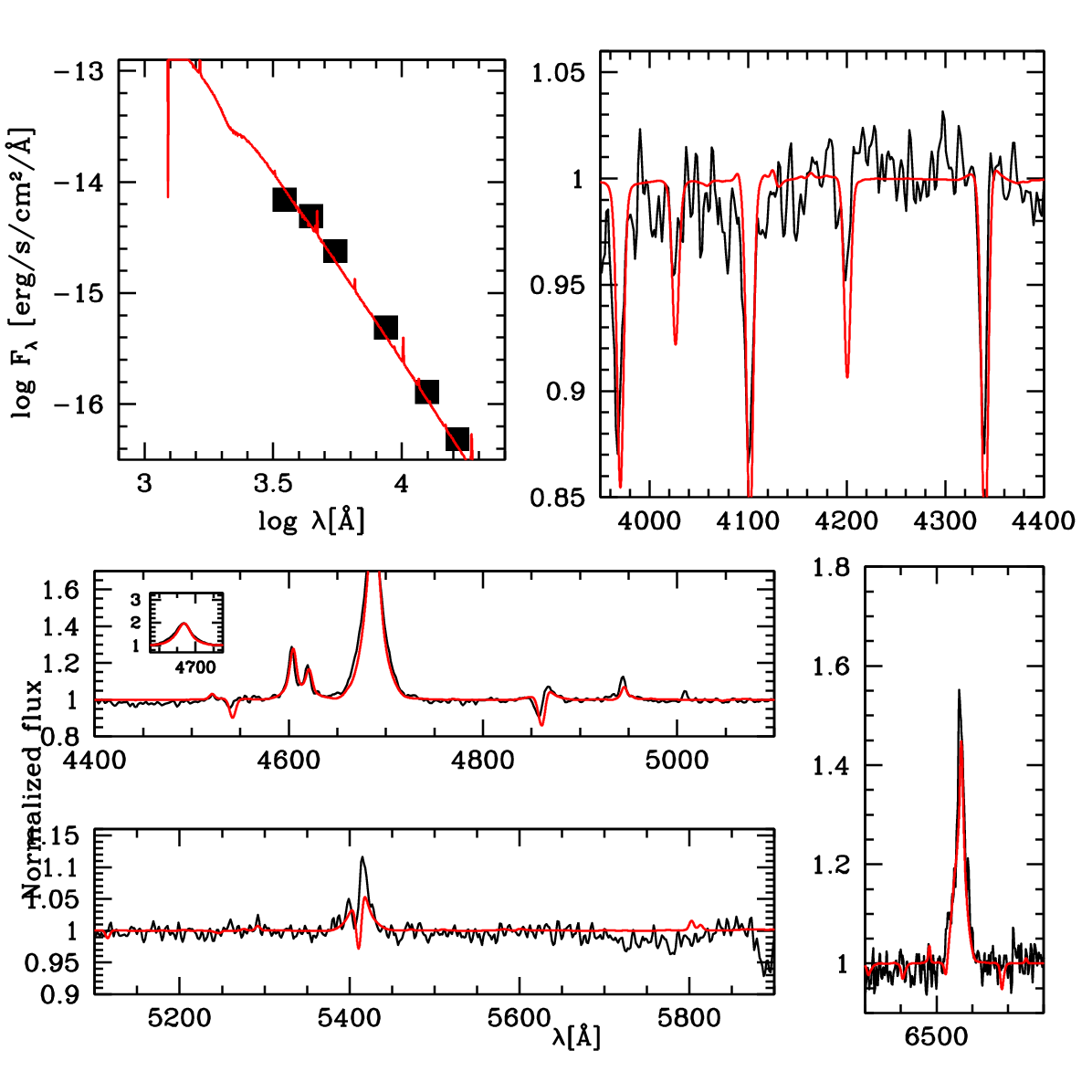}
\caption{Comparison of the observed spectrum of AB9 (black) with the best-fit model (red).}
\label{fig_ab9}
\end{figure}

\begin{figure}[]
\centering
\includegraphics[width=0.45\textwidth]{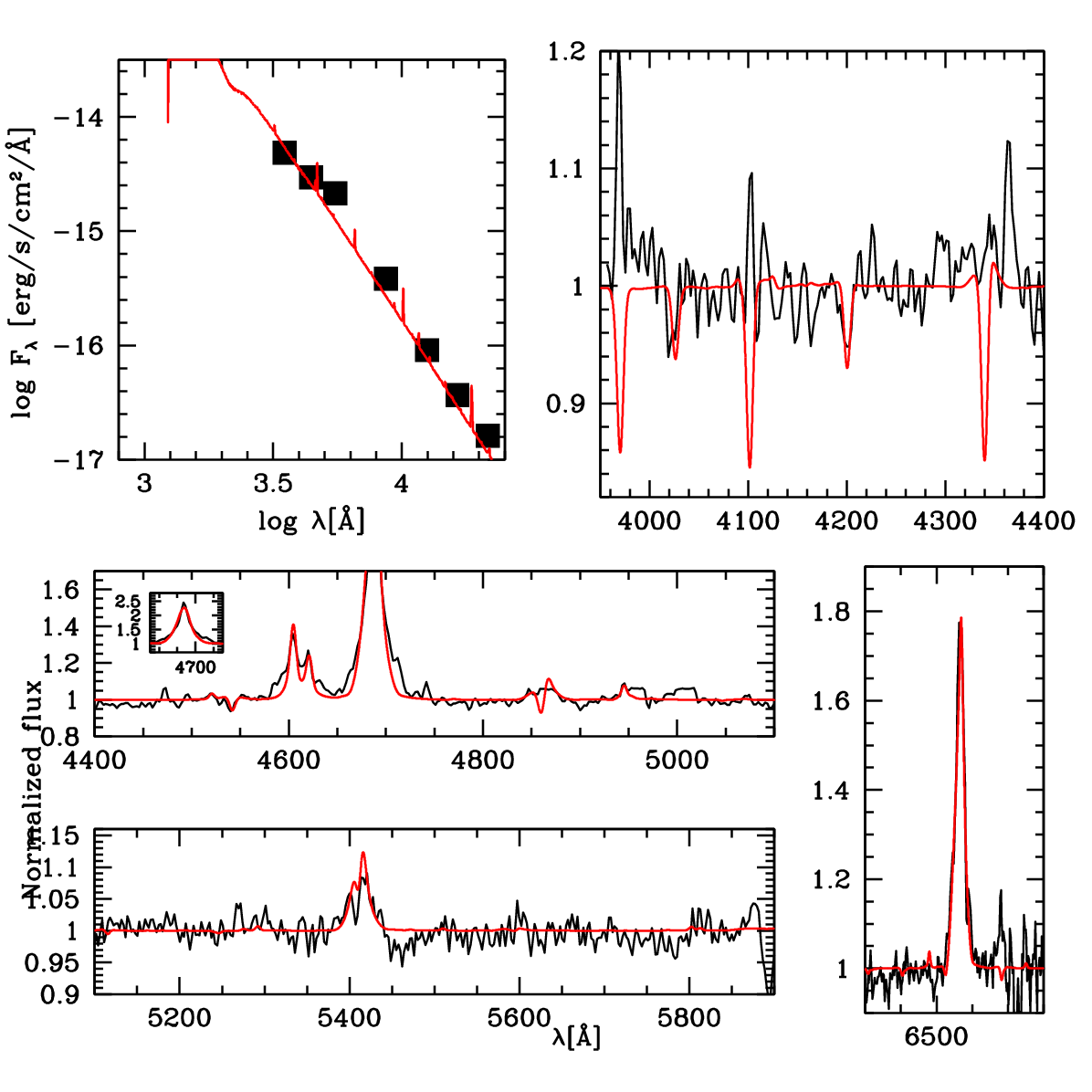}
\caption{Comparison of the observed spectrum of AB10 (black) with the best-fit model (red).}
\label{fig_ab10}
\end{figure}

\begin{figure}[]
\centering
\includegraphics[width=0.45\textwidth]{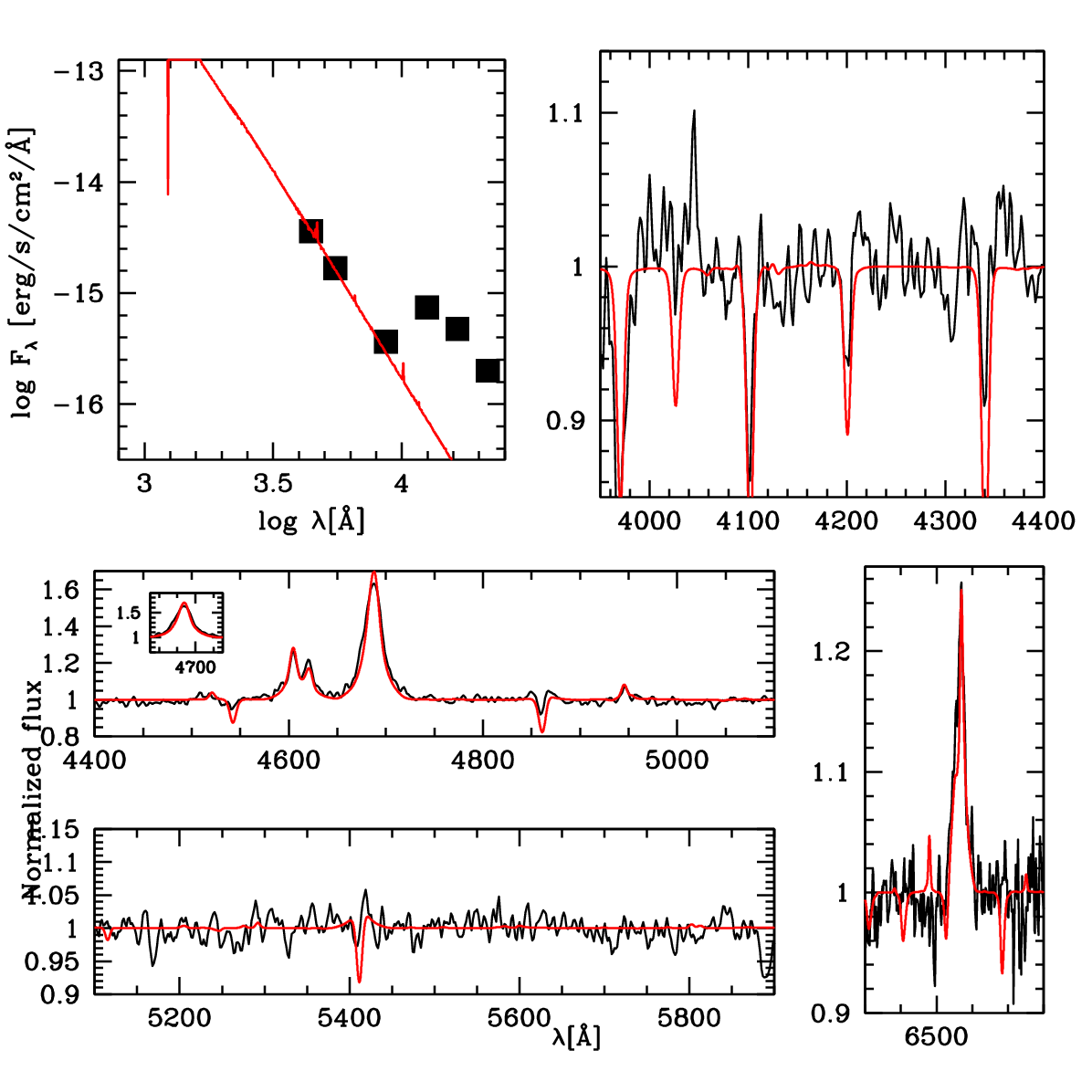}
\caption{Comparison of the observed spectrum of AB11 (black) with the best-fit model (red).}
\label{fig_ab11}
\end{figure}

\end{appendix}

\end{document}